\documentclass[]{emulateapj}
\usepackage{natbib}
\usepackage{apjfonts}
\usepackage{epsfig}

\newcommand{\gas}{\mathrm{gas}}

\newcommand{\Vvir}{V_{\mathrm{vir}}}

\newcommand{\vcirc}{v_{\mathrm{circ}}}

\newcommand{\Rds}{R_{\mathrm{d},\star}}
\newcommand{\Rdg}{R_{\mathrm{d},g}}
\newcommand{\abs}{a_{\mathrm{b},\star}}
\newcommand{\tcool}{t_{\mathrm{cool}}}  

\newcommand{\dd}{\mathrm{d}}

\newcommand{\Msun}{M_{\sun}}

\newcommand{\Mbulge}{M_{\mathrm{bulge}}}
\newcommand{\Mdisk}{M_{\mathrm{disk}}}
\newcommand{\Mds}{M_{\mathrm{d},\star}}
\newcommand{\Mdg}{M_{\mathrm{d},g}}
\newcommand{\Mbaryon}{M_{\mathrm{baryon}}}

\newcommand{\NDM}{N_{\mathrm{DM}}}

\newcommand{\Ndisks}{N_{\mathrm{d},\star}}
\newcommand{\Ndiskg}{N_{\mathrm{d},g}}
\newcommand{\Ngas}{N_{\mathrm{gas}}}
\newcommand{\Nbulge}{N_{\mathrm{bulge}}}
\newcommand{\kpc}{{\mathrm{kpc}}}
\newcommand{\Mpc}{{\mathrm{Mpc}}}
\newcommand{\pc}{{\mathrm{pc}}}
\newcommand{\yr}{{\mathrm{yr}}}

\newcommand{\Gyr}{\mathrm{Gyr}}

\newcommand{\Nneigh}{N_{\mathrm{neigh}}}
\newcommand{\mJeans}{m_{\mathrm{Jeans}}}
\newcommand{\NJeans}{N_{\mathrm{Jeans}}}
\newcommand{\lambdaJeans}{\lambda_{\mathrm{Jeans}}}
\newcommand{\mSPH}{m_{\mathrm{SPH}}}
\newcommand{\mgas}{m_{\mathrm{gas}}}
\newcommand{\hJeans}{h_{\mathrm{Jeans}}}
\newcommand{\Pext}{P_{\mathrm{ext}}}
\newcommand{\SigmaHmol}{\Sigma_{\mathrm{H2}}}
\newcommand{\SigmaSFR}{\Sigma_{\SFR}}
\newcommand{\Sigmag}{\Sigma_{\mathrm{gas}}}
\newcommand{\SigmaA}{\Sigma_{A}}
\newcommand{\SigmaQ}{\Sigma_{Q}}
\newcommand{\Qsg}{Q_{\mathrm{sg}}}
\newcommand{\Qs}{Q_{\mathrm{s}}}
\newcommand{\Qg}{Q_{\mathrm{g}}}
\newcommand{\Sigmagas}{\Sigma_{\mathrm{gas}}}

\newcommand{\Hmol}{\mathrm{H}_{2}}
\newcommand{\SigmaHI}{\Sigma_{\mathrm{HI}}}
\newcommand{\fHmol}{f_{\mathrm{H2}}}
\newcommand{\fgas}{f_{\mathrm{gas}}}
\newcommand{\SFR}{\mathrm{SFR}}

\newcommand{\fHmolave}{\left<f_{\mathrm{H2}}\right>}
\newcommand{\hgas}{h_{\mathrm{gas}}}

\newcommand{\hSFR}{h_{\mathrm{SFR}}}
\newcommand{\alphaL}{\alpha_{\mathrm{L}}}
\newcommand{\alphaDM}{\alpha_{\mathrm{DM}}}
\newcommand{\alphaHI}{\alpha_{\mathrm{HI}}}
\newcommand{\nH}{n_{\mathrm{H}}}
\newcommand{\ntot}{n_{\mathrm{tot}}}
\newcommand{\nmol}{n_{\mathrm{mol}}}
\newcommand{\Uisrf}{U_{\mathrm{isrf}}}
\newcommand{\K}{\mathrm{K}}
\newcommand{\rhog}{\rho_{g}}
\newcommand{\tstar}{t_{\star}}
\newcommand{\sech}{\mathrm{sech}}
\newcommand{\Zg}{Z_{g}}
\newcommand{\vc}{v_{\mathrm{circ}}}
\newcommand{\cc}{\mathrm{cm}^{-3}}
\newcommand{\Heating}{\mathcal{H}}

\newcommand{\kms}{\,\mathrm{km}\,\,\mathrm{s}^{-1}}

\shorttitle{Molecular Hydrogen and Star Formation}
\shortauthors{Robertson and Kravtsov}

\begin{document}

\title{Molecular Hydrogen and Global Star Formation Relations in Galaxies}
\author{Brant E. Robertson\altaffilmark{1,2,3} and Andrey V. Kravtsov\altaffilmark{1,2}}

\altaffiltext{1}{Kavli Institute for Cosmological Physics, and Department of Astronomy and
Astrophysics, University of Chicago, 933 East 56th Street, Chicago, IL 60637, USA}
\altaffiltext{2}{Enrico Fermi Institute, 5640 South Ellis Avenue, Chicago, IL 60637, USA}
\altaffiltext{3}{Spitzer Fellow}

\begin{abstract}
We use hydrodynamical simulations of disk galaxies to study relations
between star formation and properties of the molecular interstellar
medium (ISM).  
We implement a
model for the ISM that includes low-temperature
($T<10^{4}$~K) cooling, directly ties the star formation rate to the
molecular gas density, and accounts for the destruction of $\Hmol$ by
an interstellar radiation field from young stars.  We demonstrate that
the ISM and star formation model simultaneously produces a
spatially-resolved molecular-gas surface density Schmidt-Kennicutt
relation of the form $\SigmaSFR \propto \SigmaHmol^{\nmol}$ with
$\nmol\approx1.4$ independent of galaxy mass, and a total gas surface
density -- star formation rate relation $\SigmaSFR \propto
\Sigmagas^{\ntot}$ with a power-law index that steepens from
$\ntot\sim2$ for large galaxies to $\ntot\gtrsim4$ for small dwarf
galaxies.  We show that deviations from the disk-averaged $\SigmaSFR
\propto \Sigmagas^{1.4}$ correlation determined by
\cite{kennicutt1998a} owe primarily to spatial trends in the
molecular fraction $\fHmol$ and may explain observed deviations from
the global Schmidt-Kennicutt relation. 
In our model, such
deviations occur in regions of the ISM where the fraction of gas mass in
molecular form is declining or significantly less than unity. 
Long gas
consumption time scales in low-mass and low surface brightness
galaxies may owe to their small fractions of molecular
gas rather than mediation by strong
supernovae-driven winds.  Our simulations also reproduce the
observed relations between ISM pressure and molecular fraction 
and between star formation rate, gas surface density, and disk
angular frequency. We show that the Toomre criterion that accounts for
both gas and stellar densities correctly predicts the onset of
star formation in our simulated disks. We examine the density
and temperature distributions of the ISM in simulated galaxies and
show that the density probability distribution function (PDF)
generally exhibits a complicated structure with
multiple peaks corresponding to different temperature phases of the
gas. The overall density PDF can be well-modeled as a sum of 
lognormal PDFs corresponding to individual, approximately isothermal phases.  
We also present a simple
method to mitigate numerical Jeans fragmentation of dense, cold gas in
Smoothed Particle Hydrodynamics codes through the adoption of a
density-dependent pressure floor.
\end{abstract}


%---------------------
\section{Introduction}
\label{section:introduction}
%---------------------

Galaxy formation presents some of the most important and challenging
problems in modern astrophysics.  A basic paradigm for the
dissipational formation of galaxies from primordial fluctuations in
the density field has been developed
\citep[e.g.][]{white1978a,blumenthal1984a,white1991a}, but many of the
processes accompanying galaxy formation are still poorly understood.  In
particular, star formation shapes the observable properties of
galaxies but involves a variety of complicated dynamical, thermal,
radiative, and chemical processes on a wide range of scales
\citep[see][for a review]{mckee2007a}.  
Observed
galaxies exhibit large-scale correlations between their 
global star formation rate (SFR) surface density $\SigmaSFR$
and average gas surface density $\Sigmagas$ \citep{kennicutt1989a,kennicutt1998a}, and
these global correlations serve as the basis for treatments of star formation 
in many models of galaxy formation.
While such models have supplied important insights,  
detailed observations of galaxies have recently 
provided evidence that the molecular, rather than the total, gas surface density  is 
the primary driver of global star formation in galaxies
\citep[e.g.,][]{wong2002a,boissier2003a,heyer2004a,boissier2007a,calzetti2007a,kennicutt2007a}.

In this study, we adopt an approach in which empirical and
theoretical knowledge of the star formation efficiency (SFE) in dense,
molecular gas is used as the basis for a star formation model in 
hydrodynamical simulations
of disk galaxy evolution.  This approach requires modeling processes
that shape properties of the dense phase of the interstellar medium
(ISM) in galaxies. The purpose of this paper is to present such a
model. 

Stellar populations in galaxies exhibit salient trends of colors and
metallicities with galaxy luminosity
\citep[e.g.,][]{kauffmann2003c,blanton2005a,cooper2007a}. In the
hierarchical structure formation scenario these trends should emerge
through the processes of star formation and/or stellar feedback in the 
progenitors of present-day galaxies. Observationally, ample
evidence suggests that the efficiency of the conversion of
gas into stars depends strongly and
non-monotonically on mass of the system. For example, the faint-end of
the galaxy luminosity function has a shallow slope
\citep[$\alphaL\approx1.0-1.3$, e.g.,][]{blanton2001a,blanton2003a}
compared to the steeper mass function of dark matter halos
\citep[$\alphaDM\approx2$, e.g.,][]{press1974a,sheth1999a}, indicating
a decrease in SFE in low-mass galaxies.  At the
same time, the neutral hydrogen (HI) and baryonic mass functions
may be steeper than the luminosity function
\citep[$\alphaHI\approx1.3-1.5$, e.g.,][]{rosenberg2002a,zwaan2003a}.
The baryonic \cite{tully1977a} relation is continuous down to
extremely low-mass dwarf galaxies
\citep[e.g.,][]{mcgaugh2005a,geha2006a}, indicating that the
fractional baryonic content of galaxies of different mass is similar.
Hence, low-mass galaxies that are
unaffected by environmental processes are gas-rich, yet 
often form stars inefficiently.

While feedback processes from supernovae and AGN
\citep[e.g.,][]{brooks2007a,sijacki2007a}, or the efficiency of gas
cooling and accretion \citep{dekel2006a,dekel2008a}, may account for
part of these trends, the SFE as a function
of galaxy mass may also owe to intrinsic ISM processes
\citep[e.g.,][]{tassis2008a,kaufmann2007a}.  To adequately explore the
latter possibility, a realistic model for the conversion of gas into
stars in galaxies is needed.

Traditionally, star formation in numerical simulations of galaxy
formation is based on the empirical Schmidt-Kennicutt (SK) relation
\citep{schmidt1959a,kennicutt1989a,kennicutt1998a}, in which star
formation rate is a {\it universal} power-law function of the total
disk-averaged or global gas surface density: $\SigmaSFR\propto
\Sigmagas^{\ntot}$ with $\ntot\approx 1.4$ describing the correlation
for the entire population of normal and starburst galaxies.  However,
growing observational evidence indicates that this relation may not be
universal on smaller scales within galaxies, especially at low surface
densities.

Estimates of the slope of the SK relation within individual galaxies exhibits
significant variations. For example, while \citet{schuster2007a} and 
\citet{kennicutt2007a} find $\ntot\approx 1.4$ for the molecular-rich
galaxy M51a, similar estimates in other large, nearby galaxies \citep[including the
Milky Way (MW);][]{misiriotis2006a} range from $\ntot\approx1.2$ to $\ntot\approx 3.5$ \citep{wong2002a,boissier2003a}
depending on dust-corrections and fitting methods.
The disk-averaged total gas SK relation for normal 
(non-starburst) galaxies also has a comparably steep slope of $\ntot\approx2.4$, with
significant scatter \citep{kennicutt1998a}. 
While the variations in the $\SigmaSFR-\Sigmagas$ correlation may 
indicate systematic uncertainties in observational measurements,
intrinsic variations or trends in galaxy properties may also
induce differences between 
the global relation determined by \cite{kennicutt1998a}
and the $\SigmaSFR-\Sigmagas$ correlation in
individual galaxies.

Galaxies with low gas surface densities, like dwarfs or bulgeless spirals, 
display an even wider
variation in their star formation relations.
\citet{heyer2004a} and
\citet{boissier2003a} show that in low-mass galaxies the
SFR dependence on the total gas surface density
exhibits a power-law slope $\ntot\approx 2-3$ that is considerably steeper
than the global \citet{kennicutt1998a} relation slope of $\ntot\approx 1.4$.
Further, star formation in the low-surface density outskirts of galaxies also
may not be universal. 
Average
SFRs appear to drop rapidly at gas surface
densities of $\Sigmagas\lesssim 5-10{\rm\, \Msun\,pc^{-2}}$
 \citep{hunter1998a,martin2001a}, indicating that  
star formation may be truncated 
or exhibit a steep dependence on the gas surface density. 
The existence of such threshold surface
densities have been proposed on theoretical grounds 
\citep{kennicutt1998a,schaye2004a}, although recent GALEX results 
using a UV indicator of star formation suggest that star formation
may continue at even lower surface densities \citep{boissier2007a}. 
Star formation rates probed by 
damped Lyman alpha absorption (DLA) systems also appear to lie 
below the \cite{kennicutt1998a} relation, by an order of magnitude 
\citep{wolfe2006a,wild2007a}, 
which may indicate that the relation between SFR
and gas surface density in DLA systems differs from the
local relation measured at high $\Sigmagas$.

In contrast, observations generally show that star formation in
galaxies correlates strongly with {\it molecular\/} gas, especially
with the highly dense gas traced by HCN emission
\citep{gao2004a,wu2005a}.  The power-law index of the
SK relation connecting the SFR to the
surface density of molecular hydrogen consistently displays a value of
$\nmol\approx1.4$ and exhibits considerably less galaxy-to-galaxy
variation
\citep{wong2002a,murgia2002a,boissier2003a,heyer2004a,matthews2005a,leroy2005,leroy2006,gardan2007a}.
Molecular gas, in turn, is expected to form in the high-pressure regions of the
ISM \citep{elmegreen1993a,elmegreen1994a}, as indicated by
observations \citep{blitz2004a,blitz2006a,gardan2007a}. 

Analytical models and numerical simulations that tie star
formation to the fraction of gas in the dense ISM are successful in
reproducing many observational trends
\citep[e.g.,][]{elmegreen2002a,kravtsov2003a,krumholz2005a,li2005a,li2006a,tasker2006a,tasker2007a,krumholz2007a,wada2007a,tassis2007a}.
Recently, several studies have explored star formation recipes based
on molecular hydrogen.  \citet{pelupessy2006a} and \citet{booth2007a}
implemented models for $\Hmol$ formation in gaseous disks and used
them to study the molecular content and star formation in galaxies.
However, these studies focused on the evolution of galaxies of a
single mass and did not address
the origin of the
SK relation, its dependence on galaxy mass or
structure, or its connection to trends in the local molecular fraction.

Our study examines the SK relation
critically, including its dependence on the structural and ISM properties
of galaxies of different masses, to explain the observed deviations from the
global SK relation, to 
investigate other connections between star formation and disk galaxy properties
such as rotation or gravitational instability, and to 
explore how the temperature and density structure of the ISM 
pertains to the star formation attributes of galaxies.

To these ends, we develop a model for the ISM and star formation whose
key premise is that star formation on the scales of molecular clouds
($\sim 10$~pc) is a function of molecular hydrogen density with a
universal SFE per free-fall time
\citep[e.g.,][]{krumholz2007b}.  Molecular hydrogen, which
we assume to be a proxy for dense, star forming gas, is accounted for
by calculating the local $\Hmol$ fraction of gas as a function of
density, temperature, and metallicity using the photoionization code
Cloudy \citep{ferland1998a} to incorporate $\Hmol$-destruction
by the UV radiation 
of local young stellar populations.  We devise a
numerical implementation of the star formation and ISM model, and
perform hydrodynamical simulations to study the role of molecular gas
in shaping global star formation relations of self-consistent galaxy
models over a representative mass range.

The results of our study show that many of the observed global star formation
correlations and trends can be understood in terms of the dependence of molecular
hydrogen abundance on the local gas volume density. We show that
the physics controlling the abundance of molecular hydrogen and its destruction
by the interstellar radiation field (ISRF) play a key role in shaping
these correlations, in agreement with earlier calculations based on 
more idealized models of the ISM \citep{elmegreen1993a,elmegreen1994a}.
While our simulations focus on the connection between the molecular ISM phase
and star formation on galactic scales, the formation of molecular hydrogen
has also been recently studied in simulations of the ISM on smaller,
subgalactic scales \citep{glover2007a,dobbs2007a}. These simulations
are complementary to the calculations presented in our study and could
be used as input to improve the molecular ISM model we
present.

The paper is organized as follows.  The simulation methodology,
including our numerical models for the ISM, interstellar radiation
field, and simulated galaxies, is presented in \S
\ref{section:methodology}.  The results of the simulations are
presented in \S \ref{section:results}, where the simulated star
formation relations in galactic disks and correlations of the
molecular fraction with the structure of the ISM are examined.  We
discuss our results in \S \ref{section:discussion} and conclude with a
summary in \S \ref{section:summary}. Details of our tests of numerical
fragmentation in disk simulations and calculations of the model
scaling between star formation, gas density, and orbital frequency
are presented in the Appendices. Throughout, we work in the context of
a dark-energy dominated cold dark matter cosmology with a Hubble
constant $H_{0} \approx 70\kms\Mpc^{-1}$.

\begin{figure*}
\figurenum{1}
\epsscale{1}
\plotone{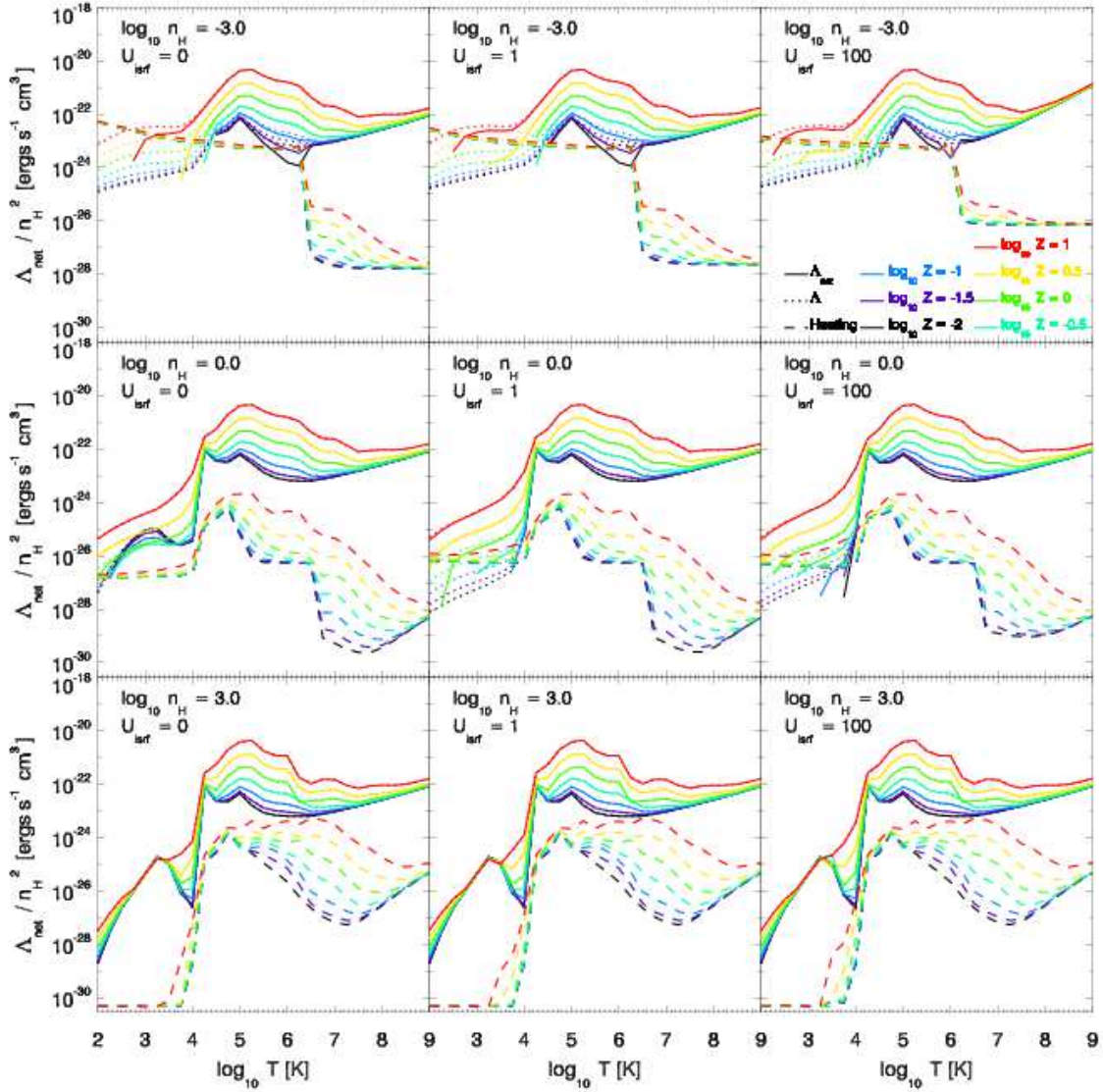}
\caption{\label{fig:cooling}
Cooling ($\Lambda$) and heating ($\Heating$) rates for interstellar and intergalactic gas as a function of gas density ($\nH$),
temperature ($T$), metallicity ($Z$), and interstellar radiation field (ISRF) strength ($\Uisrf$), in units of
the ISRF strength in the MW at the solar circle, as calculated 
by the code Cloudy \citep{ferland1998a}.  Shown are the cooling (dotted lines), heating (dashed lines) and net 
cooling (solid) functions over the temperature range $T=10^{2}-10^{9}~\K$.  Dense gas can efficiently cool via 
atomic and molecular coolants below $T=10^{4}\K$, depending on the gas density and the strength of the ISRF.  A strong
ISRF can enable the destruction of $\Hmol$ gas and 
thereby reduce the SFR.
}
\end{figure*} 

\section{Methodology}
\label{section:methodology}

To examine molecular gas and star formation in 
disk galaxies, we develop a model 
that accounts for low temperature coolants ($T<10^{4}\,\K$), calculates
the equilibrium abundance of $\Hmol$ in dense gas, and estimates the local
SFR based on the local $\Hmol$ density with a universal
SFE per free fall time.
This ISM model is detailed below in \S \ref{section:methodology:ism},
and is referred to as the $\Hmol$D-SF model (for ``$\Hmol$ density-star formation'').
We extend the $\Hmol$D-SF model to account for the destruction 
of $\Hmol$ by an
interstellar radiation field powered by local star formation, and refer
to the extended model as the $\Hmol$D-SF+ISRF model 
(for ``$\Hmol$ density-star formation plus interstellar radiation field'').
The $\Hmol$D-SF+ISRF model is also detailed below in 
\S \ref{section:methodology:ism}.
The new molecular gas ISM and star formation models are contrasted against
a simple ISM model that includes only atomic cooling with a temperature
floor $T=10^{4}\K$ and a
SFR calculated from the total gas density above a density
threshold, which we refer to as
the GD-SF model (for ``gas density-star formation'').  The GD-SF model
has been commonly used in previous
galaxy formation simulations.
Each star formation and ISM model is numerically implemented in the 
smoothed particle hydrodynamics (SPH) /
N-body code GADGET2 \citep{springel2001a,springel2005c}, which is used to
perform the simulations presented in this work.
General issues of numerical fragmentation in the
disk simulations of the kind presented here are described in 
\S \ref{section:methodology:jeans_resolution}, \S \ref{section:methodology:median_eos},
and in the Appendix.
The ISM 
models are applied to simulations of the isolated evolution
of disk galaxies with a representative range of circular velocities and structural properties.
The galaxy models are designed as analogues of the well-studied nearby galaxies
DDO154, M33, and NGC 4501, and are detailed in \S \ref{section:methodology:galaxy_models}.  The results of the simulations are presented in \S \ref{section:results}.

\subsection{ISM Properties}
\label{section:methodology:ism}

The thermal properties of the ISM are largely determined
by the radiative heating $\Heating$ and cooling $\Lambda$
rates of interstellar gas.  Given $\Heating$ and $\Lambda$,
the rate of change of the internal energy $u$ of gas
with a density $\rhog$ evolves as
\begin{equation}
\label{equation:thermal_evolution}
\rhog\frac{\dd u}{\dd t} = \Heating-\Lambda,
\end{equation}
\noindent
with additional terms to describe the energy input
from feedback 
or hydrodynamical interactions.
The temperature dependence of $\Heating$ and $\Lambda$ will depend on
the metallicity and density of gas, and on the
nature of the external radiation field.

To calculate the above heating and cooling processes,
as well as additional cooling processes from
molecules and ionic species other than $\Hmol$,  
we use the photoionization code Cloudy version C06.02a,
last described by
\citet{ferland1998a}.  
Cloudy calculates 
heating and cooling processes for H, He, and
metals, as well as absorption
by dust grains, photoelectric absorption, and molecular
photoabsorption.  The chemical network and cooling
includes molecules and atomic ions
that allow gas to cool to temperatures $T<10^4$~K.  For details on the molecular
network and cooling, the reader is encouraged to
examine \cite{hollenbach1979a,hollenbach1989a}, 
\cite{ferland1998a}, and the Cloudy documentation
suite Hazy\footnote{http://www.nublado.org}.
For purposes of calculating the efficiency of 
molecular line cooling
using Cloudy, only the thermal line width is considered.  
At temperatures
below our minimum temperature of ($T\sim100 \K$), the 
influence
of nonthermal motions on molecular line cooling will
become increasingly important and should be considered when
modeling lower temperature gas.

For the heating term $\Heating$,
our Cloudy calculations include heating from the cosmic 
ultraviolet background at $z=0$
following \cite{haardt1996a}, extended to include 
the contribution to the UV background from galaxies
(see the Cloudy documentation for details).
Cosmic ray heating and ionization are incorporated following
\cite{ferland1984a}, and are important 
at densities where the ISM becomes 
optically-thick to UV radiation.  
The presence of a blackbody cosmic microwave background 
radiation with a temperature of
$T\approx2.7\K$ is included.
Milky Way-like ($R_{V}=3.1$) dust, including both graphite and
silicate dust, is modeled with the local ISM abundance. 
Grain heating and cooling mechanisms are included, following \cite{van_hoof2001a}
and \cite{weingartner2001a}.

The $\Hmol$D-SF model of the ISM includes all the cooling and heating
processes described above. 
In addition, the $\Hmol$D-SF+ISRF model includes  the following 
treatment of the ISRF, and is used to examine the consequences 
of soft UV radiation from star formation for 
the molecular phase of the ISM.

\cite{mathis1983a} 
modeled the ISRF needed to power the emission spectrum
of dust both in the diffuse ISM and in molecular clouds
in the Milky Way.  The short-wavelength 
($\lambda\lesssim0.3\micron$) ISRF spectral energy distribution (SED)
was inferred to scale roughly exponentially with a scale length
$R_{u}\sim4\, \kpc$ in the MW disk (for a solar galactocentric radius of 
$R_{\sun}=10\,\kpc$),
tracing the stellar population, while the
strength of the long-wavelength ISRF SED varied non-trivially with
radius owing to the relative importance of emission from 
dust.  
The short-wavelength ISRF SED determined by 
\cite{mathis1983a} (their Table A3) for 
$R = 0.5-1.3R_{\sun}$, normalized to a fixed stellar surface
density,  
is roughly 
independent of radius, with a shape similar to the 
radiation field considered by \cite{draine1978a} and
\cite{draine1996a}.  For simplicity, we fix the ISRF spectrum
to have its inferred SED in the solar vicinity 
and scale the intensity
with the local SFR density normalized by the
solar value as
\begin{equation}
\label{equation:uisrf}
\Uisrf \equiv \frac{u_{\nu}}{u_{\nu,\sun}} = \frac{\SigmaSFR}{\Sigma_{\mathrm{SFR},\sun}}
\end{equation}
\noindent
where
$\Sigma_{\mathrm{SFR},\sun}\approx(2-5)\times10^{-9}\Msun\,\yr^{-1}\,\pc^{-2}$
\citep[e.g.,][]{smith1978a,miller1979a,talbot1980a,rana1987a,rana1991a,kroupa1995a}.
For our model, we adopt $\Sigma_{\mathrm{SFR},\sun} =
4\times10^{-9}\Msun\,\yr^{-1}\,\pc^{-2}$.    When included, the ISRF is
added to the input spectrum for the Cloudy calculations as an
additional source of radiation.

Absorption of soft UV ($\lambda\sim0.1\micron$)
photons can enable the destruction of $\Hmol$ through transitions to the
vibrational continuum or to excited states that can be photoionized or
photodissociated \citep{stecher1967a}.
The ISRF can supply the soft UV photons that lead to $\Hmol$ 
dissociation and can subsequently regulate the $\Hmol$ abundance at low
gas densities ($\nH\sim1\cc$).  For values of $\Uisrf\gtrsim0.01$, the
ionizing flux of the ISRF dominates over the 
\cite{haardt1996a} UV background 
\citep[see, e.g.,][]{sternberg2002a}.

Figure \ref{fig:cooling}  shows the cooling rate $\Lambda$ (dotted line), the
heating rate $\Heating$ (dashed line), and net cooling rate 
$\Lambda_{\mathrm{net}}$ (solid lines) calculated 
for a range of temperatures ($T$), densities ($\nH$), 
metallicities ($Z$), and ISRF strengths ($\Uisrf$).
For each value of $T$, $\nH$, $Z$, and $\Uisrf$ a Cloudy
simulation is performed assuming a plane-parallel radiation
field illuminating a $10\,\pc$ thick slab chosen to be comparable
to our hydrodynamical spatial resolution.  
Each simulation
is allowed to iterate until convergence, after which the heating rate,
cooling rate, ionization fraction, molecular fraction, electron
density, and molecular weight is recorded from the center of the
slab.
These quantities are tabulated on a grid over the range 
$\log_{10} T=\{2,9\} \log_{10} \K$ with $\Delta \log_{10} T=0.25$, 
$\log_{10} \nH=\{-6,6\}\log_{10}\cc$ with $\Delta \log_{10} \nH=1.5$, 
$\log_{10} Z=\{-2,1\}\log_{10}Z_{\sun}$ with $\Delta \log_{10} Z=1.5$, 
and
$\Uisrf = \{0, 1, 10, 100, 1000\}\,U_{\mathrm{isrf}, \sun}$, and log-linearly
interpolated
according to the local gas properties.  
We note that 
since the slab thickness determines the typical particle
column density for a given volume density,
the tabulated quantities can depend on
the slab thickness.  
For reference,
varying the slab thickness between $100\pc$ and $1\pc$ for a density of
$\nH\sim30\cc$ changes the molecular fraction by less than $30\%$ and the
heating and cooling rates by less than $50\%$.

The left panels of Figure \ref{fig:cooling} with $\Uisrf=0$ correspond
to net cooling functions used in the $\Hmol$D-SF model, while the
$\Hmol$D-SF+ISRF model includes all the net cooling functions presented
in the figure.  The GD-SF model that includes only atomic cooling incorporates
the regions of the net cooling functions in the left ($\Uisrf=0$) panels of 
Figure \ref{fig:cooling} at temperatures $T>10^{4}\K$.
In addition to well known cooling and heating processes operating at 
high temperatures ($T>10^{4}\K$), the results of the Cloudy 
simulations show that cooling of gas near $\nH\sim1\cc$ is
regulated by the presence of the ISRF (Figure \ref{fig:cooling}, middle
left and center panels) and an ISRF strength of $\Uisrf\sim1$ contributes a
low-temperature heating rate $\Heating$ more than an 
order of magnitude larger than that supplied by the cosmic UV
background.  At higher densities ($\nH\sim10^{3}\cc$), the gas remains
optically-thin to cosmic ray heating but becomes insensitive to the presence
of either the ISRF or the cosmic UV background.
The correlation
between the ISRF field strength and the local SFR limits the
applicability of the low-$\nH$ / high-$\Uisrf$ region of the cooling 
function but we include such regimes for completeness.

The simple description of the ISRF used in our modeling is designed
to replicate average conditions in systems like the Milky Way.  However,
the ISRF in special locations in the ISM, such as near photoionized regions,
may have a different character than the average SED
we employ.  Also, changes in the composition of dust or the initial mass function
of stars relative to Milky Way properties could alter the frequency-dependent ISRF.
We therefore caution that the detailed, frequency-dependent connection
between SFR and ISRF strength used in our modeling does not
capture every condition within the ISM of a single galaxy or between galaxies.
The model we present does realistically capture the physics that regulate the 
abundance of molecular hydrogen and local SFR for a given 
spectral form of the ISRF.

\subsection{Star Formation and Feedback}
\label{section:methodology:sf_and_feedback}

Modeling the SFR in the ISM as a power-law function 
of the gas density $\rhog$ dates to at least \cite{schmidt1959a}, who
modeled the past SFR of the Galaxy needed to 
produce the observed luminosity function of main sequence stars.
We assume that star formation occurs in molecular clouds 
in proportion to the local molecular gas density $\fHmol\rhog$,
as suggested by a variety of observations
\citep[e.g.,][]{elmegreen1977a,blitz1980a,beichman1986a,lada1987a,young1991a}. 
The model thus assumes that $\Hmol$ is a good proxy for star forming gas. 

On the scale of individual gas particles the SFR
is then determined by the density of molecular hydrogen, which 
is converted into stars on a time scale $\tstar$: $\rho_{\star}\propto \fHmol\rhog/\tstar$. 
Observations indicate that at high densities $\tstar$ scales with the local free fall
time of the gas, $t_{\mathrm{ff}}\propto \rhog^{-0.5}$, as
\begin{equation}
\label{equation:sfr_ff}
\tstar\approx t_{\mathrm{ff}}/\epsilon_{\mathrm{ff}},
\end{equation}
with the SFE per free fall time, $\epsilon_{\mathrm{ff}}\approx 0.02$,
approximately independent of density at $n\approx 10^2-10^4~\cc$ \citep{krumholz2005a,krumholz2007b}.

We adopt $\tstar=1$~Gyr for gas at density $\nH =10 h^{2}\cc$
($t_{\mathrm{ff}}=2.33\times 10^7$~yrs), which corresponds to
$\epsilon_{\mathrm{ff}}=0.023$.  
By calibrating the SFE
to observations of dense molecular clouds through 
 $\epsilon_{\mathrm{ff}}$, our model differs
from the usual approach of choosing gas consumption time scale to fit
the global SFR and the SK relation in
galaxies.  As we show below, the chosen efficiency results in global
SFR for the entire galaxies consistent with
observations (see \S \ref{section:results}).

Incorporating the assumption that a
mass fraction $\beta$ of young stars promptly explode as supernovae,
the SFR in our models is given by
\begin{equation}
\label{equation:star_formation_rate}
\dot{\rho}_{\star} = (1-\beta) \fHmol\frac{\rhog}{\tstar}\left(\frac{\nH}{10\,h^{2}\cc}\right)^{0.5}.
\end{equation}
\noindent
We assume $\beta\approx0.1$, appropriate for the \cite{salpeter1955a} initial mass function.
Implicit in Equation \ref{equation:star_formation_rate}
is that the molecular fraction $\fHmol$
may vary with a variety of local ISM properties as
\begin{equation}
\label{equation:fH2_function}
\fHmol = \fHmol\left(\rhog,T,\Zg,\Uisrf\right).
\end{equation}
\noindent
The $\Hmol$D-SF model includes the dependence on the gas density, 
temperature $T$,
metallicity $\Zg$. The $\Hmol$D-SF+ISRF model additionally includes
the dependence of the molecular fraction on the strength of the 
interstellar radiation 
field $\Uisrf$ parameterized as a fraction of the local interstellar
field energy density.  
In both the $\Hmol$D-SF and $\Hmol$D-SF+ISRF models, the 
equilibrium molecular fraction is tabulated by using the Cloudy 
calculations
discussed
in \S \ref{section:methodology:ism}. 
Variations in the cosmic ray ionization
rate or the dust model could affect the form of
Equation \ref{equation:fH2_function} but we do not consider them here.
For the simple GD-SF model that does not track the molecular gas abundance, 
we set $\fHmol=1$ in gas with densities above the star formation
threshold density $\nH\gtrsim0.1\cc$ \citep[see, e.g.,][]{governato2007a}.  

Figure \ref{fig:cooling} shows that the feature in the cooling
rate at $T\sim 10^3$~K for low metallicity, intermediate 
density gas (e.g., middle
left panel), which owes to molecular hydrogen, disappears
in the presence of the interstellar radiation 
field.
Equations 
\ref{equation:thermal_evolution} and 
\ref{equation:star_formation_rate} then imply that
destruction of $\Hmol$ by soft UV photons from young
stars regulates star formation in the $\Hmol$D-SF+ISRF model.  
The dissociation of $\Hmol$ 
by the ISRF means that ISM gas may be cold and dense but its SFR may
be suppressed if the local ISRF is strong.  
We discuss this effect in the context of the simulated
galaxy models in \S \ref{section:results}.

The energy deposition from supernovae into interstellar gas is treated
as a thermal feedback, given by
\begin{equation}
\label{equation:sn_feedback}
\rhog\frac{\dd u}{\dd t} = \epsilon_{\mathrm{SN}} \dot{\rho}_{\star}
\end{equation}
\noindent
where $\epsilon_{\mathrm{SN}}$ is the energy per unit mass of formed stars
that is deposited into the nearby ISM.  We choose 
$\epsilon=1.4\times10^{49} \mathrm{ergs}\,\,\Msun^{-1}$ or, in the language of
\cite{springel2003b}, an effective supernovae temperature of 
$T_{\mathrm{SN}}=3\times10^{8}$.  This value for the energy deposition for
supernovae has been used repeatedly in simulations of galaxy mergers
\citep[e.g.,][]{robertson2006c,robertson2006b,robertson2006a}.  Given
that extremely short cooling time of dense gas at low resolutions, prescribing 
supernovae feedback as thermal input into the gas has long been known to
have a weak effect on the global evolution of simulated galaxies 
\citep[e.g.,][]{katz1992b,steinmetz1995a}. For this reason, the results of this paper 
also do 
not depend strongly on the value of $\epsilon_{\mathrm{SN}}$ or the
supernovae mass fraction $\beta$.  Metal enrichment of the ISM
is treated in the instantaneous approximation, with a mass fraction 
$y=0.02$ of the supernovae ejecta being returned into the gas as metals
\citep[see][]{springel2003a}.

\subsection{Avoiding Numerical Jeans Fragmentation}
\label{section:methodology:jeans_resolution}

Perturbations in a self-gravitating medium of mean density 
$\rho$ and sound speed $c_{s}$ can grow only
if their wavelength exceeds the Jeans length $\lambdaJeans$ \citep{jeans1928a}. 
The corresponding Jeans mass contained within $\lambdaJeans$ is
\begin{equation}
\label{equation:jeans_mass}
\mJeans = \frac{\pi^{5/2}c_{s}^{3}}{6G^{3/2}\rho^{1/2}}
\end{equation}
\noindent
\citep[see, e.g., \S 5.1.3 of][]{binney1987a}.
\cite{bate1997a} identified a resolution requirement for
smoothed particle hydrodynamics simulations such that the
number of SPH neighbors $\Nneigh$ and gas particle mass 
$\mgas$ should satisfy
$2\Nneigh\mgas < \mJeans$ to capture the pressure forces
on the Jeans scale and avoid numerical fragmentation.
\cite{klein2004a} suggest that the effective resolution of
SPH simulations is the number of SPH smoothing lengths per
Jeans scale
\begin{equation}
\label{equation:jeans_hsml}
\hJeans = \frac{\pi^{5/2}c_{s}^{3}}{6G^{3/2} \Nneigh \mgas \rho^{1/2}},
\end{equation}
\noindent
which is very similar to the number of Jeans masses per 
SPH kernel mass, if we define the kernel mass as
\begin{equation}
\label{equation:m_SPH}
\mSPH \equiv \sum_{i=1}^{\Nneigh} m_{i}.
\end{equation}
\noindent
Our simulations include low temperature coolants that allow ISM
gas to become cold and dense, and for the typical number of
gas particles used in our models ($\Ngas \approx 400,000$) the
Jeans mass is not resolved at the lowest temperatures and
largest galaxy masses.
Motivated by techniques used to avoid numerical Jeans fragmentation in 
grid codes \citep[e.g.,][]{machacek2001a}, a density-dependent
pressure floor is introduced into our SPH calculations.  For every gas particle, 
the kernel mass is monitored to ensure it resolves some number $\NJeans$ 
of Jeans masses.  If not,
the particle internal energy is altered following 
\begin{equation}
\label{equation:pressure_floor}
 u = u \times \left\{ \begin{array} {c@{\quad:\quad}l}
\left(\frac{\NJeans}{\hJeans}\right)^{2/3} & \hJeans<\NJeans \\ 
\left(\frac{2\NJeans\mSPH}{\mJeans}\right)^{2/3} & \mJeans < 2\NJeans\mSPH
\end{array} \right.,
\end{equation}
\noindent
to provide the largest local pressure and to assure
the local Jeans mass is resolved.
We discuss the purpose and effect of this
pressurization in more detail in the first Appendix,
but we note that this requirement scales
with the resolution and naturally allows 
for better-resolved SPH simulations to
follow increasingly lower temperature gas.
For the simulations results presented in this paper, we use $\NJeans=15$,
which is larger than the effective $\NJeans=1$ used by \cite{bate1997a}.
We have experimented with simulations with $\NJeans=1-100$ and find 
$\NJeans\sim15$ to provide sufficient stability over the time evolution
of the simulations for the structure of galaxy models we use.  Other
simulations may require different values $\NJeans$ or a different 
pressurization than the $u\propto\mJeans^{-2/3}$ scaling suggested
by Equation \ref{equation:jeans_mass}.

\begin{figure}
\figurenum{2}
\epsscale{1.25}
\plotone{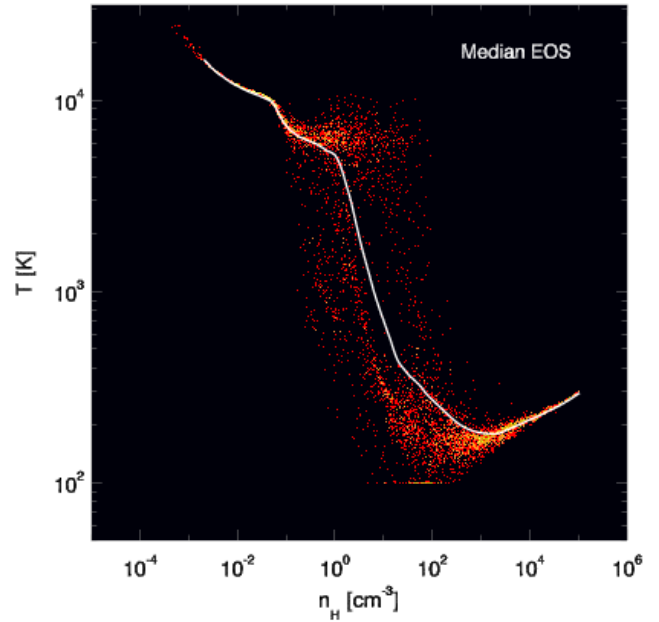}
\caption{\label{fig:median_eos}
Median equation-of-state (EOS) of gas in a Milky Way-sized galactic 
disk in our ISM model.  Shown is the histogram of particles in the hydrogen 
number density $\nH$ -- temperature $T$ plane and the median temperature 
measured in bins of density (white line).  The median EOS is used to 
determine the properties of gas pressurized to avoid numerical 
Jeans fragmentation (see \S \ref{section:methodology:jeans_resolution}
and \S \ref{section:methodology:median_eos}).
At high densities, the particles follow the loci where heating 
balances cooling for their 
densities, temperatures, metallicities,
and local interstellar radiation field strengths.
}
\end{figure}

\subsection{Median Equation-of-State}
\label{section:methodology:median_eos}

In the absence of a numerical pressure floor, the equation-of-state (EOS) of 
gas will follow the loci where heating balances cooling in the temperature-density
phase space. 
However, the gas pressurization requirements to avoid numerical Jeans fragmentation
change the EOS of the gas.
For purposes of calculating the molecular or 
ionization fraction of dense ISM gas that has been numerically pressurized,
an estimate of the EOS in the absence of resolution restraints is required.
Figure \ref{fig:median_eos} shows the temperature-density phase diagram of 
ISM gas in a simulation of a Milky Way-sized galaxy without using the SPH
pressure floor described in \S \ref{section:methodology:jeans_resolution}
and detailed in the Appendix.

The median
EOS of this gas, determined by the heating and cooling processes calculated
by the Cloudy code for a given density, temperature, metallicity, and 
interstellar radiation field strength, as well as heating from supernovae
feedback, is used in our simulations to 
assign an effective temperature to dense gas
pressurized at the Jeans resolution limit.
The behavior of gas above a density of $\nH\sim 100\cc$ is especially
influenced by the balance of available cooling mechanisms and supernovae
heating, and in the absence of supernovae feedback such dense gas would
cool to the minimum temperature treated in our simulations ($T\sim 100\K$).

\subsection{Molecular Fraction vs. Gas Density}
\label{section:methodology:fH2_vs_density}

The molecular content of the ISM in models that account
for the abundance of $\Hmol$ and other molecular
gas depends on the physical properties of interstellar
gas through Equation \ref{equation:fH2_function}.  Since
the temperature and density phase structure of the ISM is 
dictated by the local cooling time, which is nonlinearly
dependent on the temperature, density, metallicity, and
ISRF strength, the calculation of the equilibrium molecular
fraction as a function of the gas properties is 
efficiently calculated using hydrodynamical simulations
of galaxy evolution.  
We have measured the 
molecular gas fraction $\fHmol$ as a function of 
gas density $\nH$ after evolving
in ISM models with 
$\Hmol$-destruction by an ISRF (the $\Hmol$D-SF+ISRF model),
and without an ISRF (the $\Hmol$D-SF model),
for $t=0.3\Gyr$ for each of the  
$\vcirc=50-300\kms$ disk galaxy simulations. 
The galaxy-to-galaxy 
and point-to-point variations in the metallicity, temperature, ISRF
strength contribute to the spread in molecular fractions 
at a given density $\nH$.
Without $\Hmol$-destruction, the molecular hydrogen content
of low-density gas calculated using Cloudy is surprisingly
large, with $\fHmol=0.1$ at $\nH=0.1\cc$ and $\fHmol\sim0.35$ at
$\nH=1\cc$.
For the $\Hmol$D-SF model, the atomic-to-molecular 
transition represents
a physical calculation of the star formation threshold density prescription
often included in coarse models of star formation in simulations of 
galaxy formation and evolution.

For the $\Hmol$D-SF+ISRF model, the destruction of $\Hmol$ by the
ISRF reduces the molecular fraction in the low-density ($\nH\lesssim1\cc$)
ISM to $\fHmol<0.05$.
The abundance of $\Hmol$ in this model is therefore greatly reduced
relative to the $\Hmol$D-SF model that does not include an ISRF.  
The decline in molecular abundance, in turn, produces much lower SFRs
in low density gas ($\nH\lesssim1\cc$) compared with either the
$\Hmol$D-SF model (which has a larger molecular fraction at low 
densities) or the GD-SF model (which allows for star formation in all
gas above a density threshold, $\nH>0.1\cc$).

For denser gas in the $\Hmol$D-SF+ISRF model (e.g, $\nH\gtrsim1\cc$), 
the numerical pressurization of 
the ISM required to mitigate artificial Jeans fragmentation prevents
much of the gas from reaching high densities where the molecular 
fraction becomes large ($\fHmol\sim1$ at $\nH\gtrsim100\cc$).  
The molecular fraction of the numerically-pressurized gas must 
be assigned at densities that exceed resolution limit of the simulations.
As Figure \ref{fig:median_eos} indicates, the transition between
the warm phase at densities $\nH\lesssim1\cc$ and the cold phase at densities 
$\nH\gtrsim30\cc$ is rapid \citep[see also][]{wolfire2003a}.
The required pressure floor can artificially limit the density of gas
to the range $\nH\lesssim30\cc$ where, in the presence of an ISRF,
the molecular fraction of the gas would be artificially suppressed (without an ISRF,
even gas at $\nH\approx1\cc$ is already mostly molecular and would be relatively
unaffected).
To account for this numerical limitation, 
gas at densities $\nH>1\cc$ that is numerically-pressurized to resolve the Jeans 
scale is assigned a minimum molecular fraction according to the median 
EOS and the high-density ($\nH>30\cc$) $\fHmol-\nH$ trend of gas in the absence of 
the pressure floor.  The $\fHmol-\nH$ relation calculated by Cloudy can
be modeled for pressurized gas in this regime as
$\fHmol = \mathrm{max}\{[0.67\left(\log_{10}\nH+1.5\right)-1], 1\}$.

The utilization of the median EOS to assign detailed properties to the
dense gas in this manner has the further benefit of adaptively scaling with the local
resolution, allowing an approximation of the ISM properties that improves
with an increase in the particle number used in a simulation.  Throughout
the paper, when simulation results include gas particles that are
pressurized to avoid numerical 
Jeans fragmentation, their reported temperatures and molecular fractions are 
determined from the median EOS presented in Figure \ref{fig:median_eos}.

While the density dependence of molecular fraction utilized here may not
be correct in every detail (i.e., the transition to $\fHmol=1$ may occur
at an inaccurately low density), for our purposes 
the primary requirement is that the model captures
the fraction of star forming gas as a function of density in a realistic
fashion.  The gas mass in disks in the $\Hmol$D-SF+ISRF model
at $\Sigmag\gtrsim 10\,\rm M_{\odot}\,pc^{-2}$ is mostly molecular 
and the molecular fraction increases realistically with the external ISM pressure
(see \S \ref{section:results:fH2_pressure}),
in good agreement with observations \citep[e.g.,][]{wong2002a,blitz2006a}.

\subsection{Isolated Galaxy Models}
\label{section:methodology:galaxy_models}
We use three cosmologically-motivated galaxy models to study global
star formation relations.  Below we describe the models and list the
observational data used to motivate their adopted structure.

The galaxy initial conditions
follow the methodology described by \cite{springel2005b}, with
a few exceptions.  The properties of the disks
are chosen to be representative of disk galaxies that form 
within the CDM structure formation paradigm \citep[e.g.,][]{mo1998a}.
The model galaxies consist of a \cite{hernquist1990a} dark matter 
halo, exponential stellar and gaseous disks with scale lengths $\Rds$
and  $\Rdg$, and an optional \cite{hernquist1990a} stellar bulge
with scale radius $\abs$.  For a given virial mass, the 
\cite{hernquist1990a}
dark matter 
halo parameters are scaled to an effective \cite{navarro1996a} halo
concentration.  The velocity distributions of the dark matter halos
are initialized from the isotropic distribution function provided in
\cite{hernquist1990a}, using the rejection technique described by 
\cite{press1992a}.  The vertical distribution of the stellar disk is 
modeled
with a $\sech^{2}(z/2z_{\mathrm{d}})$ function with 
$z_{\mathrm{d}} = 0.2-0.3 \Rds$.  The stellar disk velocity 
field is initialized by using a Gaussian with the local 
velocity dispersion $\sigma_{\mathrm{z}}^{2}$ determined 
from the potential and
density via the Jeans equations.  The stellar disk rotational velocity 
dispersion and streaming velocity are set with the epicyclical approximation
\citep[see \S 3.2.3 and \S 4.2.1(c) of ][]{binney1987a}.
The bulge velocity field is also modeled by a Gaussian
with a velocity dispersion determined from the Jeans equations
but with no net rotation.  The gas disk is initialized as an
isothermal medium of temperature of $T=10^{4}\K$
in 
hydrostatic equilibrium with the total potential, including the
disk self-gravity, which determines vertical distribution of the gas
self-consistently \citep[see][]{springel2005b}.
The gravitational softening lengths for the dark matter and baryons
are set to $\epsilon_{\mathrm{DM}}=100h^{-1}\pc$ and $\epsilon_{\mathrm{baryon}}=50h^{-1}\pc$,
respectively.

The galaxy models are designed to roughly match the observed rotation curves,
and HI, $\Hmol$, and stellar surface mass distributions of DDO154 
($\vcirc\approx50\kms$), 
M33 ($\vcirc\approx125\kms$), and NGC4501 ($\vcirc\approx300\kms$).  These 
systems cover a wide range in circular velocity, gas fractions, disk scale
lengths, molecular gas fractions, gas surface densities, and gas volume densities.
The parameters of the galaxy models are provided in Table \ref{table:models}.

\begin{deluxetable*}{lccccccccc} 
\tiny
\tablecolumns{10} 
\tablewidth{0pc} 
\tablecaption{Galaxy Models} 
\tablehead{
\multicolumn{10}{c}{Structural Parameters} \\
\cline{1-10}
\colhead{Galaxy} & \colhead{$V_{200}$} & \colhead{$\vcirc$} & \colhead{c} & \colhead{$\Mdisk$} & 
\colhead{$\fgas$} & \colhead{$\Rds$} & \colhead{$\Rdg$} & \colhead{$\Mbulge$}  & \colhead{$\abs$}\\
\colhead{Analogue} & \colhead{$\kms$} & \colhead{$\kms$} & & \colhead{$h^{-1}\Msun$} & & \colhead{$h^{-1}\kpc$} & \colhead{$h^{-1}\kpc$}
& \colhead{$h^{-1}\Msun$}& \colhead{$h^{-1}\kpc$}
\label{table:models}}
\startdata 
DDO 154  & 50  & 50  & 6  & $2.91\times10^{8}$  & 0.99 & 0.38 & 1.52 & \nodata & \nodata \\
M33      & 100 & 125 & 10 & $4.28\times10^{9}$  & 0.40 & 0.98 & 2.94 & \nodata & \nodata \\
NGC 4501 & 180 & 300 & 14 & $1.02\times10^{11}$ & 0.04 & 3.09 & 2.16 & $1.14\times10^{10}$ & 0.62 \\
\cline{1-10}
\multicolumn{10}{c}{Numerical Parameters}\\
\cline{1-10}
\colhead{Galaxy} & \colhead{$\NDM$} & \colhead{$\Ndisks$} & \colhead{$\Ndiskg$} & 
\colhead{$\Nbulge$} & \colhead{$\epsilon_{\mathrm{DM}}$} & \colhead{$\epsilon_{\mathrm{baryon}}$} & \colhead{$\Nneigh$}  & \colhead{$\Nneigh\mgas$} \\
\colhead{Analogue} & & & & 
& \colhead{$h^{-1}\kpc$} & \colhead{$h^{-1}\kpc$} & & \colhead{$h^{-1}\Msun$} & Data Refs.\\
\cline{1-10}
DDO 154  & 120000 &  20000 & 400000 & \nodata & 0.1 & 0.05 & 64 & $4.60\times10^{4}$ &1,2,3,4,5\\  
M33      & 120000 & 120000 & 400000 & \nodata & 0.1 & 0.05 & 64 & $2.74\times10^{5}$ &5,6,7\\  
NGC 4501 & 120000 & 177600 & 400000 &   22400 & 0.1 & 0.05 & 64 & $6.51\times10^{5}$ &8,9,10,11,12\\ 
\enddata 
\tablerefs{\small
(1) \cite{carignan1989a}; (2) \cite{hunter2004a}; (3) \cite{hunter2006a}; (4) \cite{lee2006a};
(5) \cite{mcgaugh2005a}; (6) \cite{corbelli2003a}; (7) \cite{heyer2004a}; (8) \cite{wong2002a};
(9) \cite{mollenhoff2001a} (10) \cite{guhathakurta1988a}; (11) \cite{rubin1999a}; (12) \cite{boissier2003a}
}
\end{deluxetable*}

\subsubsection{DDO 154 Analogue}
\label{section:methodology:galaxy_models:ddo154}

DDO 154 is one of the most gas rich systems known and
therefore provides unique challenges to models of the ISM
in galaxies.
For the DDO 154 galaxy analogue, we used the \cite{carignan1989a}
HI map and rotation curves as the primary constraint on the mass 
distribution.  The total gas mass is 
$\Mdg \approx 2.7\times10^{8}h^{-1}\Msun$ \citep{carignan1989a,hunter2004a}.
The stellar disk mass has been estimated at 
$\Mds \approx3.4\times10^{6}h^{-1}\Msun$ \citep{lee2006a} with a
scale length of about $\Rds \approx 0.38h^{-1}\kpc$ at $\lambda=3.6\micron$.
The gaseous disk scale length was determined by approximating the HI surface
density profile, with $\Rdg \approx 1.52^{-1}\kpc$ providing a decent
mimic of the HI data.  These numbers are similar to those compiled by
\cite{mcgaugh2005a} for DDO 154 and, combined with the dark matter virial
velocity $V_{200}=50\kms$ and concentration $c=6$, produce a 
rotational velocity of
$\vcirc\approx50\kms$ (where $\vcirc$ is defined as the maximum of rotation
velocity profile) similar to the value of $\vcirc\approx54\kms$
reported by \cite{karachentsev2004a}.

\subsubsection{M33 Analogue}
\label{section:methodology:galaxy_models:M33}

The nearby spiral M33 has well measured stellar, HI, $\Hmol$, and
SFR distributions, and serves nicely as an example disk galaxy with
an intermediate rotational velocity.
For the M33 galaxy analogue, we used the \cite{heyer2004a}
HI and $\Hmol$ map as guidance for determining the
gas distribution with a total gas mass of
$\Mdg\approx1.68\times10^{9}h^{-1}\Msun$ 
\citep{corbelli2003a,heyer2004a}.  According to this
data, the gas disk scale length is roughly $\Rdg\approx2.7h^{-1}\kpc$.
We set the total stellar disk mass to $\Mds\approx2.6\times10^{9}h^{-1}\Msun$
to match the total baryonic mass of $\Mbaryon\approx4.3\times10^{9}h^{-1}\Msun$
reported by \cite{mcgaugh2005a}, and set the disk scale length to 
$\Rds\approx0.9h^{-1}\kpc$.
These numbers are similar to those compiled by
\cite{mcgaugh2005a} for M33 and, combined with the dark matter virial
velocity $V_{200}=100\kms$ and concentration $c=10$, produce a 
rotational velocity of
$\vcirc\approx125\kms$. 

\begin{figure*}
\figurenum{3}
\epsscale{1.0}
\plotone{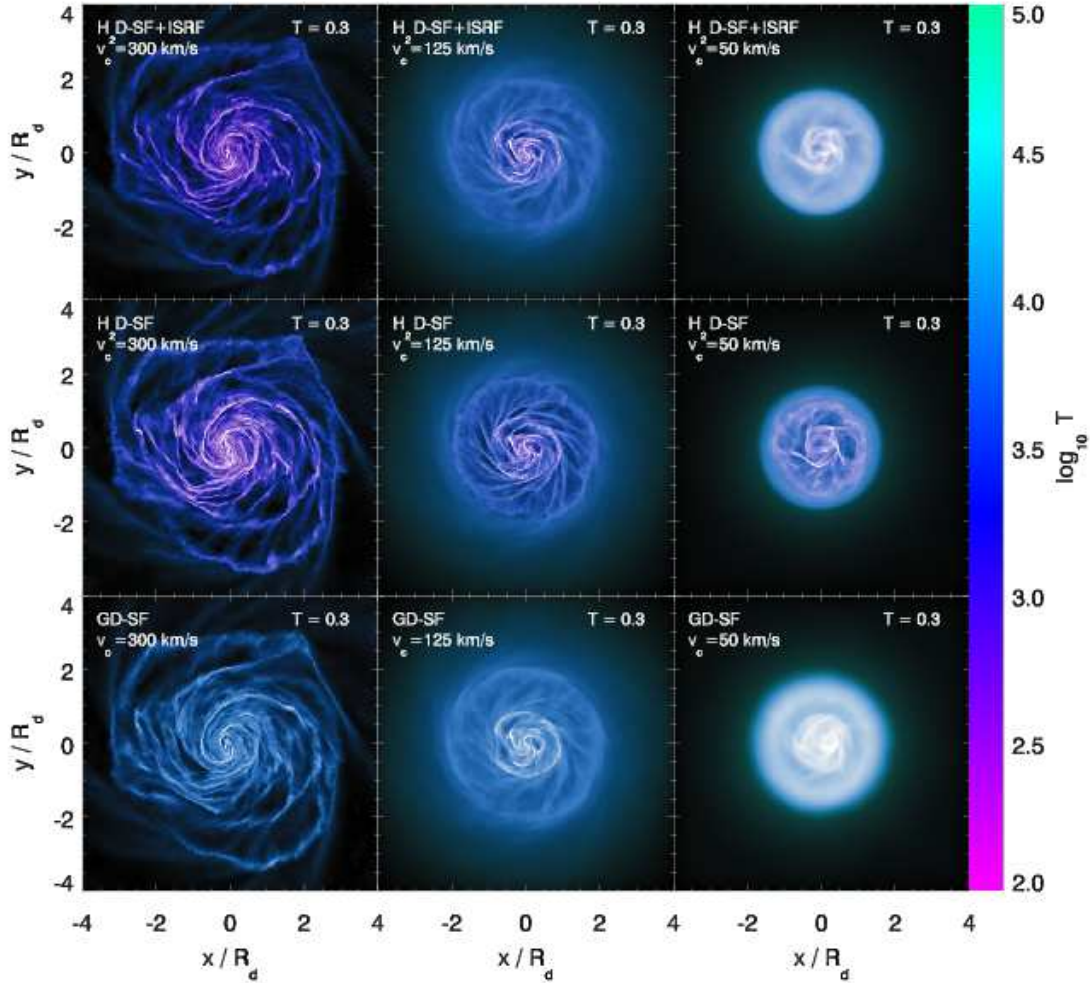}
\caption{\label{fig:disks}
\small
Gas distribution in galaxies with circular velocities of $\vc = 300\kms$ (left column), $\vc = 125 \kms$ (middle column), 
and $\vc=50\kms$ (right column) for differing models of the ISM and star formation.  Shown are an 
atomic cooling model with a $T=10^{4}\K$ temperature floor and star formation based on the gas density above a
threshold of $\nH=0.1\cc$ ("GD-SF", bottom row), an atomic+molecular cooling model with a $T=10^{2}\K$ temperature floor 
and star formation based on the molecular gas density ("$\Hmol$D-SF", middle row), and 
an atomic+molecular cooling model with a $T=10^{2}\K$ 
temperature floor, star formation based on the molecular gas density, and $\Hmol$-destruction by an interstellar
radiation field ("$\Hmol$D-SF+ISRF", top row).  The intensity of the images reflects the gas surface density and 
is color-coded by the local effective gas temperature.  The diffuse morphologies of low-mass galaxies in the GD-SF
and $\Hmol$D-SF+ISRF models owe to the abundance of warm atomic gas with temperatures comparable to the velocity
scale of the local potential.
}
\end{figure*}

\subsubsection{NGC 4501 Analogue}
\label{section:methodology:galaxy_models:NGC4501}

The massive spiral NGC 4501 is a very luminous 
disk galaxy \citep[$M_{K}=-25.38$,][]{mollenhoff2001a}
that is gas poor by roughly a factor of $\sim3$ for 
its stellar mass \citep[][]{giovanelli1983a,kenney1989a,blitz2006a},
and serves as an extreme contrast to the dwarf galaxy DDO 154. 
To construct an analogue to NGC 4501, we used the 
rotation curve data compiled by \cite{boissier2003a} 
from the observations of \cite{guhathakurta1988a} and 
\cite{rubin1999a} as a guide for designing the potential.
The stellar disk scale length $\Rds\approx3.1h^{-1}\kpc$
and bulge-to-disk ratio $\Mbulge/\Mdisk= 0.112$ were 
taken from the $K-$band observations of \cite{mollenhoff2001a},
and the total stellar mass was chosen to be consistent with
the \cite{bell2001a} $K-$band Tully-Fisher relation.
The gas distribution was modeled after the surface mass
density profile presented by \cite{wong2002a}.
These structure properties, combined with a 
virial velocity $V_{200}=180\kms$ and dark matter concentration
$c=14$, provide a circular velocity of $\vcirc\approx300\kms$
appropriate for approximating the observed rotation curve.

\begin{figure*}
\figurenum{4}
\epsscale{1.1}
\plotone{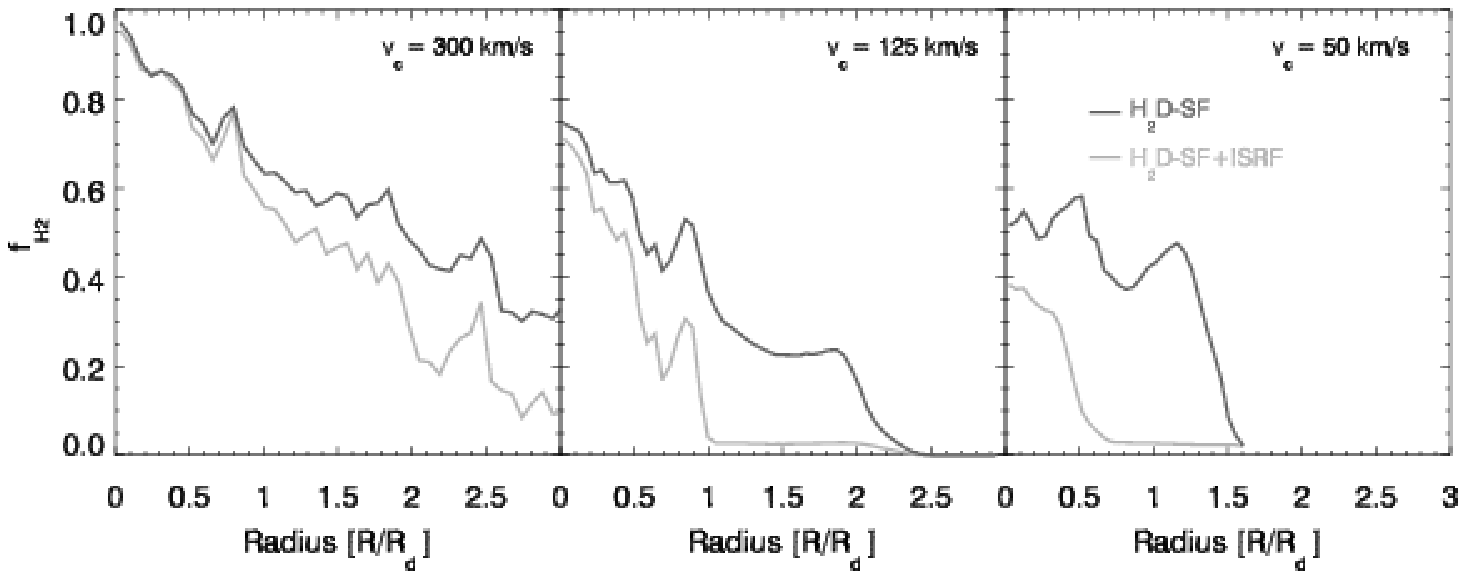}
\caption{\label{fig:fH2_vs_radius}
Molecular fraction as a function of radius in three simulated disk galaxies with ISM models with (light gray 
lines) and without (dark gray lines) an interstellar radiation field.  The general
trend is for molecular fraction to decrease and its dependence on radius
to steepen with decreasing mass of the system. This trend is more 
pronounced in the model with an interstellar radiation field. 
}
\end{figure*} 

\begin{figure*}
\figurenum{5}
\epsscale{1.15}
\plotone{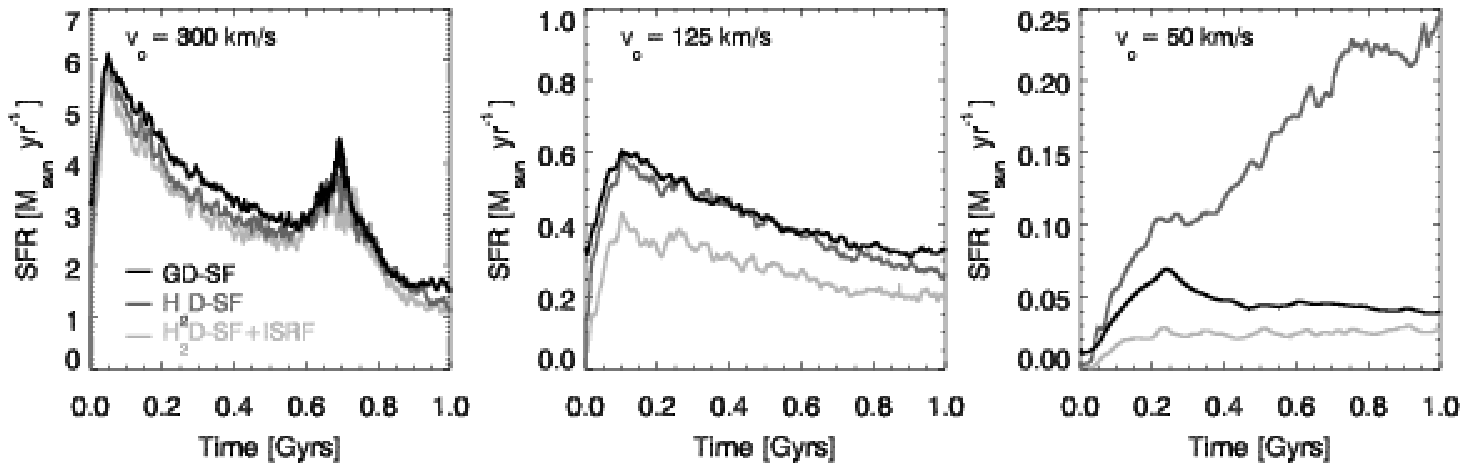}
\caption{\label{fig:sfr}
Star formation rates (SFRs) as a function of time in isolated disks.  Shown are the
SFRs for galaxy models with  $\vcirc=50\kms$ (right panel), $\vcirc=125\kms$ 
(middle panel), and $\vcirc=300\kms$ (left panel) calculated from interstellar 
medium models that allow for cooling at temperatures $T>10^{4}\K$ (black lines) 
or
cooling at temperatures $T>10^{2}\K$ with (light gray lines) and without (dark gray lines) destruction of
$\Hmol$ by a local interstellar radiation field (ISRF).  The star formation efficiency of 
the low-mass systems is strongly influenced by $\Hmol$-destruction from the ISRF.
}
\end{figure*} 

\section{Results}
\label{section:results}

The isolated evolution of each of the three
disk galaxy models is simulated for 
$t=1.0\,\Gyr$ using the three separate ISM 
and star formation models described in 
\S \ref{section:methodology}: the model 
$\Hmol$D-SF with $\Hmol$-based star formation, 
the model $\Hmol$D-SF+ISRF which in addition includes the
gas heating by local interstellar radiation,
and the GD-SF model with a $T=10^{4}\K$ temperature floor
and constant threshold density for star formation. In all
cases, the ISM of galaxies is assumed to have initial 
metallicity of $10^{-2}\,\rm Z_{\odot}$ and is self-consistently
enriched by supernovae during the course of simulation.

\subsection{Galactic Structure and Evolution}
\label{section:results:structure}

The gas distributions in the simulated disk galaxies after $t=0.3\Gyr$ of
evolution in isolation are shown in Figure \ref{fig:disks}.  The image
intensity reflects the disk surface density while the color indicates
the local mass-weighted temperature of the ISM.  A variety of
structural properties of the galaxies are evident from the images,
including ISM morphologies qualitatively similar to those observed in
real galaxies.  The large-scale structure of each disk model changes
slowly over
gigayear time-scales, excepting only the $\Hmol$D-SF model for the
$\vcirc=50\kms$ galaxy.  In this case, the ISM 
fragments owing to the large cold gas
reservoir that becomes almost fully molecular in the absence of
$\Hmol$-destruction (see \S
\ref{section:results:total_gas_schmidt_law} and \S
\ref{section:results:molecular_gas_schmidt_law} for further discussion).

The simulated disks can develop spiral structure and
bar-like instabilities during their evolution,  
especially when modeled with $T<10^4$~K coolants.
The detailed
structures of the spiral patterns and bar instabilities are likely
influenced by numerical fluctuations in the potential, but they
illustrate the role of spiral structure in the ISM of real galaxies.
As the ISM density is larger in these structures, 
the equilibrium gas temperature is lower compared with the
average disk properties.  These structures also contain higher 
molecular fractions (see \S \ref{section:results:fH2_pressure})
and will be sites of enhanced star formation
in our ISM model, similar to real galaxies.

The small-mass systems evolved with the GD-SF model are more diffuse and have
less pronounced structure than those evolved with low-temperature coolants, owing
to the higher ($T>10^4$~K) ISM temperature (and hence pressure) in
high-density regions.  The diffuse morphologies of low-mass galaxies
in the GD-SF and $\Hmol$D-SF+ISRF models owe to the abundance of warm
atomic gas with thermal energy comparable to the binding energy of the
gas.  The tendency of low-mass galaxy disks to be more
diffuse and extended than more massive systems owing to inefficient
cooling was shown by \cite{kravtsov2004a} and \cite{tassis2008a} to
successfully reproduce the faint end of the luminosity function of
galactic satellites and scaling relations of dwarf galaxies.
\cite{kaufmann2007a} used hydrodynamical simulations of the
dissipative formation of disks to demonstrate that the temperature
floor of $T=10^4$~K results in
the equilibrium disk scale height-to-scale length ratio in systems
with $\vcirc\sim40\kms$ to be roughly three times larger than for
galaxies with $\vcirc\sim80\kms$.  The disk morphologies in our GD-SF runs are
qualitatively consistent with their results.

\subsubsection{Molecular Fraction vs. Disk Radius}
\label{section:results:fH2_vs_radius}

The variations in the molecular fraction $\fHmol$ as a function of the
gas volume density $\nH$ discussed in \S
\ref{section:methodology:fH2_vs_density} map into $\fHmol$ gradients with
surface density or disk radius on the global scale of a galaxy model.
Figure \ref{fig:fH2_vs_radius} shows azimuthally-averaged molecular
fraction $\fHmol$ as a function of radius normalized to the disk scale length
in the simulated disk galaxies with
$\vcirc=50-300\kms$ for molecular ISM models $\Hmol$D-SF+ISRF and
$\Hmol$D-SF, after $t=0.3\,\Gyr$ of evolution.  The average molecular
fraction decreases with decreasing rotational velocity, from $\fHmol >
0.6$ within a disk scale length for the $\vc = 300\kms$ system (left
panel) to $\fHmol<0.4$ within a disk scale length for the $\vc = 50
\kms$ system (right column) in the $\Hmol$D-SF+ISRF model.  The radial
dependence of the molecular fraction steepens in smaller mass galaxies
relative to the $\vcirc=300\kms$ model since a larger fraction 
of gas has densities below the
atomic-to-molecular transition.  
The impact of this trend on the global star
formation relation is discussed below in \S
\ref{section:results:total_gas_schmidt_law}.
We note that details of the model initial conditions can 
influence the radial distribution of the molecular gas by changing
the typical gas density in the disk.  In particular, the $\vc = 125 \kms$
model galaxy has a more centrally concentrated molecular gas distribution than does 
M33.  

 Figure
\ref{fig:fH2_vs_radius} clearly shows that the ISRF in the
$\Hmol$D-SF+ISRF model steepens the dependence of molecular fraction
$\fHmol$ on the radius $R$ for each galaxy compared with the
$\Hmol$D-SF model.

\subsubsection{Star Formation Histories}
\label{section:results:sfr}

The star formation histories of the simulated galaxies
provide a summary of the efficiency of 
gas consumption for systems with different disk structures
and models for ISM physics. Figure \ref{fig:sfr} shows the
SFR over the time interval $t=0-1\,\Gyr$
for the isolated galaxy models evolved with the GD-SF,
$\Hmol$D-SF, and $\Hmol$D-SF+ISRF models.  In most simulations, the SFR
experiences a general decline as the gas is slowly converted
into stars.  Substantial increases in the SFR with time are 
driven either by the formation of bar-like structures
(e.g., after $t\approx 0.7\Gyr$ for the $\vcirc=300\kms$
galaxy) or, in one case,
fragmentation of the cold ISM (after $t\approx 0.4\Gyr$ for 
the $\vcirc=50\kms$ galaxy evolved with the $\Hmol$D-SF
model).

For the most massive galaxy,
differences in the
relative star formation efficiencies of models 
that tie the SFR to the total gas density (the GD-SF model)
or to the molecular gas density (the $\Hmol$D-SF and
$\Hmol$D-SF+ISRF models) are not large.  The ISM in the
massive galaxy is dense, and when $\Hmol$ abundance is modeled
much of the gas becomes molecular 
(see Figure \ref{fig:fH2_vs_radius}).  When 
$\Hmol$-destruction is included the molecular content
of the ISM in this galaxy is reduced at large radii and 
small surface densities, but much of the mass of the ISM (e.g., interior 
to a disk scale length) retains a molecular fraction comparable 
to the ISM model without $\Hmol$-destruction. The
effect of the ISRF on the SFR of this galaxy is fairly small.

The SFR history of the intermediate-mass ($\vcirc=125\kms$) galaxy model
is more strongly influenced by the $\Hmol$-destroying ISRF.
While the GD-SF and $\Hmol$D-SF ISM models produce similar
star formation efficiencies, reflecting the substantial molecular fraction
in the ISM for the $\Hmol$D-SF model, the effects of 
$\Hmol$-destruction in the $\Hmol$D-SF+ISRF model
reduce the SFR by $\sim30-40\%$.  The time derivative of the SFR
in the 
$\Hmol$D-SF+ISRF 
 model 
is also shallower than for
the 
GD-SF or 
$\Hmol$D-SF models, reflecting a reservoir 
of non-star-forming gas.
In each model, the SFR of the whole system 
is similar to that estimated for M33 
\citep[SFR$\sim0.25-0.7\,\Msun\yr^{-1}$,][]{kennicutt1998a,hippelein2003a,blitz2006a,gardan2007a}.

The effects of $\Hmol$-destruction play
a crucial role in determining the star
formation history of the $\vcirc=50\kms$ dwarf system.
The single phase ISM (GD-SF) model produces a
SFR history qualitatively similar to the
more massive galaxies, but in the
$\Hmol$D-SF model the star formation
history changes dramatically.   The colder ISM
temperature allows the gaseous disk to become 
substantially
thinner in this case compared to the GD-SF model, leading
to higher densities and SFRs.
After $t\approx0.4\Gyr$ of evolution, the
cold, dense gas in the $\Hmol$D-SF model
allows the ISM in the dwarf to undergo large-scale
instabilities and, quickly thereafter, fragmentation.
The dense knots of gas rapidly increase their
SFE compared with the
more diffuse ISM of the GD-SF model and
the SFR increases to a rate comparable to
that calculated for the $\vcirc=125\kms$ model.
In the $\Hmol$D-SF+ISRF model,
the equilibrium temperature of most diffuse (and metal
poor) gas in the dwarf is increased and the
vertical structure of the disk becomes 
thicker than in the $\Hmol$D-SF model.  The molecular content
of the disk is thus substantially reduced
(see the right panel of 
Figure \ref{fig:fH2_vs_radius}) and the
disk remains diffuse.  As a result, the
SFR drops to a low and roughly constant
level.  The constancy of the SFR in the
$\Hmol$D-SF+ISRF model is a result of
the large gas reservoir that remains
atomic and is therefore unavailable
for star formation.  As the molecular gas,
which comprises a small fraction of the 
total ISM in the dwarf ($\fHmol\lesssim0.1$
by mass),
is gradually consumed it is continually
refueled by the neutral gas
reservoir.  

The star formation history of the dwarf galaxy system in the
$\Hmol$D-SF+ISRF model demonstrates that physics other than energetic
supernova feedback can regulate star formation and produce long gas
consumption time scales in dwarf galaxies.  
Star formation rates in
real galaxies similar to the $\vcirc=50\kms$ model fall in the wide
range $\SFR\sim10^{-4}-0.1\Msun\yr^{-1}$ \citep[e.g.,][]{hunter2004a}.
While the $\Hmol$D-SF+ISRF simulation of the dwarf galaxy produces a
steady $\SFR\sim0.025\Msun\yr^{-1}$, we note that the galaxy DDO 154
that served as a model for the $\vcirc=50\kms$ system has a star
formation rate of only $\SFR\sim0.004\Msun\yr^{-1}$
\citep{hunter2004a}.

\subsection{Star Formation Relations in Galactic Disks}
\label{section:results:sfr_laws}

The disk-averaged SK relation in galaxies, determined by
\cite{kennicutt1998a}, is well described by the power-law
\begin{eqnarray}
\SigmaSFR &=& (2.5\pm0.7)\times 10^{-4}\nonumber\\
&&\times \left(\frac{\Sigmagas}{1\Msun\pc^{-2}}\right)^{1.4\pm0.15} \Msun \yr^{-1}\kpc^{-2}.
\end{eqnarray}
\noindent
However, as discussed in the Introduction, spatially-dependent determinations of the
total gas SK relation slope vary in the range
$\ntot=1.7-3.55$.  While 
the slope of the spatially-resolved molecular gas
SK relation is consistently measured to be
$\nmol\approx1.4$ in the same systems, with a two-sigma
variation of $\nmol\sim1.2-1.7$
\citep{wong2002a,boissier2003a,heyer2004a,boissier2007a,kennicutt2007a}.
Below, we examine SK relations in the simulated disks and compare
our results directly with these observations.

\begin{figure*}
\figurenum{6}
\epsscale{1.0}
\plotone{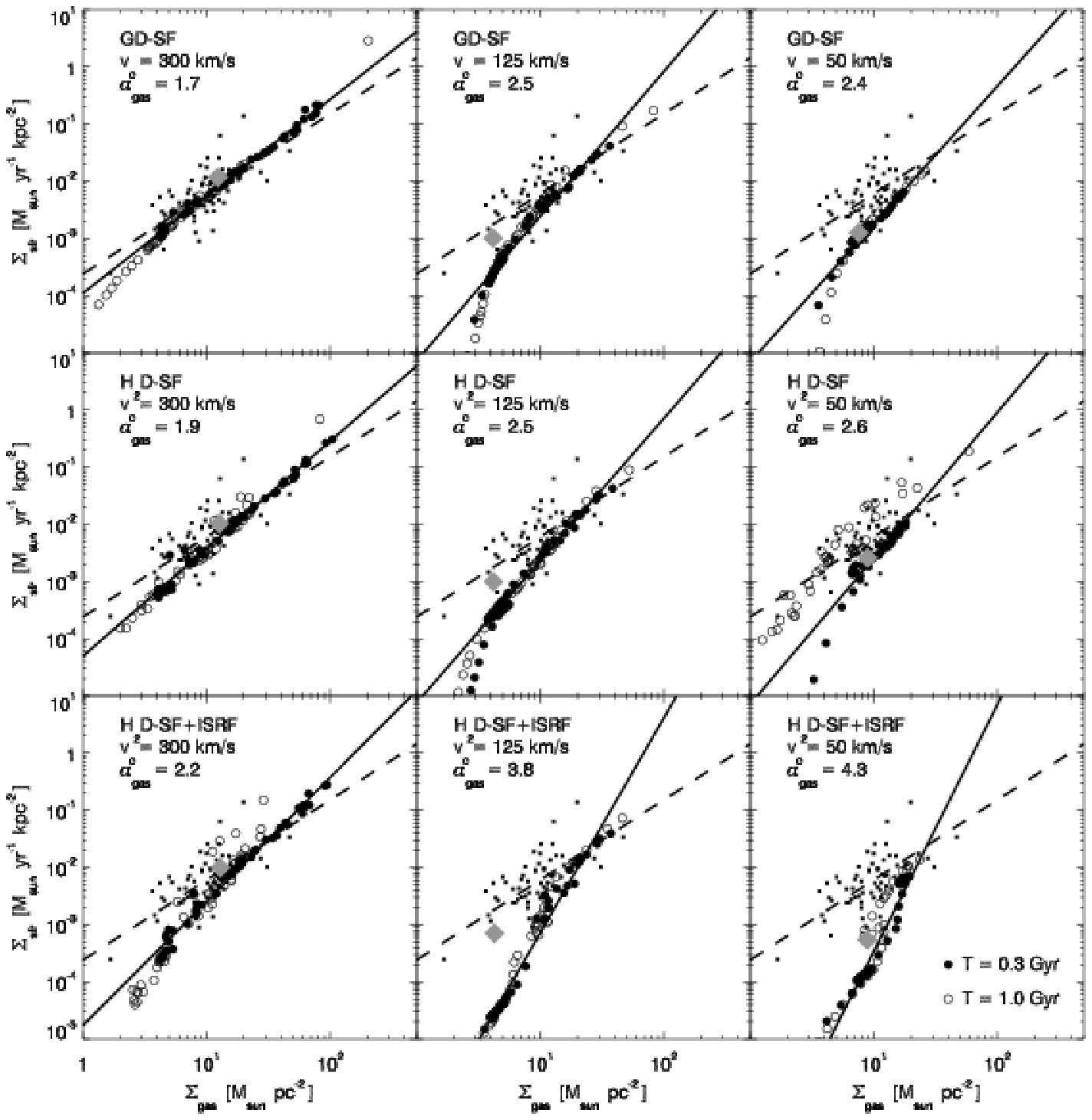}
\caption{\label{fig:kennicutt.gas}
\small
Schmidt-Kennicutt relation for total gas surface densities.  Shown is the SFR surface density
$\SigmaSFR$ as a function of the total gas surface density $\Sigmag$ for galaxy models with circular
velocities of $\vcirc=50\kms$ (right column), $\vcirc=125\kms$ (middle column), and $\vcirc=300\kms$ 
(left column) for 
ISM models that include atomic coolants (upper row), atomic+molecular coolants with (lower row) 
and without (middle row)
$\Hmol$-destruction from a local interstellar radiation field.  The surface densities $\SigmaSFR$ and
$\Sigmagas$ are measured in annuli from the simulated disks after evolving for $t=0.3\,\Gyr$ (solid circles)
and $t=1\,\Gyr$ (open circles) in 
isolation.  
The disk-averaged data from \cite{kennicutt1998a} is shown for comparison (small dots).
The best power-law fits to the individual galaxies have indices in the range $\alpha\approx1.7-4.3$ (solid lines).
Deviations from the $\SigmaSFR \propto \Sigmagas^{1.5}$ relation are caused by the radial variation in
the molecular gas fraction $\fHmol$, the scale height of star-forming gas $\hSFR$, and the scale height
of the total gas distribution $\hgas$.
The total disk averaged surface densities (grey diamonds) are consistent with
the \cite{kennicutt1998a} normalization for normal, massive star-forming galaxies (dashed lines).
}
\end{figure*}

\begin{figure*}
\figurenum{7}
\epsscale{1.1}
\plotone{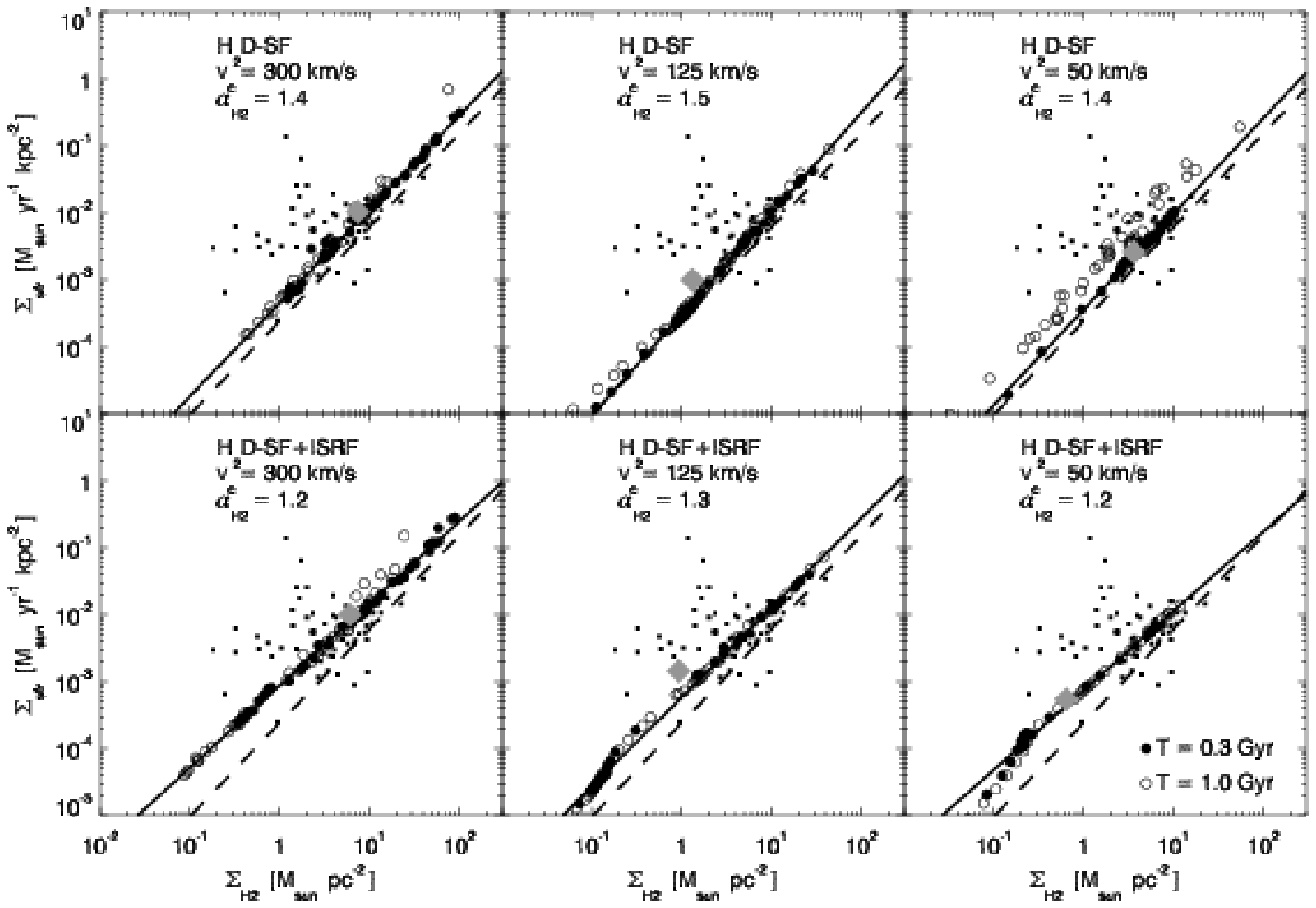}
\caption{\label{fig:kennicutt.H2}
Schmidt-Kennicutt relation for molecular gas surface densities.  Shown
is the star formation rate surface density $\SigmaSFR$ as a function
of the molecular gas surface density $\SigmaHmol$ for galaxy models
with circular velocities of $\vcirc=50\kms$ (right panels),
$\vcirc=125\kms$ (middle panels), and $\vcirc=300\kms$ (left panels)
and ISM models that include atomic+molecular coolants with (lower
panels) and without (upper panels) $\Hmol$-destruction from a local
interstellar radiation field.  The surface densities $\SigmaSFR$ and
$\SigmaHmol$ are measured in annuli (dark circles) from the simulated
disks after evolving for $t=0.3\Gyr$ (solid circles) and $t=1\Gyr$
(open circles) in isolation.  The disk-averaged data from
\cite{kennicutt1998a} is shown for comparison (small dots).  The best
power-law fits to the individual galaxies have indices in the range
$\alpha\approx1.2-1.5$ (solid lines).  The total disk averaged surface
densities (grey diamonds) are consistent with the
\cite{kennicutt1998a} relation for normal, massive star-forming
galaxies (dashed lines).  }
\end{figure*}

\subsubsection{The Total Gas Schmidt-Kennicutt Relation}
\label{section:results:total_gas_schmidt_law}

A dichotomy between the total gas and molecular gas SK
relations is currently suggested by the data and should be a feature
of models of the ISM and star formation.  Figure
\ref{fig:kennicutt.gas} shows the simulated total gas
SK relation that compares the SFR
surface density $\SigmaSFR$ to the total gas surface density $\Sigmag$
for different galaxy models.  The quantities $\SigmaSFR$ and $\Sigmag$
are azimuthally-averaged in annuli of width $\Delta r =150h^{-1}\,\pc$
for the $\vcirc=300\,\kms$ and $\vcirc=125\,\kms$ systems and $\Delta r
=75h^{-1}\,\pc$ for the smaller $\vcirc=50\,\kms$ disk.  Regions within
the radius enclosing $95\%$ of the gas disk mass are shown.  The
SK relation is measured in each disk at times
$t=0.3\,\Gyr$ and $t=1.0\,\Gyr$ to monitor its long-term evolution.
Power-law fits to the SFR density as a function of the total gas
surface mass densities in the range $\Sigmag=5-100\Msun\pc^{-2}$
measured at $t=0.3\,\Gyr$ are indicated in each panel, with the range
chosen to approximate regimes that are currently observationally
accessible.  The simulated relations between $\SigmaSFR$ and $\Sigmag$
are not strict power-laws, and different choices for the range of
surface densities or simulation time of the fit may slightly change
the inferred power-law index, with lower surface densities typically
leading to larger indices and, similarly, later simulation times
probing lower surface densities and correspondingly steeper indices.
Exceptions are noted where appropriate, but the conclusions of this
paper are not strongly sensitive to the fitting method of the
presented power-law fits.  Disk-averaged SFR and surface mass
densities are also measured and plotted for comparison with the
non-starburst galaxies of \cite{kennicutt1998a}.

Each row of panels in Figure \ref{fig:kennicutt.gas} shows results
for the three galaxy models
evolved with different models for the 
ISM and star formation.  
For the traditional GD-SF model,  the $\SigmaSFR-\Sigmag$ correlation 
in massive, thin-disk systems 
tracks the input relation for this model, 
$\dot{\rho}_{\star}\propto\rho_{g}^{1.5}$, at intermediate surface
densities.  At low surface densities, massive disks deviate from the
$\SigmaSFR\propto\Sigmag^{1.5}$ relation as the imposed density threshold
is increasingly probed in the disk interior and disk flaring begins to
steepen the relation between the volume density $\rho_{g}$
and the gas surface density $\Sigmag$.    In lower mass galaxies,
the disks are thicker at a given gas surface density, as dictated
by hydrostatic equilibrium \citep[e.g.,][]{kaufmann2007a}, with a
correspondingly lower average three-dimensional density.  The SK
relation therefore steepens somewhat between galaxies with $\vcirc=300\kms$ and
$\vcirc=125\kms$ in this star formation model, but does not fully
capture the mass-dependence of the SK relation 
apparent in the observations.

The middle row of Figure \ref{fig:kennicutt.gas} shows the total gas
SK relation measured in disks evolved with the $\Hmol$D-SF model,
which are generally similar to the more prescriptive GD-SF model in
terms of the relation slope.  The arbitrary density threshold for star
formation included in the GD-SF model roughly mimics the physics of
the atomic-to-molecular gas transition calculated by Cloudy for the
$\Hmol$D-SF model, as molecular transition begins at densities of
$\nH\sim0.1\cc$ when the effects of $\Hmol$-destruction by an ISRF are
\it not \rm included.

Note that the $\Hmol$D-SF model allows gas to cool to considerably
lower temperatures than in the atomic cooling-only GD-SF model. This
allows the ISM gas in the $\vcirc=50\kms$ galaxy to become cold enough
to settle into a thin, dense disk and reach high molecular
fractions. Indeed, the increase in the molecular content of the disk
contributes to the increase in the SFE in the dwarf galaxy at $t=1$~Gyr during its
evolution seen in the middle right panel of
Figure~\ref{fig:kennicutt.gas}.  However, observed
$\vcirc\approx50\kms$ galaxies have \it very low \rm molecular
fractions and the high molecular content of the dwarf at $t=1\,\Gyr$
is a potential deficit of the $\Hmol$D-SF model.

The bottom row of Figure \ref{fig:kennicutt.gas} shows the
SK relation for the most complete "$\Hmol$D-SF+ISRF" model.
In this model, the
SK relation power-law index increases systematically from
$\ntot=2.2$ for the $\vcirc=300\kms$ galaxy to $\ntot=4.3$ for the
$\vcirc=50\kms$ dwarf.  The decreased abundance of $\Hmol$ at low
densities has steepened the SK relation at low gas
surface densities compared with either the GD-SF or $\Hmol$D-SF
models.  The ISM in the dwarf galaxy is prevented from becoming fully
molecular and maintains an extremely steep SK relation
over its evolution from $t=0.3\,\Gyr$ to $t=1\,\Gyr$. Consequently,
the global SFR of the model dwarf galaxy in this case
stays low and approximately constant. The global gas consumption time
scale is also much longer than in the other two models.  

While the disk-averaged SK relation
determination for the galaxies have the lowest SFE
in the $\Hmol$D-SF+ISRF model, the
values are still consistent with the observational determinations by
\cite{kennicutt1998a}, \cite{wong2002a}, \cite{boissier2003a}, and
\cite{boissier2007a}.

\subsubsection{The Molecular Gas Schmidt-Kennicutt Relation}
\label{section:results:molecular_gas_schmidt_law}

 As discussed in \S \ref{section:results:sfr_laws}, the two-sigma
range in observational estimates of the molecular gas SK relation
power-law index is $\nmol\approx1.2-1.7$, and is smaller than the
variation in the slope of the $\SigmaSFR-\Sigmag$ relation.  Figure
\ref{fig:kennicutt.H2} shows the molecular gas SK relation measured in
galaxies evolved with the two molecular ISM models considered in this
paper.  The SFR and molecular gas surface densities are measured from
the simulations as described in \S
\ref{section:results:total_gas_schmidt_law} and the reported power-law
indices were fit to molecular gas surface densities in the range
$\SigmaHmol=0.1-100 \Msun\pc^{-2}$. Note that the correlation of 
the SFR 
surface density with the $\Hmol$ surface
density is not an entirely trivial consequence of 
our H$_2$-based implementation
of star formation, as we implement our star formation model 
on the scale
of individual gas particles in three dimensions 
and the observational correlations we consider are
azimuthal surface density averages in annuli. The two are not equivalent, as we
discuss below in the next subsection \citep[see also][]{schaye2007a}.

 The disk-averaged SFR and molecular
gas surface densities  in the $\Hmol$D-SF model are consistent with observations
of the molecular gas SK relation by
\cite{kennicutt1998a}.  The molecular gas SK relation
power-law index of this model is similar for galaxies of different
mass $\nmol=1.4-1.5$. Unlike the $\SigmaSFR-\Sigmag$ relation, the
molecular gas SK relation is well-described by a
power-law over a wide range in surface densities in the $\Hmol$D-SF
model.  As discussed previously, the low-mass dwarf galaxy
experiences an increase in the SFE at a fixed
surface density during the evolutionary period between $t=0.3\,\Gyr$ and
$t=1.0\,\Gyr$ owing to a nearly complete conversion of its ISM
to molecular gas in the $\Hmol$D-SF model.

Panels in the bottom row of Figure \ref{fig:kennicutt.H2} show 
that the addition of a local interstellar radiation field
in the $\Hmol$D-SF+ISRF model controls the amplitude of the
$\SigmaSFR-\SigmaHmol$ relation, but does not strongly affect its slope. 
The normalization and power-law index of the molecular gas
SK relation in this model are quite stable with time,
reflecting how the equilibrium temperature and molecular fractions of
the ISM in the $\Hmol$D-SF+ISRF are maintained between simulation
times $t=0.3\Gyr$ and $t=1.0\Gyr$.  Only the most massive galaxy
experiences noticeable steepening of the $\SigmaSFR-\SigmaHmol$
correlation at high surface densities.  The disk-averaged observations
of the molecular gas SK relation by
\cite{kennicutt1998a} are well-matched by the simulations and remain
approximately constant over their duration.  

While the scale height of the cold, dense gas in the disks is
typically comparable to our gravitational softening, the slope of the
molecular gas SK relation appears to be insensitive to modest changes
in the simulation resolution.  Additional resimulations of the
calculations presented here with an increased resolution of
$N_{\mathrm{d,g}}=1.2 \times 10^{6}$ gas particles show that the slope
of the molecular gas SK relation changes by less than $5\%$.

\begin{figure*}
\figurenum{8}
\epsscale{1.15}
\plotone{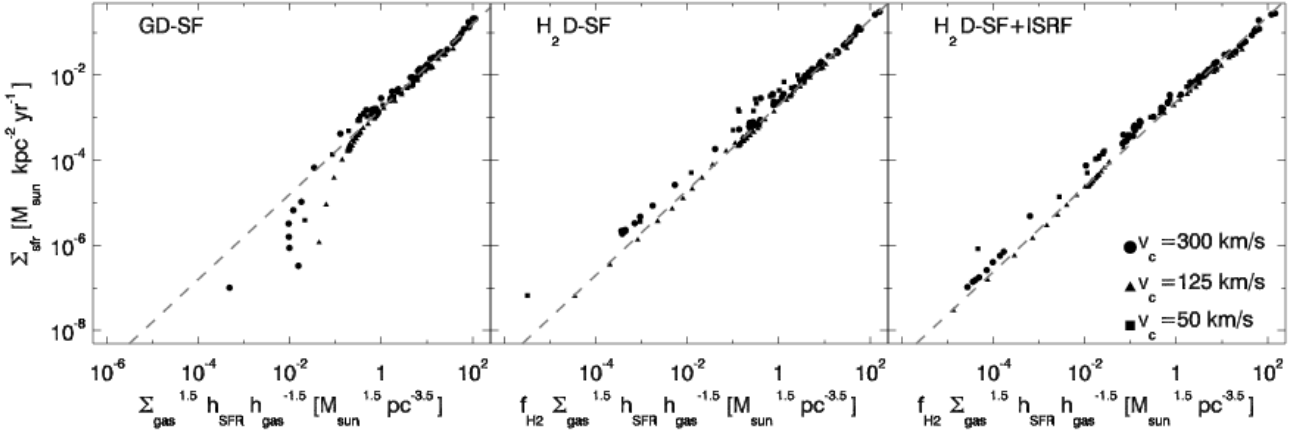}
\caption{\label{fig:new_sfr_law}
A predicted star formation correlation for disk galaxies.  The star formation rate (SFR) density in the simulations
is calculated as $\dot{\rho}_{\star} = \fHmol \rhog / \tstar \propto \fHmol(\rhog) \rhog^{1.5}$, 
where $\fHmol$ is the molecular fraction, $\rhog$ is the gas density, and $\tstar$ is a star formation
time-scale that varies with the local gas dynamical time.  The local SFR surface density
should scale as $\SigmaSFR\propto\dot{\rho}_{\star}\hSFR$, where $\hSFR$ is the scale height of
star-forming gas.  Similarly, $\Sigmagas\propto\rhog \hgas$, where $\hgas$ is the total gas scale height.
The simulations predict that the SFR density in disks should correlate with the
gas surface density as $\SigmaSFR \propto \fHmol \hSFR \hgas^{-1.5} \Sigmagas^{1.5}$ (dashed lines).
}
\end{figure*} 

\subsubsection{Structural Schmidt-Kennicutt Relation}
\label{section:results:structural_schmidt_law}

While the observational total and molecular gas SK
relations are well-reproduced by the simulated
galaxy models, local deviations from a power-law
$\SigmaSFR-\Sigmag$ or $\SigmaSFR-\SigmaHmol$
correlation are abundant.  Although these deviations
are physical, exhibited by real galaxies, and
are well-understood within the context of the
simulations, a question remains as to whether a
relation that trivially connects the
SFR density $\SigmaSFR$ and the properties of
the disks exists.

The primary relation between the local star formation
rate and the gas properties in 3D  in the simulations
are encapsulated by Equation \ref{equation:star_formation_rate},
which relates the three-dimensional SFR
density to the 
molecular gas density.  We can relate the three-dimensional properties in the
ISM to
surface densities 
through the characteristic scale heights of the disks.
The SFR surface density is related to an average
local SFR volume density 
$\left<\dot{\rho}_{\star}\right>$ through the scale height
of star-forming gas $h_{\SFR}$ as
\begin{equation}
\SigmaSFR \propto \left< \dot{\rho_{\star}} \right> h_{\SFR}
\end{equation}
\noindent
Here, the averaging of the three-dimensional density is understood
to operate over the region of the disk where the two-dimensional
surface density is measured.
Similarly, the gas surface and volume densities are related by
the proportionality
\begin{equation}
\Sigmagas \propto \left<\rhog\right> h_{g}.
\end{equation}
\noindent
If Equation \ref{equation:star_formation_rate} is averaged over the scale on which
$\SigmaSFR$ is measured as
\begin{equation}
\left<\dot{\rho}_{\star} \right>\propto \fHmolave \left<\rhog\right>^{1.5},
\end{equation}
\noindent
then the proportionalities can be combined to give
\begin{equation}
\label{equation:structural_schmidt_law}
\SigmaSFR \propto  \fHmolave \frac{h_{\SFR}}{h_{g}^{1.5}} \Sigmag^{1.5}.
\end{equation}
\noindent
Equation \ref{equation:structural_schmidt_law} can be interpreted
as a SK-like correlation, which depends {\it both\/}
on the molecular gas abundance and the structure of the disk.

Figure \ref{fig:new_sfr_law} shows this expected structural SK
relation for the simulated disk galaxies.  The star formation,
and surface densities are azimuthally-averaged in annuli, as before, and
limited to regions of the disk where star formation
is active.
The characteristic gas scale height
$h_{\mathrm{gas}}$ is determined as a
mass-weighted average of the absolute value of
vertical displacement for the annular gas distributions.
The characteristic star-forming gas scale height
$h_{\mathrm{SFR}}$ is determined in a similar
manner, but for gas that satisfies the condition
for star formation: namely, $\fHmol>0$ for the $\Hmol$D-SF and
$\Hmol$D-SF+ISRF models or $\nH>0.1\cc$ for the GD-SF model.   
Measurements
from the simulations are presented after evolving the
systems for $t=0.3\Gyr$ with the GD-SF, 
$\Hmol$D-SF,
and $\Hmol$D-SF+ISRF ISM models.
In each case,  as expected, $\SigmaSFR$ in the disks linearly 
correlates with 
quantity
$\fHmol \hSFR \hgas^{-1.5} \Sigmagas^{1.5}\,\,$\footnote{Small 
deviations in the GD-SF model at low surface densities 
owe to the sharp density threshold for star formation adopted
in this model \citep[see][for a related discussion]{schaye2007a}.}.

This relation shows explicitly that the behavior of the $\SigmaSFR-\Sigmagas$
relation can be understood in terms of the dependence of $\fHmol$ and
$\hgas$ on $\Sigmagas$. For example, the SK
relation can have a
fixed power law $\SigmaSFR\propto\Sigmagas^{1.5}$ when $\fHmol$ and
$\hgas$ are roughly independent of $\Sigmagas$ \citep[e.g., M51a,][]{kennicutt2007a}, 
but then
steepen when the mass fraction of star forming gas becomes a strong function of gas
surface density in a manner characteristic of dwarf galaxies, large low
surface brightness galaxies, or the outskirts of normal
galaxies. Deviations from the $\Sigmagas^{1.5}$ relation can also
occur when $\hgas$ changes quickly with decreasing $\Sigmagas$ in the
flaring outer regions of the disks. Such flaring, for example, is observed
for the Milky Way at $R\gtrsim 10$~kpc
\citep[e.g.,][and references therein]{wolfire2003a}.

\begin{figure}
\figurenum{9}
\epsscale{1.1}
\plotone{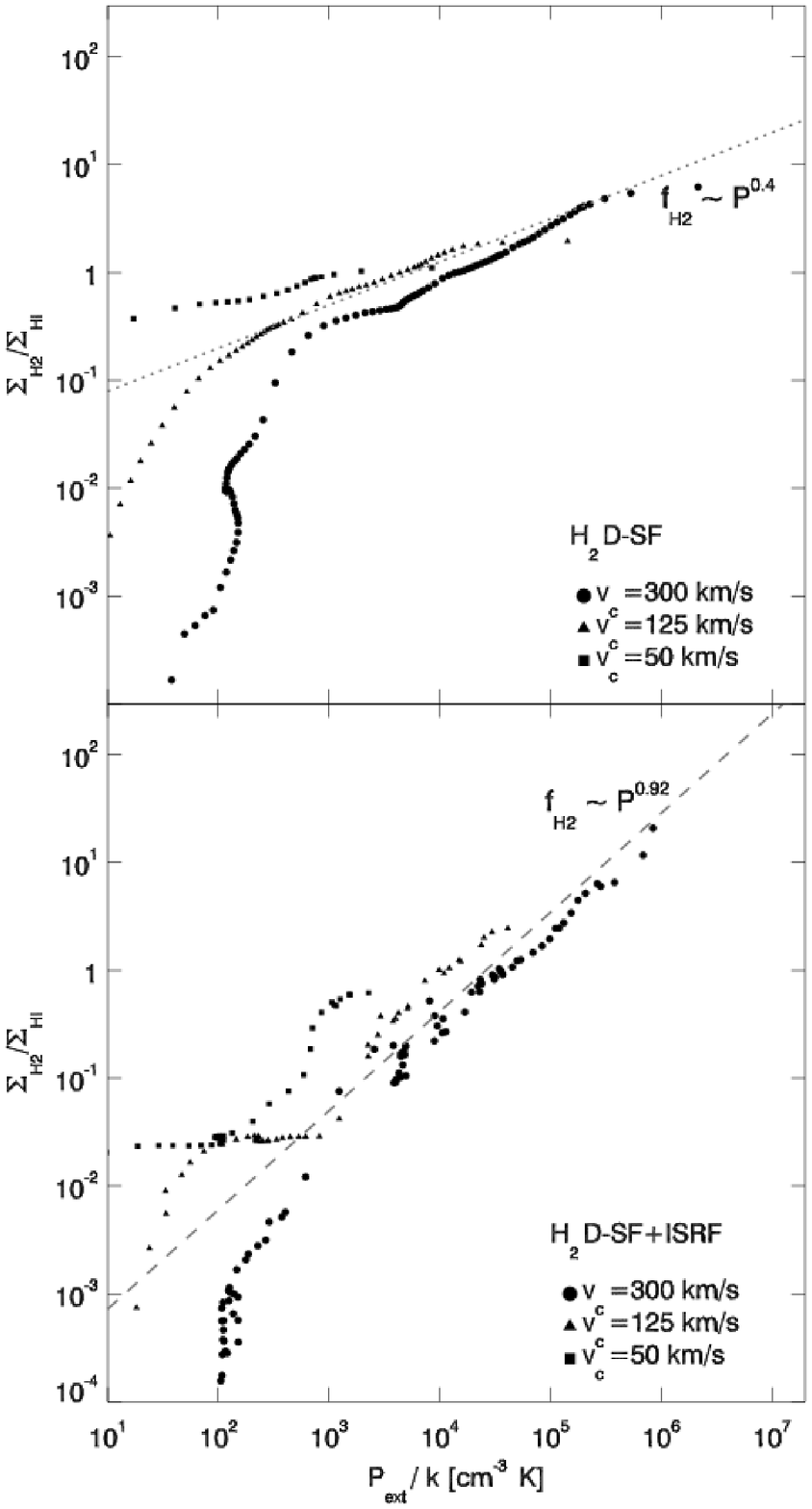}
\caption{\label{fig:fH2_pressure}
\small
Molecular fractions as a function of external pressure in isolated disks.  Shown
are the molecular-to-atomic gas surface density ratios $\SigmaHmol/\SigmaHI$
measured in radial bins as a function of the local external pressure $\Pext$
defined by \cite{blitz2004a} for galaxies with $\vcirc\approx300\,\kms$
(circles), $\vcirc\approx125\,\kms$ (triangles), and
$\vcirc\approx50\,\kms$ (squares) circular velocities.  The simulations that include $\Hmol$-destruction
by an ISRF (lower panel) follow the observed $\SigmaHmol/\SigmaHI\propto\Pext^{0.9}$ power-law scaling (dashed line),
while the simulations that do not include an ISRF (upper panel) have a weaker scaling than
that found observationally ($\SigmaHmol/\SigmaHI\sim\Pext^{0.4}$, dotted line).
\\
}
\end{figure}

\subsection{Local Pressure, Stellar Radiation, and $\Hmol$-Fraction}
\label{section:results:fH2_pressure}

\cite{elmegreen1993b} calculated the dependence of the
ISM molecular fraction $\fHmol$ on the local pressure
and radiation field strength, including the effects of $\Hmol$
self-shielding and dust extinction.  The calculations
of \cite{elmegreen1993b} suggested that
\begin{equation}
\fHmol \propto \Pext^{2.2}j^{-1},
\end{equation}
\noindent
where $\Pext$ is the external pressure and $j$ is the volume
emissivity of the radiation field.  
\cite{wong2002a} used observations to
determine the correlation between molecular fraction $\fHmol$
and the hydrostatic equilibrium mid-plane pressure in galaxies, given by
\begin{equation}
\Pext\approx\frac{\pi}{2} G \Sigmagas\left(\Sigmagas + \frac{\sigma}{c_{\star}}\Sigma_{\star}\right),
\end{equation}
where $c_{\star}$ is the stellar velocity dispersion and $\sigma$ is the gas
velocity dispersion 
\citep[see][]{elmegreen1989a}.
From observations of seven disk galaxies and the Milky Way, \cite{wong2002a}
determined that the observed molecular fraction scaled with pressure as
\begin{equation}
\SigmaHmol/\SigmaHI \propto \Pext^{0.8}.
\end{equation}
Subsequently,
\cite{blitz2004a} and \cite{blitz2006a}
studied these correlations in a larger
sample of 28 galaxies.  \cite{blitz2006a}
found that if the mid-plane pressure was
defined as 
\begin{eqnarray}
\label{equation:pext}
\frac{\Pext}{k} &=& 272\,\,\cc\,\K\,\left(\frac{\Sigma_{g}}{\Msun \pc^{-2}}\right)\left(\frac{\Sigma_{\star}}{\Msun \pc^{-2}}\right)^{0.5}\nonumber\\
&\times&\left(\frac{\sigma}{\kms}\right)\left(\frac{h_{\star}}{\pc}\right)^{-0.5},
\end{eqnarray}
\noindent
where $h_{\star}$ is the stellar scale height, 
and if the local mass density was dominated by
stars, 
then
the molecular fraction scaled with $\Pext$ as
\begin{equation}
\SigmaHmol/\SigmaHI \propto \Pext^{0.92}.
\end{equation}
\noindent
\cite{wong2002a} state that if the 
ISRF volume emissivity 
$j\propto\Sigma_{\star}\propto\Sigma_{g}$ 
and the stellar velocity dispersion was constant,
then the \cite{elmegreen1993b} calculations 
would provide a relation $\fHmol\propto \Pext^{1.7}$
much steeper than observed.  \cite{blitz2006a}
suggested that if the stellar velocity dispersion
scales as $c_{\star}\propto\Sigma_{\star}^{0.5}$
and $j\propto\Sigma_{\star}$, then the predicted
scaling is closer to the observed scaling.  
These assumptions provide $\fHmol\propto \Pext^{1.2}$.

The relation between $\SigmaHmol/\SigmaHI$ and $\Pext$ 
can be measured for the simulated disks.
Figure~\ref{fig:fH2_pressure} shows $\SigmaHmol/\SigmaHI$
as a function of $\Pext$, measured from the simulations
according to Equation \ref{equation:pext} in annuli, 
for simulated disks with (bottom panel) and without (top
panel) $\Hmol$-destruction by an ISRF.  The simulations
with an ISRF scale in a similar way to the systems
observed by \cite{blitz2006a}, while the simulated disks
without an ISRF have a weaker scaling.  The simulated
galaxies have $c_{\star}\propto\Sigma_{\star}^{0.5}$ and
roughly constant gas velocity dispersions.

Given the scaling of the ISRF strength in the simulations, for which
$\Uisrf\propto\SigmaSFR\propto\Sigmag^{\ntot}$ with $\ntot\sim2-4$ according
to the total gas SK relations measured in 
\S \ref{section:results:total_gas_schmidt_law},
and the scale-dependent SFE,
which will instill the proportionality $\Sigma_{\star}\propto\Sigmag^{\beta}$
with $\beta\ne1$ in general,
the scaling between the molecular fraction
and pressure
$\fHmol\propto \Pext^{0.9}$ is not guaranteed to
be satisfied in galaxies with greatly varying SK
relations.
However, given the relation $\fHmol\propto \Pext^{2.2}j^{-1}$
calculated by \cite{elmegreen1993b}, a
general relation between the molecular fraction
and interstellar pressure can be calculated in
terms of the power-law indices $n$ of the total
gas SK relation and $\beta$ that describes the
relative distribution gas and stars in the disk.

Assuming a steady state, the
volume emissivity in the simulated ISM then 
scales with the gas surface density
as
\begin{equation}
j \propto \Uisrf \propto \Sigmag^{\ntot}.
\end{equation}
\noindent
Given the ISRF strength in the simulations
and a proportionality of the stellar surface
density with the total gas surface density
of the form $\Sigma_{\star}\propto\Sigmag^{\beta}$,
we find that, with
the scaling between the molecular
fraction, pressure, and ISRF volume emissivity calculated
by \cite{elmegreen1993b}, the $\fHmol-\Sigmag$ relation 
should follow
\begin{equation}
\fHmol \propto \Sigmag^{2.2(1+\beta/2) - n}.
\end{equation}
\noindent
If the molecular fraction scales with the mid-plane
pressure as $\fHmol\propto P^{\alpha}$, then
the index $\alpha$ can be expressed in terms of
the power-law dependences of the SK relation
and the $\Sigma_{\star}-\Sigmag$ correspondence as
\begin{equation}
\label{equation:general_fH2_pressure}
\alpha = \frac{2.2(1+\beta/2) - n}{1+\beta/2}.
\end{equation}
\noindent
The largest galaxy simulated, with a circular velocity
of $\vcirc\approx 300 \kms$, has a total gas SK
relation with a scaling $\SigmaSFR \propto \Sigmag^{2}$ (see Figure
\ref{fig:kennicutt.gas}) and a nearly linear scaling
of the stellar and gas surface densities ($\beta\approx0.97$).
According to Equation \ref{equation:general_fH2_pressure},
power-law index for this system is then expected to follow
$\alpha=0.85$.  At the opposite end of the
mass spectrum, the  $\vcirc=50\kms$ dwarf galaxy, which 
obeys a very different SK relation ($\ntot\approx4$) and
$\Sigma_{\star}-\Sigmag$ scaling ($\beta\sim4$), has an expected 
$\alpha\approx0.87$.
These $\alpha$ values are remarkably close to the relation observed
by \cite{blitz2006a}, who find $\alpha\approx0.92$.
That the ISRF model
reproduces the observed correlation in galaxies of different
masses then owes to the opposite compensating effects of the scaling
of $j$ with $\Sigmagas$ and the $\Sigma_{\star}-\Sigmag$ scaling. 

For the simulations that do not include an ISRF, the
scaling between molecular fraction and pressure must
be independent of the ISRF volume emissivity.  We
find that $\fHmol \sim \Sigmag^{0.6}$ in these galaxies,
with significant scatter, which provides the much weaker
relation $\SigmaHmol/\SigmaHI\sim \Pext^{0.4}$.  The
simulations without an ISRF have $\fHmol-\Pext$ scalings
similar to this estimate or intermediate between this
estimate and the observed relation.

The good agreement between the $\fHmol-\Pext$
scaling in the simulations with an ISRF and the
observations provide an \it a posteriori \rm justification
for the chosen scaling of the ISRF strength in the
simulations (Equation \ref{equation:uisrf}), even as
the ISRF strength was physically motivated by the
generation of soft UV photons by young stellar populations.
The robustness
of the $\fHmol-\Pext$ may be a consequence of the 
regulatory effects of $\Hmol$-destruction
by the ISRF, which motivated in part the original
calculations by \cite{elmegreen1993b}, in concert
with the influence of the external pressure \citep{elmegreen1994a}.
However, we note that systems with a steep total gas SK relation,
such as the $\vcirc\sim50\kms$ and $\vcirc\sim125\kms$
systems, do break from the $\fHmol-\Pext$ relation in
the very exterior of the disk where the molecular fraction
quickly declines.
Observations of M33, the real galaxy analogue to the $\vcirc\sim125\kms$
simulated galaxy,
indicate that the system may deviate from the 
observed $\fHmol-\Pext$
relation determined at higher surface densities in the very exterior
of the disk 
\citep{gardan2007a}.
The surface density of M33 in these exterior regions is dominated by   
gas rather than stars, and the \cite{blitz2006a} relation may not
be expected to hold under such conditions.
The high-surface density regions of M33
do satisfy the $\fHmol-\Pext$ relation \citep{blitz2006a}, as
does the simulated M33 analogue.
We delay a more thorough examination of the exterior disk regions 
until further work.  

Recently, \citet{booth2007a} presented a model of star formation based
on a subgrid model for the formation of molecular clouds from
thermally unstable gas in a multiphase ISM.  In their model, molecular
clouds are represented as ballistic particles which can coagulate
through collisions. This is quite different from the treatment of ISM
in our model, as we do not attempt to model molecular clouds on small
scales via a subgrid model, but calculate the equilibrium abundance of
molecular hydrogen using the local gas properties.  \citet{booth2007a}
show that their simulations also reproduce the molecular
fraction-pressure relation of \citet{blitz2004a,blitz2006a}. However,
the interpretation of this relation in their model must be different
from our interpretation involving the relation between the gas surface
density, the stellar surface density, and the ISRF strength, because
their model does not include the dissociating effects of the interstellar
radiation field.

In addition, in contrast to our model, the global SK relation in
simulations of \citet{booth2007a} does not show a break or steepening
down to very small surface densities ($\Sigma\approx 10^{-2}{\rm\
M_{\odot}\,pc^{-2}}$; see their Fig. 15). Given the differences, it
will be interesting to compare results of different models of
$\Hmol$-based star formation in more detail in future studies.

\begin{figure*}
\figurenum{10}
\epsscale{1}
\plotone{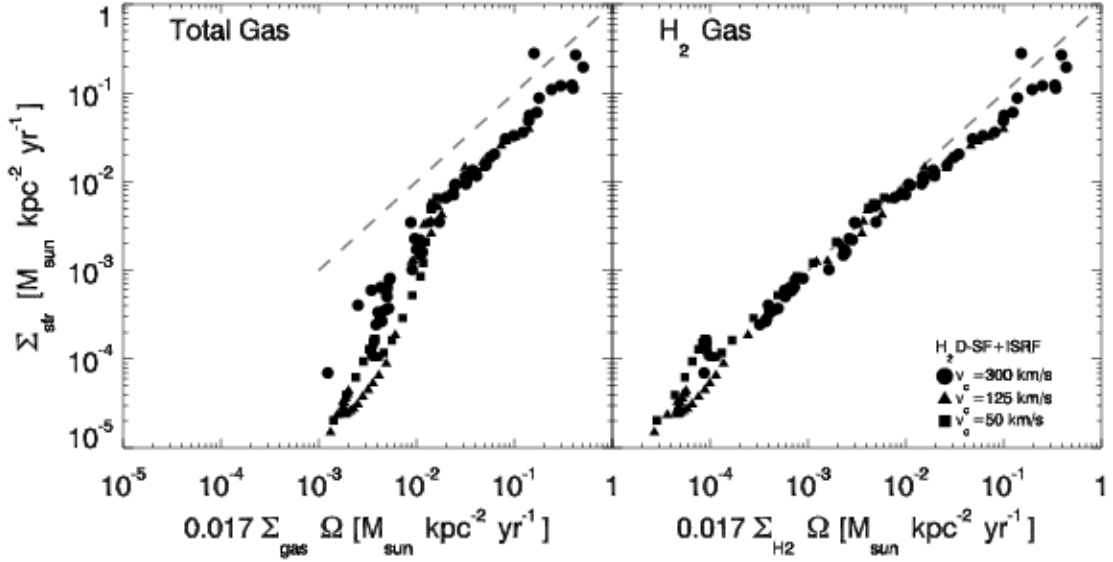}
\caption{\label{fig:sfr_rotation}
Star formation rate surface density $\SigmaSFR$ as a function of the total gas 
surface density consumed per orbit $C_{\Sigma\Omega}\Sigmagas\Omega$ (left panel) and molecular gas
surface density consumed per orbit $C_{\Sigma\Omega}\SigmaHmol\Omega$ (right panel), where $\Omega$ is
the orbital frequency and the efficiency $C_{\Sigma\Omega}=0.017$ was determined from 
observations by \cite{kennicutt1998a} (represented by the {\it dashed\/} line which extends
over the range probed by observations).  Shown are the local values of $\Sigmagas$, 
$\SigmaHmol$ and $\Omega$ in annuli for galaxy models with circular velocities of
$\vcirc=50\kms$ (squares), $\vcirc=125\kms$ (triangles), and $\vcirc=300\kms$
(circles) evolved with atomic+molecular cooling and destruction of $\Hmol$
by a local interstellar radiation field for $t=0.3\Gyr$. The star formation rate 
in the simulations display the correlations $\SigmaSFR\propto\SigmaHmol\Omega$ and, in the
high surface density region where the molecular gas and total gas densities are comparable, 
$\SigmaSFR\propto\Sigmagas\Omega$.
}
\end{figure*}

%-----------------------------------------------
\subsection{Star Formation and Angular Velocity}
\label{section:results:sfr_rotation}
%-----------------------------------------------

The SFR surface density $\SigmaSFR$ has been shown to
correlate with the product of the gas surface density $\Sigmagas$ and
the disk angular frequency $\Omega$ as
\begin{equation}
\SigmaSFR \simeq C_{\Sigma\Omega} \Sigmagas\Omega
\end{equation}
\noindent
with the constant $C_{\Sigma\Omega}=0.017$
\citep{kennicutt1998a} and where the angular
frequency is defined as  
\begin{equation}
\Omega^{2} = \frac{1}{R}\frac{\dd \Phi}{\dd R}
\end{equation}
\citep[e.g., \S 3.2.3 of ][]{binney1987a}.  Star formation relations
of this form have been forwarded to connect the SFR to
the cloud-cloud collision time-scale
\citep[e.g.,][]{wyse1986a,wyse1989a,tan2000a}, to tie the gas
consumption time-scale to the orbital time-scale
\citep[e.g.,][]{silk1997a,elmegreen1997a}, or to reflect the
correlation between the ISM density and the tidal density
\citep{elmegreen2002a}.  These ideas posit that the
$\SigmaSFR-\Sigmagas\Omega$ correlation provides a reasonable physical
interpretation of the $\Omega\propto\sqrt{\rho}$ scaling.

Figure \ref{fig:sfr_rotation} shows the SFR surface
density $\SigmaSFR$ as a function of the total gas surface density
consumed per orbit $C_{\Sigma\Omega}\Sigmagas\Omega$ and molecular gas surface density
consumed per orbit $C_{\Sigma\Omega}\SigmaHmol\Omega$, compared to the observed
$\SigmaSFR-\Sigmagas\Omega$ relation \cite{kennicutt1998a}.  The
local values of $\Sigmagas$ and $\SigmaHmol$ are measured in radial
annuli from the gas properties, while $\Omega$ is measured from the
rotation curve at the same location.  
The SFR surface density scales as $\Sigmagas\Omega$ at 
$\SigmaSFR\gtrsim 5\times 10^{-3}\,\rm M_{\odot}yr^{-1} kpc^{-2}$. 
The offset between the disk-averaged $\SigmaSFR-\Sigmagas\Omega$
relation found by \cite{kennicutt1998a} and the simulated relation may
owe in part to the typically higher values of $\Omega$ in the disk
interior that can affect the normalization of the azimuthally-averaged
$\SigmaSFR-\Sigmagas\Omega$ relation.  This offset is 
comparable to the observed offset between the disk-averaged relation
of \cite{kennicutt1998a} and the spatially-resolved
$\SigmaSFR-\Sigmagas\Omega$ relation found in M51a by
\cite{kennicutt2007a}. As in the previous sections, 
scaling $\SigmaSFR-\SigmaHmol\Omega$ works throughout the disk, while the
$\SigmaSFR-\Sigmagas\Omega$ scaling steepens significantly 
at lower surface densities where the
molecular fraction is declining.

That the SFR in the simulated disks 
correlates well with the quantity $\Sigmagas\Omega$, even though the
SFR in the simulations is determined 
directly from the {\it local} molecular gas density and 
dynamical time without direct knowledge of $\Omega$, may
be surprising. 
The correlation suggests that the local gas density 
should scale with $\Omega$, 
although the physical reason for such a correlation in our
simulations is unclear. 
While previous examinations of the
$\SigmaSFR-\Sigmagas\Omega$ correlation have developed
explanations of why $\Omega \propto \sqrt{\rho_{d}}$, these 
simulations have a coarser treatment of the ISM than is
invoked in such models;
the molecular fraction and SFR
are primarily functions of the local density.
The $\Omega\sim\sqrt{\rho_{d}}$ correlation should then
reflect the structural properties 
galaxy models rather than the detailed properties of the
ISM that are somehow influenced by global processes in the
disk.

The galaxy models used in the simulations remain roughly
axisymmetric after their evolution, modulo spiral structure
and disk instabilities.  Hence, the fundamental equation
that describes the gravitational
potential $\Phi$ of the galaxies is the Poisson
equation
\begin{equation}
\label{equation:poisson}
\frac{\dd^{2} \Phi}{\dd z^{2}} + \frac{\dd^{2} \Phi}{\dd R^2} + \Omega^{2} = 4\pi G \rho_{\mathrm{total}}. 
\end{equation}
\noindent
where the density $\rho_{\mathrm{total}}$ reflects the total density
of the multicomponent system.  If the form of the potential generated
by $\rho_{\mathrm{total}}$ has the property that $\Omega^2 =
(1/R)\dd\Phi/\dd R \propto \rho_{d}$, then the quantity
$\Sigmagas\Omega$ will mimic the scaling of $\SigmaSFR$.  In the
second Appendix, we examine the solutions for $\Omega$ determined from
Equation \ref{equation:poisson} for the limiting cases of locally
disk-dominated ($\rho_{\mathrm{total}}\approx\rho_{\mathrm{disk}}$)
and halo-dominated ($\rho_{\mathrm{total}}\approx\rho_{\mathrm{DM}}$)
potentials, and show that typically $\Omega \propto \sqrt{\rho_{d}}
B(\rho_{d})$ where $B(\rho_{d})\sim\mathcal{O}(1)$ is a weak function
of density if the disk density is exponential with radius.  This
behavior suggests that whether the disk dominates the local potential,
as is the case for massive galaxies, or the dark matter halo dominates
the potential as it does in low-mass dwarfs, the angular frequency
scales as $\Omega\propto\sqrt{\rho_{d}}$ throughout most of the disk.
Based on these calculations, we conclude that the correlation
$\SigmaSFR\propto\Sigmagas\Omega$ in the simulations is likely a
consequence of the density-dependent star formation relation (Equation
\ref{equation:star_formation_rate}) and the galaxy structure inducing
the correlation $\Omega\propto\sqrt{\rho_{d}}$. 
This
correlation may be established during the formation of the exponential
disk in the extended dark matter halo, and the correlation observed in
real galaxies therefore may have a similar origin.  Star formation may
then correlate with $\Sigmagas\Omega$, but not be
directly controlled or
influenced by the global kinematics of the disk.

\begin{figure}
\figurenum{11}
\epsscale{1.15}
\plotone{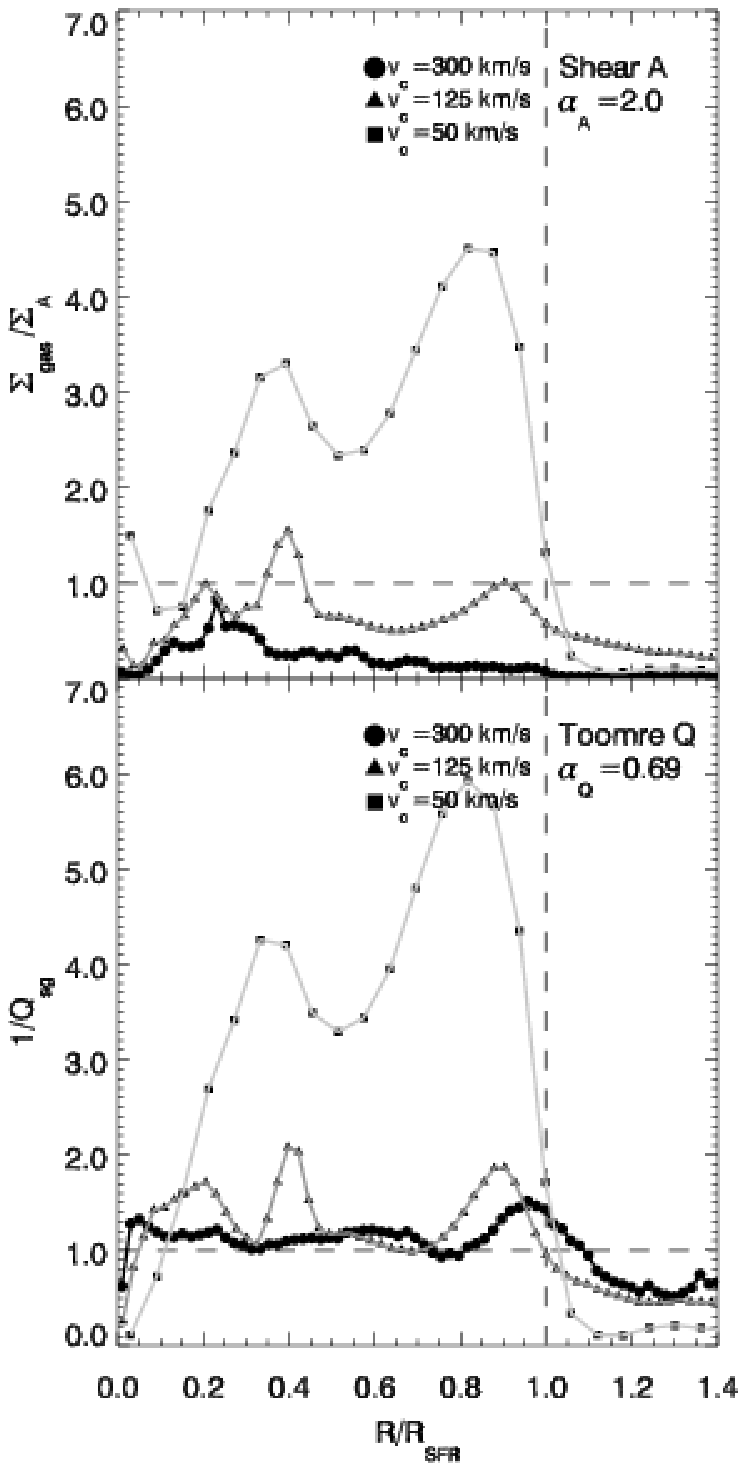}
\caption{\label{fig:crit_mu}
\small
Stability parameters for
two-fluid Toomre instability  
($\Qsg^{-1}$, bottom panel) or shear 
instability ($\Sigmag/\SigmaA$, top panel) 
as a function of
radius normalized to the radius $R_{\SFR}$ containing
$99.9\%$ of the total active disk star formation.  Shown are the
parameters $\Qsg^{-1}$ and $\Sigmag/\SigmaA$ for galaxy models with 
circular velocities
of $\vcirc = 50\kms$ (squares, light gray lines), 
$\vcirc = 125\kms$ (triangles, dark gray lines), 
and $\vcirc = 300\kms$ (circles, black lines).  The proportionality constants 
$\alpha_{Q}=0.69$ and $\alpha_{A}=2$ are taken to match 
the values used by \cite{martin2001a}. For each
simulated galaxy, the parameter $\Qsg^{-1}$
approaches the transition to stability at the radius $R_{\SFR}$
beyond which only very little star formation operates.
An increase in the stability against growing shearing modes also
occurs near $R_{\SFR}$ for the simulated disks, but 
the transition is less uniform than for Toomre instability.
\\
}
\end{figure}

\subsection{Star Formation and Critical Surface Density Thresholds}
\label{section:results:critical_surface_densities}

Disk surface density, dynamical
instabilities, and the formation of molecular clouds and
stars may be connected.
A single-component thin disk will experience growing axisymmetric 
instabilities if its surface mass density exceeds the critical
value $\Sigma_{Q}$, given by
\begin{equation}
\label{equation:sigma_crit_Q}
\Sigma_{Q}(\sigma) \approx  \frac{\alpha_{Q} \kappa \sigma}{\pi G},
\end{equation}
\noindent
where $\kappa$ is the epicyclical frequency
\begin{equation}
\kappa^2 = 2\left(\frac{v^2}{R^2} + \frac{v}{R}\frac{\dd v}{\dd R}\right),
\end{equation}
\noindent
$\sigma$ is the characteristic velocity dispersion 
of the fluid,
 and the parameter
$\alpha_{Q}\sim\mathcal{O}(1)$
\citep{safronov1960a,toomre1964a}.
For a two component system, consisting of stars with surface
mass density $\Sigma_{\star}$ and velocity dispersion
$c_{\star}$, and gas with
surface mass density $\Sigmag$ and velocity dispersion
$\sigma$, the instability 
criterion for axisymmetric modes with a wavenumber $k$ can
be expressed in terms of the normalized wavenumber
$q\equiv k c_{\star}/\kappa$ and the ratio of
velocity dispersions $R\equiv\sigma/c_{\star}$ as
\begin{equation}
\label{equation:general_toomre_criterion}
\Qsg^{-1} \equiv \frac{2}{\Qs}\frac{q}{1+q^{2}} + \frac{2}{\Qg} R\frac{q}{1+q^{2}R^{2}} > 1
\end{equation}
\noindent
\citep{jog1984a,rafikov2001a}, where $\Qs = \Sigma_{Q}(c_{\star})/\Sigma_{\star}$
and $\Qg =\Sigma_{Q}(\sigma)/ \Sigmag$.
The most unstable wavenumber $q$ can be determined by maximizing
this relation.

\cite{elmegreen1993a} 
considered another
critical surface density $\Sigma_{A}$, given by
\begin{equation}
\label{equation:sigma_crit_A}
\Sigma_{A} \approx  \frac{\alpha_{A} A \sigma}{\pi G}
\end{equation}
\noindent
where the parameter $\alpha_{A}\sim\mathcal{O}(1)$,
and 
the Oort constant $A$ is
\begin{equation}
A = -0.5R\frac{\dd \Omega}{\dd R}
\end{equation}
\noindent
\citep[e.g.,][]{binney1987a}, above which density perturbations 
can grow through self-gravity before being disrupted by shear.
This instability criterion has been shown to work well in 
Milky Way spiral arms \citep[e.g.,][]{luna2006a} and 
ring galaxies \citep{vorobyov2003a}. In regions of low shear, the
magneto-Jeans instability may induce the growth of perturbations
\citep[e.g.,][]{kim2002a}.

Observational 
evidence for connection between global disk instabilities
and star formation has been a matter 
of considerable debate \citep[e.g.,][]{kennicutt1989a,martin2001a,hunter1998a,boissier2003a,boissier2007a,kennicutt2007a,yang2007a}. In particular, the existence of global star formation thresholds related
to Toomre instability \cite[e.g.,][]{kennicutt1989a,martin2001a} has been 
challenged \citep{boissier2007a}. It is
thus interesting to use
model galaxies to explore the role of critical surface densities 
in star formation.  While proper radiative transfer
calculations are needed to assign either H$\alpha$ or UV emissivities
to the simulated disk galaxies, an indirect comparison between
simulations and observations can be made by measuring surface mass
densities, velocity dispersions, dynamical properties, and a
characteristic extent of star formation in the model disks.  These
comparisons are presented in Figure \ref{fig:crit_mu}, where the
instability parameters $\Qsg^{-1}$ and $\Sigmag/\Sigma_{A}$ are
plotted as a function of radius in the disk galaxies simulated with
the $\Hmol$D-SF+ISRF molecular ISM and star formation model.  For
comparison with the two-fluid Toomre axisymmetric instability
criterion, the parameter $\Qsg$ is measured by determining the locally
most unstable wavenumber $q$ from $\Qs$ and $\Qg$ with $\alpha=0.69$
to approximate the observational results of \cite{kennicutt1989a} and
\cite{martin2001a}.  When examining the shearing instability
criterion, the parameter $\alpha_{A}=2$ is chosen for comparison with
the observational estimation of critical surface densities by
\cite{martin2001a} and only the gas surface density is utilized.  
A ``critical'' radius
$R_{\mathrm{SFR}}$ containing $99.9\%$ of the total star formation 
in each simulated disk is measured and used as a proxy
for the star formation threshold radius observationally
determined from, e.g., the H$\alpha$ emission. 

The shearing instability criterion (upper panel Figure \ref{fig:crit_mu})
is supercritical ($\Sigmag/\Sigma_{A}>1$) over the extent of star formation 
in the low-mass galaxy, but the larger galaxies are either marginally supercritical
or subcritical.  Given
the large gas content in this dwarf system and that this comparison only uses
the gas surface mass density $\Sigmag$, the result is unsurprising. 
While the $\vcirc\sim125\kms$
and $\vcirc\sim300\kms$ galaxies are subcritical at 
$R=R_{\mathrm{SFR}}$, we note that 
there is still a noticeable drop in $\Sigmag/\Sigma_{A}$
near $R=R_{\SFR}\sim1$ for these systems. 

Interestingly, with the chosen definition of $R_{\SFR}$, the bottom panel of
Figure \ref{fig:crit_mu} shows that the two-component Toomre
instability parameter $\Qsg$ is a remarkably accurate indicator of
where the majority of star formation operates in the simulated disk galaxies.
Each galaxy approaches the $\Qsg\sim1$ transition near 
$R_{\SFR}$, demonstrating that much of the star formation is operating
within Toomre-unstable regions of the disks.  These results appear most consistent
with the observations by \cite{kennicutt1989a}, \cite{martin2001a}, and \cite{yang2007a},
who find that star formation operates most efficiently in regions that are gravitationally
unstable.

Our simulations are also qualitatively
consistent with the theoretical findings of \cite{li2005b,li2005a,li2006a}, who used
hydrodynamical simulations of isolated galaxies with sink particles 
to represent dense molecular gas and star clusters.  They found that if
the accretion onto the sink particles resulted in star formation with
some efficiency, the
two-component Toomre parameter was $\Qsg<1.6$ in regions of active star formation.
Although a direct comparison is difficult given their use of sink particles,
we similarly find that the majority of star formation in our simulations 
occurs in regions of the disk where $\Qsg\lesssim1$.

In related theoretical work, a gravithermal instability criterion for star formation was explored by 
\cite{schaye2004a},
who suggested that the decrease in thermal velocity dispersion associated with 
the atomic-to-molecular transition in the dense ISM may induce gravitational instability
and the corresponding connection between critical surface densities and star formation.
The simulated galaxies do experience a decline in thermal sound speed near the
$\Qsg\sim1$ transition, consistent with the mechanism advocated by \cite{schaye2004a}.
However, of our $\Hmol$D-SF+ISRF simulations (that include physics necessary for
such instability), 
only the dwarf galaxy model displays
a sharp transition in $\Qsg$ near the critical radius while the stability of the other, 
more massive galaxies
are more strongly influenced by the turbulent gas velocity dispersion and have more
modest increases in $\Qsg$.  Further, we note that the gravithermal mechanism suggested by
\cite{schaye2004a} should not operate in isothermal treatments of the ISM such as those
presented by \cite{li2005b,li2005a,li2006a}.  Similarly our GD-SF simulations, which
are effectively isothermal ($T\approx10^{4}\K$) at ISM densities, display similar transitions
to $\Qsg\lesssim1$ at the critical radius even as such simulations do not include a cold 
phase to drive gravitational instability. 

We stress that the definition of $R_{\SFR}$ may affect 
conclusions about the importance of threshold surface densities.
For instance, if a $90\%$ integrated
star formation threshold was used $R_{\SFR}$ would be
lowered to $0.3-0.6$ of the chosen value and the
subsequent interpretations about the applicability of
the instability parameter $\Qsg$ could be quite different.
Detailed radiative transfer calculations may therefore be necessary to
determine the integrated star formation fraction that brings
$R_{\SFR}$ in the best agreement with
observationally-determined critical radii.

\begin{figure*}
\figurenum{12}
\epsscale{1}
\plotone{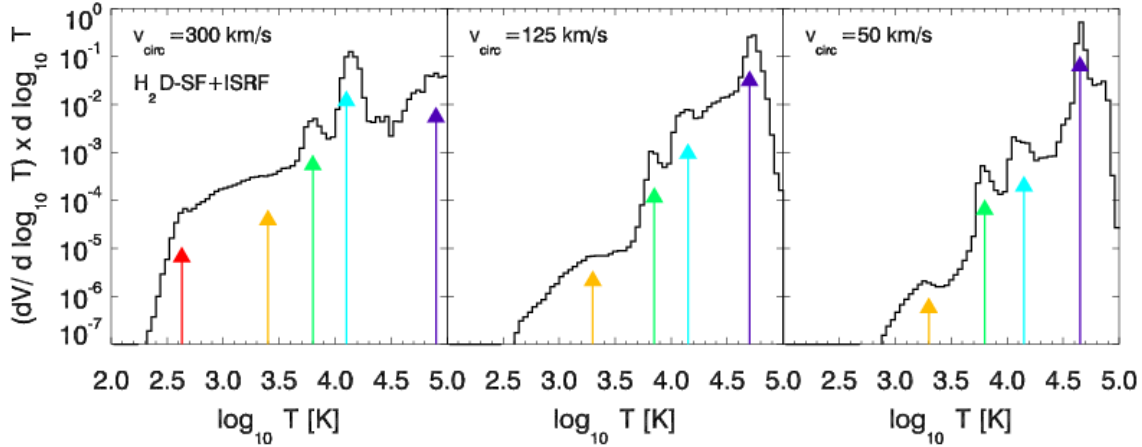}
\caption{\label{fig:temperature_pdf}
Temperature structure of the interstellar medium in galaxies
of different mass.  Shown are the probability distribution
functions (PDF) of the fractional volume of the ISM with a temperature
$T$ (black histograms), for systems with circular velocities of
$\vcirc=300\kms$ (left panel),
$\vcirc=125\kms$ (middle panel), and $\vcirc=50\kms$ (right panel).
The simulations use the $\Hmol$D-SF+ISRF model to evolve the systems in isolation for $t=0.3\Gyr$.
The peaks in the temperature PDF correspond to regions of 
density-temperature phase space where the cooling time $\tcool$ has local
maxima (i.e., where $\tcool$ is locally a weak function of density).
These peaks in the temperature PDF correspond to features in the density
PDF, and 
the colored arrows highlight regions that can be modeled approximately as 
separate isothermal phases (i.e., lognormals) in the density PDF
(see Figure \ref{fig:density_pdf} and 
\S \ref{section:results:pdfs} of the text for details).
} 
\end{figure*}

\begin{figure*}
\figurenum{13}
\epsscale{1}
\plotone{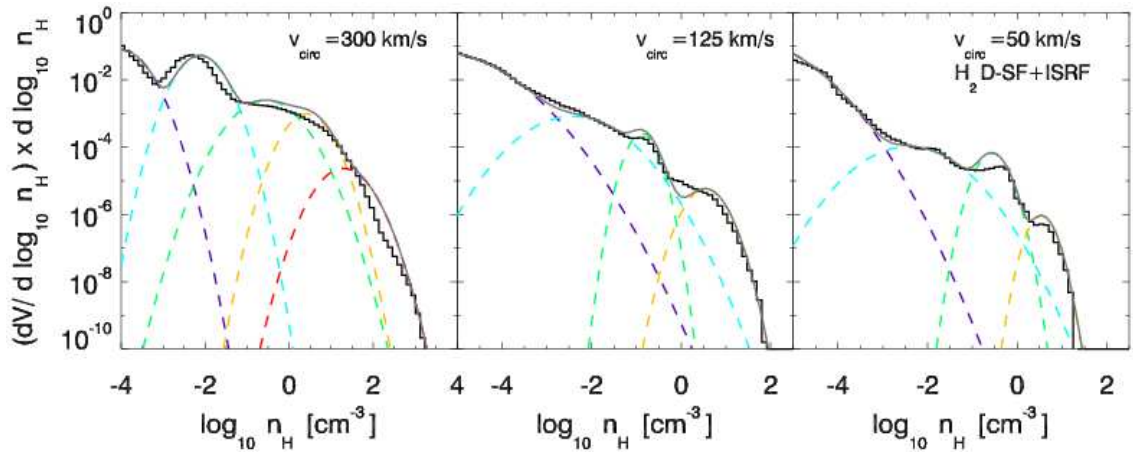}
\caption{\label{fig:density_pdf}
Density structure of the interstellar medium in galaxies
of different mass.  Shown are the probability distribution
functions (PDF) of the fractional volume of the ISM with a density 
$\nH$ (black histograms), for systems with circular velocities of
$\vcirc=300\kms$ (left panel),
$\vcirc=125\kms$ (middle panel), and $\vcirc=50\kms$ (right panel)
for the $\Hmol$D-SF+ISRF model at $t=0.3\Gyr$.
The features in the density PDF correspond to regions of 
density-temperature phase space where the cooling time $\tcool$ has local
maxima (i.e., where $\tcool$ is locally a weak function of density).
The density PDF of the ISM in these simulated galaxies
is not well-modeled by a single log-normal, except at the high-density tail.
The figure shows that the density PDF may be modeled approximately as a sum of
isothermal phases (i.e., lognormal PDFs in density, colored dashed lines),
where the  
characteristic density of each phase is determined by using
the ISM equation-of-state to find the typical density
of peaks in the temperature PDF 
(see Figure \ref{fig:temperature_pdf} and 
\S \ref{section:results:pdfs} of the text for details).
}
\end{figure*}

%--------------------------------------------------------
\subsection{Temperature and Density Structure of the ISM}
\label{section:results:pdfs}
%-------------------------------------------------------

The temperature and density structure of interstellar gas is
determined by a variety of processes including radiative 
heating, cooling, and, importantly, 
turbulence 
\citep[see, e.g.,][]{elmegreen2002a}.  
The calculation of the density
probability distribution function (PDF) of an isothermal gas, defined
as a volume fraction of gas with a given density, suggests that the
PDF will resemble a lognormal distribution with a
dispersion $\sigma$ that scales linearly with the mean gas Mach number
$\sigma \propto \beta \mathcal{M}$.  The origin of
the lognormal form is discussed in detail in
\citet{padoan1997a} and \citet{passot1998a}.
For a
non-isothermal gas the density PDF may more generally follow
a power-law distribution \citep{passot1998a,scalo1998a}.

Although the lognormal PDF was initially discussed in the context
of molecular clouds,  \cite{wada2001a}
found that two-dimensional numerical simulations of a multiphase ISM on galactic scales
produced a lognormal density PDF for dense gas, while at low densities gas
followed a flatter distribution in $\log \rho$ \citep[see][for more
recent investigations of this issue]{wada2007a,wada2007b}. A similar
PDF was found in a cosmological simulation of galaxy formation at
$z\approx 4$ by \citet{kravtsov2003a}. Motivated in part by the result of 
\cite{wada2001a}, \cite{elmegreen2002a}
suggested that the SK relation could be explained in terms of the density
PDF of ISM gas if a fraction $f_{c}$ of gas that resides above some 
critical density threshold $\rho_{c}$ converts
into stars with a constant efficiency and $\rho_{c}$ scales linearly 
with the average ISM density $\left<\rho\right>$.
Similarly, \cite{kravtsov2003a} found 
that the fraction of dense gas in cosmological simulations of
disk galaxy formation at redshifts $z\gtrsim4$ varied with the
surface density in a manner to produce the 
$\SigmaSFR \propto \Sigmag^{1.4}$ relation at high surface 
densities. The gas density distribution in these simulations
also followed a log-normal PDF at large densities and 
flattened at lower densities \citep[see also][]{kravtsov2005a}. 

\cite{joung2006a} used three-dimensional models of the
stratified ISM to demonstrate that supernovae explosions can
act to regulate star formation by inputting energy on small
scales. Their scheme produced temperature and density structures
consistent with those observed in the real ISM, with
density and temperature PDFs that contain multiple peaks. 
\cite{tasker2007a} found the density PDF in their
multiphase simulations could be approximated by a single lognormal
at high densities, but exhibited deviations from the lognormal
in the form of peaks and dips at smaller densities.

Figure \ref{fig:temperature_pdf} shows the probability distribution
function (PDF) of the fractional volume in the ISM at a temperature
$T$ in the $\Hmol$D-SF+ISRF simulations at $t=0.3\,\Gyr$.  The
temperature PDF is measured by assigning an effective volume to each
particle based on its mass and density, counting the particles in bins
of fixed width in $\log_{10} T$, and then normalizing by the total
effective volume of the ISM.  The temperature PDF displays structure
corresponding to gas at several phases.  These temperature peaks
correspond to features in the density-temperature phase diagram of the
ISM, set by the cooling and heating processes.  The abundance of gas
in different regions of this diagram depend on the local cooling time,
with regions corresponding to long local cooling times becoming more
populated. The overall gravitational potential of the system also
influences the density of gas in the disks and thus indirectly
influences cooling and abundance of cold, dense gas, as denser ISM gas
can reach lower equilibrium temperatures.
Figure~\ref{fig:temperature_pdf} shows that the temperature PDFs of
the galaxies do differ, with more massive galaxies containing more
volume at low ISM temperatures.  In smaller mass galaxies, the ISM
densities are lower, gas distributions are more extended, and the
abundance of cold gas is greatly decreased.

 Figure \ref{fig:density_pdf}
shows the PDF of gas densities for each model galaxy in the
$\Hmol$D-SF+ISRF model at $t=0.3\Gyr$.  The density PDFs are measured
in the same manner as the temperature PDFs, but the particles are
binned according to the $\log_{10} \nH$.  Each galaxy has a PDF with
noticeable features that change depending on the galaxy mass scale.
None of the density PDFs are well represented by a single lognormal
distribution, reflecting the multiple peaks apparent in the
temperature PDFs (Figure \ref{fig:temperature_pdf}).

Motivated by the multiphase structure of the ISM displayed
in Figure \ref{fig:temperature_pdf}, a model of the density PDFs
can be constructed by identifying characteristic temperature peaks in
the temperature PDF and assigning separate lognormal distributions to
represent each approximately isothermal gas phase.
The colored arrows in Figure
\ref{fig:temperature_pdf} identify characteristic temperatures in the
ISM of each galaxy.  The characteristic density corresponding to each
selected temperature, determined by the median equation-of-state
calculated using Cloudy, is employed as the mean of a lognormal
distribution used to model a feature in the density PDF (colored
dashed lines in Figure \ref{fig:density_pdf}).  The heights and widths
of the lognormals are constrained to provide the same relative volume
fraction in the density PDFs as the phase occupies in the temperature
PDF, resulting in a model shown by the solid gray lines in Figure
\ref{fig:density_pdf}.
  
Clearly the ISM is best understood in terms of a continuous
distribution of temperatures, represented in the simulations by Figure
\ref{fig:temperature_pdf}, but the multiphase model of the density
PDFs provides some conceptual insight into the ISM phase structure.
First, the model of separate isothermal ISM phases can capture
the prominent features of the density PDFs.  Second, the galaxies
share common features in the density (and temperature) PDFs, such as a
warm neutral phase ($\log_{10} T\sim3.8$) and the mix of cold neutral
and molecular gas at low temperatures ($\log_{10} T\lesssim3.5$).
Third, the relative widths of the lognormal distributions representing
the dense ISM in each galaxy increase with increasing galaxy mass
(i.e., the orange lognormals centered near $\log_{10} \nH \sim 0.5$),
as do their normalizations.  The increasing widths of the lognormals
mimic the increasing velocity dispersions (and turbulent Mach
numbers) of the dense gas in the more massive galaxy models.

Note that for the ISM of the $\vcirc=300\kms$ system the cold dense
phase contributes the majority of the ISM by mass, while shock heated
gas at $T\sim10^{5}\K$ contributes the largest ISM fraction by volume.
We note that this result qualitatively resembles the effect of the
volume-averaged equation-of-state utilized in subgrid models of the
multiphase ISM \citep[e.g.][]{springel2003a}.

For purposes of calculating the overall SFE
of galaxies, our results suggest that approximating the density PDF by
a single lognormal distribution is overly simplistic.  The detailed
shape of the PDF is set by global properties of galaxies and heating
and cooling processes in their ISM and can therefore be expected to
vary from system to system. Nevertheless, a single PDF modeling may be
applicable for the cases when most of the ISM mass is in one dominant
gas phase (e.g., dense cold gas), as is the case for the massive
galaxy in our simulations.  We leave further exploration of this
subject for future work.

%------------------------
\section{Discussion}
\label{section:discussion}
%------------------------

Results presented in the previous sections indicate that
star formation prescriptions based on the local abundance
of molecular hydrogen lead to interesting features of
the global star formation relations. 

We show that the inclusion of an interstellar radiation field is
critical to control the amount of diffuse $\Hmol$ at low gas
densities.  For instance, without a dissociating ISRF the low mass
dwarf galaxy eventually becomes almost fully molecular, in stark
contrast with observations. We also show that without the dissociating
effect of the ISRF our model galaxies produce a much flatter relation
between molecular fraction $\fHmol$ and pressure $\Pext$, as can be
expected from the results of \cite{elmegreen1993a}.  Including
$\Hmol$-destruction by an ISRF results in a $\fHmol-\Pext$ relation in
excellent agreement with the observations of \citet{wong2002a} and
\citet{blitz2004a,blitz2006a}.

Our model also predicts that the relation between $\Sigma_{\rm
SFR}$ and $\Sigma_{\rm gas}$ should not be universal and can be
considerably steeper than the canonical value of $n_{\rm tot}=1.4$,
even if the three-dimensional Schmidt relation in molecular clouds is
universal. The slope of the relation
is controlled by the dependence of molecular fraction (i.e., fraction
of star forming gas) on the total local gas
surface density. This relation is non-trivial because the
molecular fraction is controlled by pressure and ISRF strength, 
and can thus vary 
between different regions with the same total gas surface density. 
The relation can also be different in the regions
where the disk scale-height changes rapidly (e.g., in flaring outer
regions of disks), as can be seen from equation~\ref{equation:structural_schmidt_law}
\citep[see also][]{schaye2007a}.   We show that the effect 
of radial variations in 
the molecular fraction $\fHmol$ and gas scale heights ($\hSFR$ and $\hgas$)
on the SFR can be accounted for
in terms of a structural, SK-like correlation, 
$\SigmaSFR\propto\fHmol \hSFR \hgas^{-1.5} \Sigmagas^{1.5}$, that trivially relates
the local SFR to the consumption of molecular gas with an
efficiency that scales with the local dynamical time.

A generic testable prediction of our model is that deviations from the
SK $\SigmaSFR-\Sigmagas$ relation are expected in 
galaxies or regions of galaxies where the molecular fraction is declining or 
much
below unity.  As we discussed above, star formation in the molecule-poor
($\fHmol\sim0.1$) galaxy M33 supports
this view as its total gas SK power-law index is
$\ntot\approx 3.3$ \citep{heyer2004a}.  While our simulations of an M33
analogue produce a steep SK power-law, the overall SFE
in our model galaxies lies
below that observed for M33.  However, as recently emphasized by
\cite{gardan2007a}, the highest surface density regions of M33 have
unusually efficient star formation compared with the normalization of
the \cite{kennicutt1998a} relation and so the discrepancy is not
surprising.
Given that dwarf galaxies generally
have low surface densities and are poor in molecular gas, it will be
interesting to examine SK relation in other small-mass
galaxies.

Another example of a low-molecular fraction galaxy close to home is
M31, which has only a fraction $\fHmol\approx0.07$ 
of its gas in molecular form within
18 kpc of the galactic center \citep{nieten2006a}.  
Our model would predict that
this galaxy should deviate from the total gas SK
relation found for molecular-rich galaxies. Observationally, the
SK relation of M31 measured by \citet{boissier2003a} is
rather complicated and even has an increasing SFR with decreasing gas
density over parts of the galaxy.  Low molecular fractions can also be
expected in the outskirts of normal galaxies and in the disks of low
surface brightness galaxies. The latter have molecular fractions of only
$\fHmol\lesssim 0.10$ \citep{matthews2005a,das2006a}, and we therefore predict
that they will not follow the total gas SK relation obeyed by
molecule-rich, higher surface density galaxies. At the same time, LSBs
do lie on the same relation between $\Hmol$ mass and far infrared
luminosity 
as higher surface brightness (HSB) galaxies \citep{matthews2005a}, 
which suggests that the dependence of star formation on
molecular gas may be the same in both types of galaxies.  

An alternative formulation of the global star formation relation is
based on the angular frequency of disk rotation: $\Sigma_{\rm
SFR}\propto \Sigmagas\Omega$. That this relation works in real
galaxies is not trivial, because star formation and dynamical
time-scale depend on the local gas density, while $\Omega$ depends on
the total mass distribution {\it within} a given radius. Although
several models were proposed to explain such a correlation
\citep[see, e.g.,][for reviews]{kennicutt1998a,elmegreen2002a}, we show in
\S \ref{section:results:sfr_rotation} and the second Appendix that the star
formation correlation with $\Omega$ can be understood as a fortuitous
correlation of $\Omega$ with gas density of $\Omega\propto \rho_{\rm
gas}^{\alpha}$, where $\alpha\approx 0.5$, for self-gravitating
exponential disks or exponential disks 
embedded in realistic halo potentials. Moreover, 
we find that $\SigmaSFR\propto \Sigmagas\Omega$ breaks down 
at low values of $\Sigmagas\Omega$ where the molecular fraction declines, 
similarly to the
steepening of the SK relation. 
Our models therefore predict that the $\SigmaSFR\propto\SigmaHmol\Omega$
relation is more robust than the $\SigmaSFR\propto\Sigmagas\Omega$ relation.

An important issue related to the global star formation in galaxies is
the possible existence of star formation thresholds
\citep{kennicutt1998a,martin2001a}.  Such thresholds are expected to
exist on theoretical grounds, because the formation of dense,
star-forming gas is thought to be facilitated by either dynamical
instabilities \citep[see, e.g.,][for comprehensive
reviews]{elmegreen2002a,mckee2007a} or gravithermal instabilities
\citep{schaye2004a}. We find that the two-component Toomre instability
threshold that accounts for both stars and gas, $Q_{\rm sg}<1$, works
well in predicting the transition from atomic gas, inert to
star formation, to the regions where molecular gas and star formation
occur in our simulations.  Our results are in general agreement with
\citet{li2005a,li2005b,li2006a}, who used sink-particle simulations of
the dense ISM in isolated galaxies to study the relation between star
formation and the development of gravitational instabilities, and with
observations of star formation in the LMC \citep{yang2007a}.  Our
simulations further demonstrate the importance of accounting for
all mass components in the disk to predict correctly which regions
galactic disks are gravitationally unstable.

Given that our star formation prescription is based
on molecular hydrogen, the fact that $Q_{\rm sg}$ is a good 
threshold indicator may imply that gravitational instabilities
strongly influence the abundance of dense, molecular gas in the disk.
Conversely, the gas
at radii where the disk is stable remains at low density and
has a low molecular fraction. We find that in our 
model galaxies, the shear instability criterion of
\cite{elmegreen1993a} does not work as well as the Toomre $Q_{\rm sg}$-based
criterion. Almost all of the star formation in our model galaxies 
occurs at surface densities $\Sigmagas\gtrsim 3{\rm\ M_{\odot}\,pc^{-2}}$, 
which is formally consistent with the \citet{schaye2004a} 
constant surface density criterion for gravithermal instability. 
However, as Figure~\ref{fig:kennicutt.gas} shows, we do not see a clear indication
of threshold at a particular surface density and
our GD-SF models that have an effectively isothermal ISM with $T\approx10^{4}\K$
(and hence do not have a gravithermal instability) still show a good correlation
between regions where $\Qsg<1$ and regions where star formation operates.

Our results have several interesting implications for interpretation
of galaxy observations at different epochs. First, 
low molecular fractions in dwarf galaxies mean that
only a small fraction of gas is participating in star formation
at any given time. This connection between SFE
and molecular hydrogen abundance may explain why dwarf galaxies are
still gas rich today compared to larger mass galaxies \citep{geha2006a}, 
without relying on
mediation of star formation or gas blowout by supernovae. 
Note that a similar
reasoning may also explain why large LSBs at low redshifts 
are
gas rich but anemic in their star formation.  
Understanding the star formation
and evolution of dwarf galaxies is critical because they serve
as the building blocks of larger galaxies at high redshifts. Such
small-mass galaxies are also expected to be the first objects
to form large masses of stars and should therefore play an important
role in enrichment of primordial gas and the
cosmic star formation
rate at high redshifts \citep{hopkins2004a,hopkins2006am}.  

The star-forming disks at $z\sim2$ that may be
progenitors of low-redshift spiral galaxies are observed to lack
centrally-concentrated bulge components \citep{elmegreen2007a}.  Given
that galaxies are expected to undergo frequent mergers at $z>2$,
bulges should have formed if a significant fraction of baryons are
converted into stars during such mergers
\citep[e.g.,][]{gnedin2000a}. The absence of the bulge may indicate
that star formation in the gas rich progenitors of these $z\sim 2$
systems was too slow to convert a significant fraction of gas into
stars. This low SFE can be understood if the
high cosmic UV background, low-metallicities, and low dust content of
high-$z$ gas disks keep their molecular fractions low
\citep{pelupessy2006a}, thereby inhibiting star formation over most of
gas mass and keeping the progenitors of the star-forming $z\sim 2$
disks mostly gaseous.  Gas-rich progenitors may also help explain the
prevalence of extended disks in low-redshift galaxies despite the
violent early merger histories characteristic of $\Lambda$CDM
universes, as gas-rich mergers can help build
high-angular momentum disk galaxies \citep[][]{robertson2006c}.  
Mergers of mostly stellar disks, on the other hand, would form spheroidal systems.

Our results may also provide insight into the interpretation of the results of
\cite{wolfe2006a}, who find that the SFR associated with neutral atomic gas
in damped Lyman alpha (DLA) systems is an order of magnitude lower than
predicted by the local \citet{kennicutt1998a} relation.  The DLAs 
in their study sample regions with column densities $N_{\mathrm{H}}\approx 2\times 10^{21}\rm\
cm^{-2}$, or surface gas densities of $\Sigma_{\rm DLA}\approx 20\rm\
M_{\odot}\,pc^{-2}$, assuming a gas disk with a thickness of
$h\approx100$~pc. Suppose the local SK relation steepens from the local
relation with the slope $n_0=1.4$ to a steeper slope $n_1$ below some
surface density $\Sigma_{\rm b}>\Sigma_{\rm DLA}$. Then for
$\Sigmagas<\Sigma_{\rm b}$, the SFR density will be
lower than predicted by the local relation by a factor of
$(\Sigmagas/\Sigma_{\rm b})^{n_1-n_0}$. For $n_0=1.4$ and $n_1=3$ the
SFR will be suppressed by a factor of $>10$ for
$\Sigmagas/\Sigma_{\rm b}<0.25$. Thus, the results \cite{wolfe2006a} can
be explained if the total gas SK relation at $z\sim3$ 
steepens below $\Sigmagas\lesssim 100\rm\ M_{\odot}\, pc^{-2}$.
We suggest that if the majority of the molecular hydrogen at these redshifts
resides in rare, compact, and dense systems \citep[e.g.,][]{zwaan2006a}, then
both the lack of star formation and the rarity of molecular hydrogen
in damped Ly$\alpha$ absorbers may be explained simultaneously.

Our results also indicate that the thermodynamics of the ISM can leave
an important imprint on its density probability distribution. Each
thermal phase in our model galaxies has its own log-normal density
distribution.   Our results thus imply that using
a single lognormal PDF to build a model of global star formation in
galaxies \citep[e.g.,][]{elmegreen2002a,wada2007a} is likely an
oversimplification. Instead, the global star formation relation may
vary depending on the dynamical and thermodynamical properties of the
ISM.  We can thus expect differences in the SFE
between the low-density and low-metallicity environments of dwarf and
high-redshift galaxies and the higher-metallicity, denser gas of many
large nearby spirals.  Note that many of the results and effects we
discuss above may not be reproduced with a simple 3D density threshold
for star formation, as commonly implemented in galaxy formation
simulations.  Such a threshold can reproduce the atomic-to-molecular
transition only crudely and would not include effects of the local
interstellar radiation field, metallicity and dust content, etc.

 A clear caveat for our work is that the simulation resolution
limits the densities we can model correctly. At high densities, the gas
in our simulations is over-pressurized to avoid numerical Jeans instability.
The equilibrium density and temperature structure of the ISM and 
the molecular fraction are therefore not correct in detail.
Note, however, that our pressurization prescription
is designed to scale with the resolution, and should converge to
the ``correct'' result as the resolution improves.  
In any event, the simulations likely do not include all the relevant
physics shaping density and temperature PDFs of the ISM in real
galaxies. The results may of course depend on other microphysics
of the ISM as they influence both the
temperature PDF and the fraction of gas 
in a high-density, molecular form.  
Future simulations of the molecular ISM
may need to account for new microphysics as
they resolve scales where such processes become important.

%----------------------
\section{Summary}
\label{section:summary}
%----------------------

Using hydrodynamical simulations of the ISM and star formation in
cosmologically motivated disk galaxies over a range of
representative masses,
we examine the connection between molecular hydrogen abundance
and destruction, observed
star formation relations, and the thermodynamical structure of
the interstellar medium.  Our simulations provide a 
variety of new insights into the mass dependence of star formation
efficiency in galaxies.
A summary of our methodology and
results follows.

\begin{itemize}
\item[1.]  A model of heating and cooling processes
in the interstellar medium (ISM), including low-temperature
coolants, dust heating and cooling processes, and heating 
by the cosmic UV background, cosmic rays, and the local
interstellar radiation field (ISRF), is calculated using the
photoionization code Cloudy \citep{ferland1998a}.
Calculating the molecular fraction of the ISM 
enables us to implement a
prescription for the star formation rate (SFR) that ties 
the SFR directly to the molecular gas density.  The
ISM and star formation model is implemented in the SPH/N-body code GADGET2
\citep{springel2005c} and used to simulate the evolution
of isolated disk galaxies.

\item[2.]  We study the correlations between gas surface density
($\Sigmagas$), molecular gas surface density ($\SigmaHmol$), and SFR
surface density ($\SigmaSFR$).  We find that in our most realistic
model that includes heating and destruction of $\Hmol$ by the
interstellar radiation field, the power law index 
of the SK relation, $\SigmaSFR\propto\Sigmagas^{\ntot}$,
(measured in annuli)
varies from $\ntot\sim2$ in massive galaxies to $\ntot\gtrsim4$ in
small mass dwarfs.  The corresponding slope of the
$\SigmaSFR\propto\SigmaHmol^{\nmol}$ molecular-gas Schmidt-Kennicutt
relation is approximately the same for all
galaxies, with $\nmol\approx 1.3$.  These results are consistent
with observations of star formation in different galaxies
\citep[e.g.,][]{kennicutt1998a,wong2002a,boissier2003a,heyer2004a,boissier2007a,kennicutt2007a}.

\item[3.] In our models, the SFR density scales as 
$\SigmaSFR \propto \fHmol h_{\SFR} h_{\gas}^{-1.5} \Sigmagas^{1.5}$,
where $h_{\gas}$ is the scale-height of the ISM and $h_{\SFR}$ is the scale-height
of star-forming gas. The different $\SigmaSFR-\Sigmagas$ relations in galaxies
of different mass and in regions of different surface density
in our models therefore owe to the dependence of molecular fraction $\fHmol$ and 
scale height of gas on the gas surface density. 

\item[4.]
We show that the $\SigmaSFR\propto\Sigmagas\Omega$ and
$\SigmaSFR\propto\SigmaHmol\Omega$ correlations describe the
simulations results well where the molecular gas and total gas
densities are comparable,  
while the simulations deviate from
$\SigmaSFR\propto\Sigmagas\Omega$ \citep[e.g.,][]{kennicutt1998a} at
low $\Sigmag$ owing to a declining molecular fraction. 
We demonstrate that these relations may
owe to the fact that the angular frequency
and the disk-plane gas density are generally related as $\Omega\propto\sqrt{\rho}$
for exponential disks if the potential is dominated by either the
disk, a \cite{navarro1996a} halo, \cite{hernquist1990a} halo, or an
isothermal sphere. The correlation of $\SigmaSFR$ with $\Omega$ is thus 
a secondary correlation in the sense that  $\Omega\propto\sqrt{\rho}$ is set 
during galaxy formation and $\Omega$ does not directly influence
star formation. 

\item[5.] The role of critical surface densities for shear
instabilities ($\SigmaA$) and \cite{toomre1964a} instabilities
($\SigmaQ$) in star formation \citep[e.g.,][]{martin2001a} is examined
in the context of the presented simulations.  We find that the
two-component Toomre instability criterion $\Qsg<1$ is an accurate
indicator of the star-forming regions of disks, and that gravitational
instability and star formation are closely related in our simulations.
Further, the $\Qsg$ criterion works even in 
simulations in which cooling is restricted to $T>10^4$~K where
gravithermal instability cannot operate. 

\item[6.] Our simulations that include $\Hmol$-destruction by an ISRF
naturally reproduce the observed scaling $\fHmol\propto\Pext^{0.9}$
between molecular fraction and external pressure
\citep[e.g.,][]{wong2002a,blitz2004a,blitz2006a}, but we find that simulations
without an ISRF have a weaker scaling $\fHmol\propto \Pext^{0.4}$.  We
calculate how the connection between the scalings of the gas surface
density, the stellar surface density, and the ISRF strength influence
the $\fHmol-\Pext$ relation in the ISM, and show how the simulated
scalings reproduce the $\fHmol-\Pext$ relation even as the power-law
index of the total gas Schmidt-Kennicutt relation varies dramatically
from galaxy to galaxy.

\item[7.] We present a method for mitigating numerical Jeans
fragmentation in Smoothed Particle Hydrodynamics simulations
that uses a density-dependent pressurization of 
gas on small scales to ensure that the Jeans mass is properly
resolved, similar to techniques 
used in grid-based simulations
\citep[e.g.,][]{truelove1997a,machacek2001a}.  The gas 
internal energy $u$ at the Jeans scale is 
scaled as $u\propto\mJeans^{-2/3}$,
where $\mJeans$ is the local Jeans mass,
to ensure the Jeans mass is resolved by some $\NJeans$ number
of SPH kernel masses $2\Nneigh m_{\mathrm{SPH}}$, where $\Nneigh$
is the number of SPH neighbor particles and $m_{\mathrm{SPH}}$ is
the gas particle mass.  
For the simulations presented here, we find the
\cite{bate1997a} criterion of $\NJeans=1$
to be insufficient to avoid numerical 
fragmentation and that $\NJeans\sim15$ 
provides sufficient stability against numerical fragmentation
over the time evolution of our simulations.
Other simulations may have more stringent resolution requirements \citep[e.g.,][]{commercon2008a}.
We also 
demonstrate that isothermal galactic disks with temperatures
of $T=10^{4}\K$ may be susceptible to numerical Jeans instabilities
at resolutions common in cosmological simulations of disk galaxy
formation, and connect this numerical effect to possible
angular momentum deficiencies in cosmologically simulated disk galaxies.
\end{itemize}

The results of our study indicate that star formation may deviate
significantly from the relations commonly assumed in models of galaxy
formation in some regimes and that these deviations can be important for the
overall galaxy evolution.  Our findings provide strong motivation for
exploring the consequences of such deviations and for developing
further improvements in the
treatment of star formation in galaxy formation simulations.\\[2mm]

\acknowledgments

BER gratefully acknowledges support from a Spitzer Fellowship through
a NASA grant administrated by the Spitzer Science Center.
AVK is supported by the NSF grants
AST-0239759, AST-0507596, AST-0708154 and by the Kavli Institute for
Cosmological Physics at the University of Chicago. AVK thanks 
the Miller Institute and Astronomy department of UC Berkeley 
for hospitality during completion of this paper. 
We thank 
Andrew Baker,
Leo Blitz, 
Bruce Elmegreen, 
Nick Gnedin,
Dan Marrone,
Chris McKee,
Eve Ostriker,
Erik Rosolowsky,
and
Konstantinos Tassis
for helpful ideas, comments, and discussions. 
We also thank
Gary Ferland and collaborators for developing and maintaining the
Cloudy code used to tabulate cooling and heating rates and molecular
fractions in our simulations, and Volker Springel for making his
hydrodynamical simulation code GADGET2 available. 
Simulations presented here were
performed at the {\tt cobalt} system at the National Center for
Supercomputing Applications (NCSA) under project TG-MCA05S017.  We
made extensive use of the NASA Astrophysics Data System and {\tt
arXiv.org} preprint server in this study.

\appendix

\cite{truelove1997a} showed that hydrodynamical 
calculations that do not 
properly resolve pressure forces on scales where
self-gravity begins to dominate over gaseous
pressure support are susceptible to a numerical
instability that leads to the rapid artificial
fragmentation of gas. \cite{truelove1997a} suggest
a constraint on the spatial resolution of hydrodynamical
mesh calculations such that the Jeans length $\lambdaJeans$ is
locally larger than the grid resolution $\Delta x$.
This resolution requirement for grid codes can be expressed
in terms of a constraint on the ``Jeans number''
$J=\Delta x/\lambdaJeans<0.25$ to properly resolve
the Jeans scale and prevent such 
artificial fragmentation.

Similarly, the inability of SPH codes to capture pressure forces
properly on scales less than the smoothing length can also allow
for numerically-induced fragmentation if the smoothing length
is of order the Jeans length \citep{bate1997a}.  As discussed
by \cite{bate1997a} and in \S \ref{section:methodology:jeans_resolution},  
the requirement that the Jeans scale is resolved an SPH calculation by
some number $\NJeans$ of SPH kernels is $2\NJeans\Nneigh\mgas<\mJeans$,
where $\mgas$ is the SPH particle mass and $\Nneigh$ is the number
of SPH neighbors.
A similar requirement can be formulated in terms of relevant
spatial scales \citep[e.g.,][]{klein2004a}.
Owing to the
temperature dependence of the sound speed ($c_{s}\propto\sqrt{T}$ 
for ideal gas) and the dependence of the Jeans scale on 
density and temperature,
it is clear that for
a given
number of gas particles $\Ngas$ used in an SPH calculation, the
lower the temperature floor allowed the more potentially
susceptible the  
simulated fluid will be to numerical Jeans fragmentation.

Simulations of the ISM in galaxies with molecular coolants
must be especially cautious given the typical numerical
resolutions, high densities, and potentially low temperatures.
For example, for the typical conditions of the average ISM in a Milky Way-size
galaxies (number density of $\nH=1\,\rm cm^{-3}$, sound speed $c_s=10\,\rm km\,s^{-1}$), the Jeans mass is $m_{\mathrm{Jeans}} \sim 6.5 \times 10^7\,\rm M_{\odot}$, 
which is comparable to or smaller than the gas mass resolution in many recent 
simulations of disk galaxy formation. However, for molecular gas ($\nH\gtrsim100\,\rm cm^{-3}$) with a sound speed 
$c_s=1\,\rm km\,s^{-1}$ (corresponding to the temperature $T=100$~K), the Jeans mass
would be only $m_{\mathrm{Jeans}}\sim 6.5 \times 10^{3}\,\rm M_{\odot}$ --- far smaller than the gas 
mass resolution of the largest MW-sized galaxy formation simulations performed to date. 

For SPH calculations in the literature, typical approaches for
addressing numerical Jeans instability include improving the numerical
resolution such that the Jeans scale is resolved with $\NJeans=1$, as
suggested by \cite{bate1997a}, or using sink particles to represent
high density regions of the ISM
\citep[e.g.][]{li2005a,li2006a,booth2007a}.  Each of these approaches
have possible drawbacks, as the \cite{bate1997a} resolution criterion
can clearly depend on the density, depending on the gas
equation-of-state, and the sink particle method, depending on its
implementation, may neither avoid numerical fragmentation nor properly
capture the complicated chemical, thermodynamical, and radiative
processes that operate in dense ISM gas \citep[see,
e.g.,][]{commercon2008a}. Another possible solution is to use a
stiffer equation-of-state for the gas
\citep[e.g.,][]{springel2005b,schaye2007a}.

\begin{figure*}
\figurenum{14}
\epsscale{1.2}
\plotone{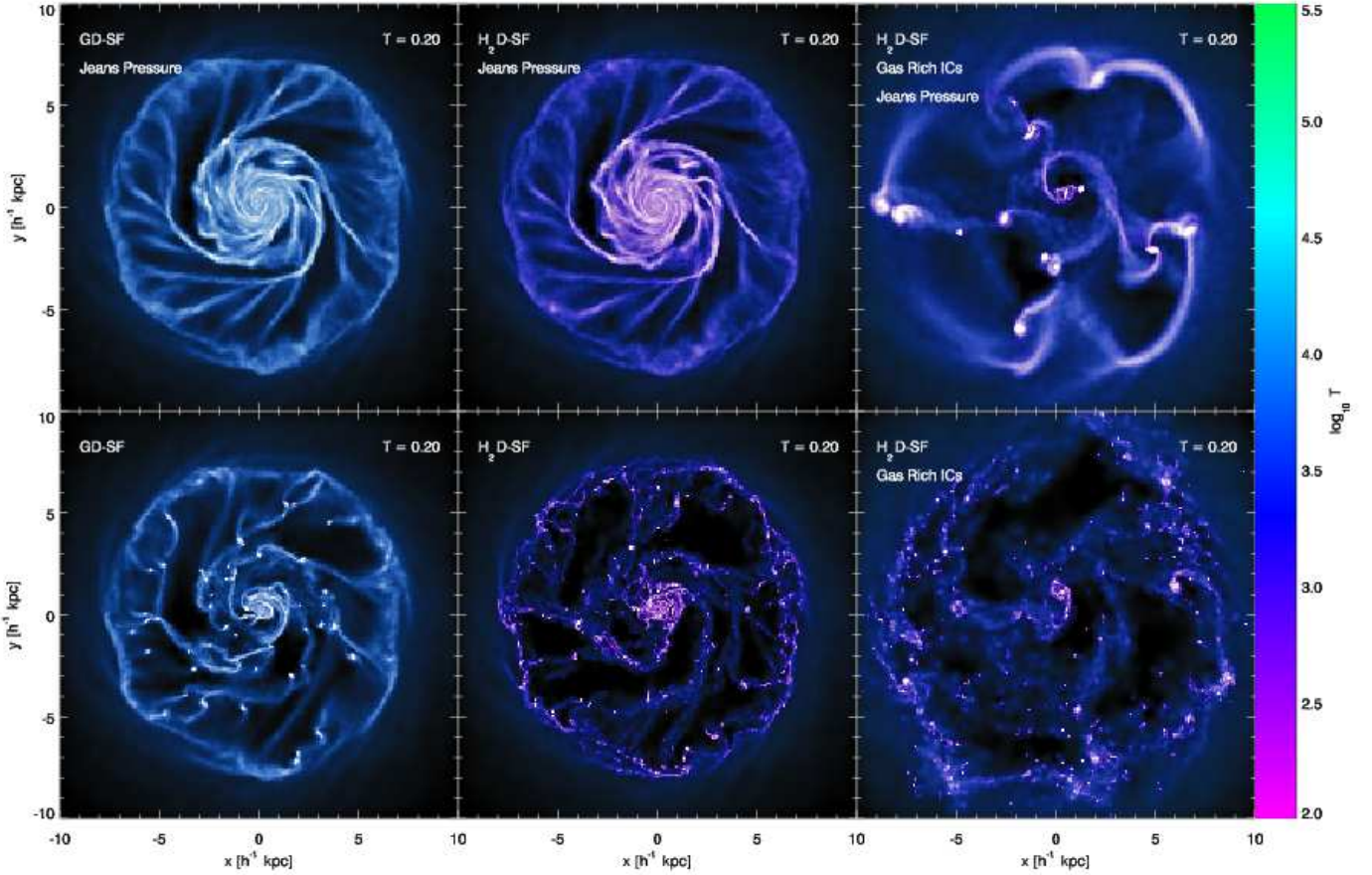}
\caption{\label{fig:fragmentation}
\small
Results of simulations of an isolated galaxy model with
a circular velocity of $\vc = 160 \kms$ 
evolved with (upper row) and without (lower row) a pressure floor to resolve the local
Jeans mass and length.  The image intensity tracks the surface density while the color reflects the gas
effective temperature.
For galaxy systems of this mass and disk structure with gas fractions of $\fgas=0.2$
and simulated with $N=400,000$ gas particles, the local Jeans mass in dense regions
is smaller than the typical mass in an
SPH kernel if ISM is modeled with either atomic cooling and a $T=10^{4}\K$ temperature floor (lower left
panel) or 
atomic and molecular cooling with a $T=10^{2}\K$ temperature floor (lower middle
panel).
These systems are susceptible to numerical fragmentation since, owing to the limited resolution,
the local gas pressure is artificially low on the scale where the gas is marginally stable against
collapse through self-gravity \citep{truelove1997a}.  To avoid this numerical artificiality, a pressure-floor
is introduced to increase the local sound speed until the Jeans scale is resolved by several SPH kernels.
Our prescription allows for galaxy models with either atomic cooling at temperatures $T>10^{4}\K$ (upper left panel) 
or atomic and molecular cooling at temperatures $T>10^{2}\K$ (upper middle panel) to be evolved stably.
Systems that are extremely gas-rich (e.g., $\fgas=0.99$, right panels) can still become globally unstable and
undergo large scale fragmentation if a pressure floor is used (upper right panel), but 
fragmentation occurs at much smaller scales and is qualitatively different
without such preventative measures (lower right panel).
}
\end{figure*}

Instead, the simulations presented in this paper adopt a pressure
floor on the scale of the Jeans length through an approach 
similar to that implemented by \cite{machacek2001a} in grid codes. 
\cite{machacek2001a} adopt an effective pressure $P_{\mathrm{eff}}$ on the scale
of the finest grid resolution $\Delta x$, given by
\begin{equation}
P_{\mathrm{eff}} = K G \rho_{b}^{2} \Delta x^{2}/\mu
\end{equation}
\noindent
where $\rho_{b}$ is the baryon density in the cell, $\mu$ is the
mean molecular mass, $G$ is the gravitational constant, and $K\approx100$
is a parameter that sets the effective resolution of their simulations.
If the effective pressure $P\propto \rho u$ on the scale of the
SPH kernel is set to ensure the Jeans scale is over-resolved by a factor
$\NJeans$, then the internal energy of gas particles with kernel 
mass $\mSPH\sim\Nneigh\mgas$
should be scaled as
\begin{equation}
 u = u \times \left\{ \begin{array} {c@{\quad:\quad}l}
\left(\frac{\NJeans}{\hJeans}\right)^{2/3} & \hJeans<\NJeans \\ 
\left(\frac{2\NJeans\mSPH}{\mJeans}\right)^{2/3} & \mJeans < 2\NJeans\mSPH
\end{array} \right.,
\end{equation}
\noindent
which is Equation \ref{equation:pressure_floor} from \S 
\ref{section:methodology:jeans_resolution}.  A factor $\NJeans=1$
would correspond to a density-dependent version of the
\cite{bate1997a} criterion, while applying the effective pressure
of \cite{machacek2001a} with $K\sim100$ and equating 
$\Delta x \sim h$ would imply $\NJeans\gtrsim1000$.  
\cite{klein2004a} provide a survey of the literature 
and find that values of $\NJeans\sim0.4-30$ for a fixed maximum density
are in common usage.

As briefly mentioned in \S \ref{section:methodology:jeans_resolution},
we have performed a variety of resolution studies where we examine the
range of $\NJeans=1-100$ and found $\NJeans\approx15$ to provide 
sufficient stability for the time interval in our simulations. 
Note, however, that more stringent criteria are likely to be required
for longer periods of evolution for galaxies in cosmological 
simulations. 
The resolution studies utilized a galaxy model initialized
with a virial velocity $\Vvir=160\kms$, spin $\lambda=0.033$, NFW
concentration $c=9$, disk mass fraction $m_{\mathrm{d}}=0.05$,
bulge mass fraction $m_{\mathrm{b}}=0.0167$, and disk gas fraction
$\fgas = 0.2$ according to the method described in 
\S \ref{section:methodology:galaxy_models}.  The dark matter halo,
bulge, stellar disk, and gaseous disk
were resolved with $N_{\mathrm{DM}}=120,000$, $N_{\mathrm{b}}=40,000$,
$N_{\mathrm{d}}=128,000$, and $N_{\mathrm{g}}=400,000$ particles
respectively, and the halo and baryonic particle gravitational
softenings were set to $\epsilon_{\mathrm{DM}}=100h^{-1}\pc$
and $\epsilon_{\mathrm{b}}=50h^{-1}\pc$.  The mass contained
within the SPH kernel with $\Nneigh=64$ for this galaxy
model was typically 
$\mSPH=2\times10^{6}\Msun$.
The galaxy model 
was evolved with two ISM models, the atomic cooling treatment
with a temperature floor of $T=10^{4}\K$ (the GD-SF model) and  
the atomic and molecular cooling treatment with a temperature floor of
 $T=10^{2}\K$ (the $\Hmol$D-SF model).
Figure \ref{fig:fragmentation} shows the disk galaxy
evolved with a density-dependent pressure floor 
described by Equation \ref{equation:pressure_floor} 
with $\NJeans=15$ (upper row)
and without a pressure floor (lower row) evolved in isolation
for $t=0.2\Gyr$.  The systems without a pressure floor
quickly fragment if the effective temperature floor is
either $T=10^{2}\K$ (lower middle panel) or  $T=10^{4}\K$ (lower
left panel), as the Jeans mass is initially not resolved at the
highest gas densities for either temperature.  By design, the
simulations that include the suggested pressure floor do not
fragment on the Jeans scale (upper row).
The pressurization on the Jeans scale used to ameliorate
artificial fragmentation does not insure that simulated
disks will avoid other instabilities, such as 
global \cite{toomre1964a} instability.  The right panels of
Figure \ref{fig:fragmentation} show the $\Vvir=160\kms$ disk
model with the disk gas fraction increased to $\fgas = 0.99$,
evolved for $t=0.2\Gyr$ in isolation.  
The gas rich model 
simulated without a pressure floor violently fragments,
but the gas rich model evolved
using a pressure floor also undergoes fragmentation on a
much larger mass scale.  As discussed in \cite{springel2003a}, \cite{robertson2004a}
and \cite{springel2005b}, systems like this gas rich model
that are
violently Toomre-unstable when using an ideal gas or 
weakly pressurized equation-of-state
can be stabilized by adopting a more
strongly pressurized
ISM equation-of-state, but we do not examine such
models here.

Separating artificial fragmentation and 
physical Jeans fragmentation owing to self-gravity may
be difficult without detailed resolution studies that
examine the convergence of the growth rate of 
perturbations on different size
scales.  
For Galaxy-mass systems with the disk structure we utilize
and gas fractions of $\fgas=0.2$
simulated with $N=400,000$ gas particles, the local Jeans mass in dense regions
is smaller than the typical mass in an
SPH kernel if either an ISM model with atomic cooling and a $T=10^{4}\K$ temperature floor or
(the GD-SF model) or
an ISM model with atomic + molecular cooling and a $T=10^{2}\K$ temperature floor
(the $\Hmol$D-SF model)
is used, by a factor of $\mJeans/2 \NJeans \mSPH\sim100$ for a $T=10^{4}\K$ temperature floor or
$\mJeans/2 \NJeans \mSPH>10,000$ for a $T=10^{2}\K$ temperature floor.  
The simulations that include a pressure floor maintain $\mJeans/2\NJeans\mSPH\approx1$
by design, with a small scatter induced by fluctuations in $\mSPH$ from changing
particle masses from star formation and the small allowed range of neighbors $\Nneigh=62-66$.
Given the values of the Jeans mass for the typical densities and temperatures of the ISM
($m_{\mathrm{Jeans}}\sim 10^3-10^7\,\rm M_{\odot}$, see above), our calculations thus suggest that simulating molecular cooling in 
Milky Way-sized disks of $M_{\rm gas}\approx 10^{10}\,\rm M_{\odot}$ 
with SPH \it without \rm using an artificial pressure 
floor would require 
$N\sim10^{9}-10^{10}$ gas particles to avoid numerical Jeans fragmentation if gas
is allowed to cool to $100$~K.  
We delay such resolution studies for future work (and faster computers).

Artificial fragmentation of the ISM in disks
may be closely related to the angular momentum problem for producing
realistic disk galaxies in cosmological simulations 
(see, e.g., \S 6 of \cite{robertson2004a} and \cite{robertson2006c}).  
The simulations of a Milky Way analogue
galaxy model presented in this appendix contain $\Ngas=400,000$,
which exceeds the numerical resolution of many disk
galaxies formed in cosmological simulations to date.
Both \cite{truelove1997a} and
\cite{bate1997a} comment on the possible implications of numerical 
Jeans fragmentation for studies of galaxy formation; based on 
these test simulations we
suggest that to reliably follow the formation and evolution of disks
using SPH simulations a pressure floor similar to our Equation 
\ref{equation:pressure_floor} should be used.
Prescriptions that mimic pressurization from the multiphase structure of the
ISM that may develop through feedback from star formation 
\citep[e.g.,][]{yepes1997a,hultman1999a,springel2000a,thacker2000a,springel2003a,stinson2006a,schaye2007a}
may also serve to ameliorate numerical Jeans fragmentation by raising the
sound speed as a function of the mean ISM density.
This affect is apparent in numerical tests examining the stability of isolated disks
by \cite{springel2005b} using various ISM equations-of-state.
We plan to revisit the
implications of numerical Jeans fragmentation on cosmological simulations
of galaxy formation in future work.
Lastly, we note that since disks that are Toomre-unstable are also
roughly Jeans-unstable \citep[see, e.g.,][]{krumholz2005a}, numerical Jeans fragmentation
will artificially reinforce any natural correspondence between gravitational instability
and star formation.  Simulations that examine the relation between star formation and
the Toomre instability parameter $\Qsg$ should therefore explicitly demonstrate that
the Jeans scale is properly resolved at all densities.

\appendix

As discussed in \S \ref{section:results:sfr_rotation}, the simulated galaxies
display a correlation between the SFR surface density $\SigmaSFR$
and the product of the total gas surface mass density $\Sigmagas$ and the angular frequency $\Omega$
in regions where the molecular gas and total gas densities are comparable.
The correlation in our simulations is similar to the observed $\SigmaSFR-\Sigmagas\Omega$ correlation 
\citep[e.g.,][]{kennicutt1998a, kennicutt2007a}.  Since the star formation relation implemented
in the simulations calculates the SFR density $\dot{\rho}_{\star}$
as proportional to the product of the molecular fraction $\fHmol$ and the gas density
scaled with the local dynamical time $\rhog^{1.5}$ (Equation \ref{equation:star_formation_rate}),
the $\SigmaSFR-\Sigmagas\Omega$ correlation would occur if the angular frequency scaled
with the density as $\Omega\propto\sqrt{\rhog}$.
Given that the gas density is a local quantity while orbital frequency
is determined by the overall mass distribution and potential of the galaxy,
it is not immediately clear why such a relation should hold. 
Here we show that the relation between $\rhog$ and $\Omega$ with a similar
scaling exists for self-gravitating exponential disks or exponential
disks embedded in realistic dark matter potentials.

Given the axisymmetry of the
disk galaxy models, the 
angular frequency is related to the density through Poisson's equation 
\begin{equation}
\frac{\dd^{2} \Phi}{\dd z^{2}} + \frac{\dd^{2} \Phi}{\dd R^2} + \Omega^{2} = 4\pi G \rho_{\mathrm{total}}. 
\end{equation}
\noindent
(Equation \ref{equation:poisson} in \S \ref{section:results:sfr_rotation}).
Below, the scaling of the angular frequency $\Omega$ with the local disk density $\rho_{\mathrm{disk}}$
is calculated in the limiting cases of potentials dominated by disks 
($\rho_{\mathrm{total}}\approx\rho_{\mathrm{disk}}$) or dark matter 
halos ($\rho_{\mathrm{total}}\approx\rho_{DM}$)
for exponential disks ($\rho_{\mathrm{disk}}\propto\exp[-R/R_{d}]$),
and we show that for the examined cases $\Omega\propto\sqrt{\rho_{\mathrm{disk}}}B(R/R_{d})$, where
$B(R/R_{d})\sim\mathcal{O}(1)$ is a weak function of the disk radius $R$ 
that depends on the form of the dominate potential.

\subsection{Exponential Disk-Dominated Potentials, $\rho_{\mathrm{disk}}>>\rho_{\mathrm{DM}}$}

For a flattened system, \cite{binney1987a}
recognized that as the vertical scale $h$ of
the density distribution decreases the 
first term on the left hand side of Equation
\ref{equation:poisson} begins to dominate
and eventually
\begin{equation}
\label{equation:poisson:vertical}
\frac{\dd^{2} \Phi}{\dd z^{2}} \simeq 4\pi G \rho.
\end{equation}
\noindent
If the surface density of the disk is exponential,
the axisymmetric density distribution of the disk
can be written as
\begin{equation}
\label{equation:disk_density}
\rho_{\mathrm{d}}(R,z) = \frac{A\Sigma_{0}}{R_{d}} \exp(-R/R_{d})f(z)
\end{equation}
\noindent
where $f(z)\propto\sech^{2}(z)$ is determined by Equation \ref{equation:poisson:vertical}, and the 
constant $A$ is determined by taking $f(0)=1$ and the integral
constraint from the disk mass.
Using the methods of \S 2.6.3 of \cite{binney1987a}, one can show that the 
rotation curve in the plane of a very thin exponential disk is
\begin{equation}
v_{c}^{2}(R) =  4\pi G \Sigma_{0}R_{d} y^{2} \left[I_{0}(y)K_{0}(y) - I_{1}(y)K_{1}(y)\right],
\end{equation}
\noindent
where $y\equiv R/R_{d}$ \citep{freeman1970a}.
Substituting for $\Sigma_{0}/R_{d}$ from Equation \ref{equation:disk_density}, as is
appropriate if the gas densities in the disk plane decrease in proportion to the 
surface density, we find
\begin{eqnarray}
\label{equation:omega:disk}
\Omega_{\mathrm{disk}}(R) &=& \left\{ 4\pi G A^{-1}\rho_{\mathrm{d}}(R,z) \exp(y)
\times \left[I_{0}(y)K_{0}(y) - I_{1}(y)K_{1}(y)\right] \right\}^{1/2}\\
&=& \sqrt{\rho_{d}} B_{\mathrm{disk}}(R/R_{d}).
\end{eqnarray}
\noindent
The function $B_{\mathrm{disk}}(x)$ is a weak function of the disk surface density $\Sigma(R)$.  Over the 
inner $95\%$ of the disk mass ($x\simeq\{0.355,4.75\}$), the fractional deviation from
the $\Omega\propto\sqrt{\rho_{d}}$ correlation, normalized to the disk properties at one 
scale length, is only $B(0.355)/B(1)-1=0.48$ near the disk center and 
$B(4.75)/B(1)-1=-0.26$ in the disk exterior.

\subsection{Halo-Dominated Potentials, $\rho_{\mathrm{DM}}>>\rho_{\mathrm{disk}}$}
In the case of $\rho_{\mathrm{total}}\approx\rho_{\mathrm{DM}}$, the angular frequency
can be determined directly from the integrated mass $M$ within a radius $R$ as
\begin{equation}
\label{equation:rotation:spherical}
\Omega^{2} = \frac{GM(<R)}{R{^3}}.
\end{equation}
\noindent
There are a wide variety of possible dark matter halo 
profiles, but we limit ourselves to three salient
models.

Dark matter halos in cosmological simulations on average 
follow
the \cite{navarro1996a} (NFW) dark matter halo profile, given by
\begin{equation}
\rho_{\mathrm{DM}}(R,z) = \frac{4\rho_{s}}{(\sqrt{R^2+z^2}/r_{s})(1+\sqrt{R^2+z^2}/r_{s})^{2}},
\end{equation}
\noindent
where $\rho_{s}$ is the density at the scale radius $r_{s}$.
If we substitute for the disk density
\begin{equation}
\rho_{s} = f_{d}^{-1} \rho_{d}(R,z) \exp(R/R_{d}-r_{s}/R_{d}),
\end{equation}
\noindent
then the angular frequency in the disk plane of a NFW profile
can be written
\begin{eqnarray}
\Omega_{\mathrm{NFW}}(R) &=& \sqrt{\rho_{d}}B_{\mathrm{NFW}}\left(R/r_{s}\right),
\end{eqnarray}
\noindent
where the function
\begin{eqnarray}
B_{\mathrm{NFW}}(x) &=& \left\{16 \pi G f_{d}^{-1}\exp\left(\frac{r_{s}(x-1)}{R_{d}}\right) 
\left[ \frac{ (1 + x)\ln(1+x) - x}{x^{3}(1+x)}\right]\right\}^{1/2}
\end{eqnarray}
\noindent
is, again, a weak function of the disk surface density $\Sigma(R)$.  Over the 
inner $95\%$ of the disk mass, the fractional deviation from
the $\Omega\propto\sqrt{\rho_{d}}$ correlation, normalized to the disk properties at one 
scale length, is only $B_{\mathrm{NFW}}(0.355R_{d}/r_{s})/B_{\mathrm{NFW}}(R_{d}/r_{s})-1=0.27$ near the disk center and 
$B_{\mathrm{NFW}}(4.75R_{d}/r_{s})/B_{\mathrm{NFW}}(R_{d}/r_{s})-1=1.45$ in the disk exterior.

Alternatively, we can consider the \cite{hernquist1990a} profile
\begin{equation}
\rho_{\mathrm{H}}(R,z) = \frac{8\rho_{a}}{(\sqrt{R^{2}+z^{2}}/a)(1+\sqrt{R^{2}+z^{2}}/a)^{3}},
\end{equation}
\noindent
where $\rho_{a}$ is the density at a scale radius $a$.  The 
\cite{hernquist1990a} profile is used as the DM halo profile
in the galaxy models presented in this paper and has a 
density and potential similar to the NFW profile within the scale radius.
Substituting for the disk density, we can write
\begin{equation}
\rho_{a} = f_{d}^{-1} \rho_{d}(R,z) \exp(R/R_{d}-a/R_{d}).
\end{equation}
\noindent
The mass profile of the \cite{hernquist1990a} halo provides
\begin{eqnarray}
\Omega_{\mathrm{H}}(R) &=& \sqrt{\rho_{d}} B_{\mathrm{H}}(R/a)
\end{eqnarray}
\noindent
where the function
\begin{eqnarray}
B_{\mathrm{H}}(x) = \left\{32 \pi G f_{d}^{-1}\exp\left[\frac{a(x-1)}{R_{d}}\right] 
\left[ \frac{1}{x(1+x)^{2}}\right]\right\}^{1/2}
\end{eqnarray}
\noindent
has a scaling nearly identical to the function $B_{\mathrm{NFW}}(x)$ when $x<<1$,
and is a correspondingly weak function of the disk surface density $\Sigma(R)$.  Over the 
inner $95\%$ of the disk mass, the fractional deviation from
the $\Omega\propto\sqrt{\rho_{d}}$ correlation, normalized to the disk properties at one 
scale length, is only $B_{\mathrm{H}}(0.355R_{d}/a)/B_{\mathrm{H}}(R_{d}/a)-1=0.26$ near the disk center and 
$B_{\mathrm{H}}(4.75R_{d}/a)/B_{\mathrm{H}}(R_{d}/a)-1=1.48$ in the disk exterior.

In real galaxies, it is thought that the dark matter halo will
adiabatically compress in response to the dissipational formation
of the disk \citep[e.g.,][]{blumenthal1986a}.
Models of adiabatic compression suggest that the dark matter 
halo response tends to drive the halo profile to resemble an
isothermal sphere over much of the disk \citep{gnedin2004a}.
The isothermal density profile can be written
\begin{equation}
\rho_{\mathrm{iso}}(R,z) = \frac{\rho_{0}}{(\sqrt{R^{2}+z^{2}}/r_{0})^{2}},
\end{equation}
\noindent
where $\rho_{0}$ is the density at an arbitrary scale radius
$r_{0}$.  For simplicity, we set $r_{0}=R_{d}$ and substitute
for the disk density as
\begin{equation}
\rho_{0} = f_{d}^{-1}\rho_{d}(R,0)\exp(R/R_{d}-1).
\end{equation}
Integrating the isothermal density profile to find the enclosed
mass provides the angular frequency in the disk plane as
\begin{equation}
\Omega_{\mathrm{iso}}(R) = \sqrt{\rho_{d}}B_{\mathrm{iso}}(R/R_{d})
\end{equation}
\noindent
where
\begin{equation}
B_{\mathrm{iso}}(x) = \left\{4\pi G f_{d}^{-1} x^{-2} \exp(x-1)\right\}^{1/2}.
\end{equation}
\noindent
The function $B_{\mathrm{iso}}(x)$ is also weak function of the 
disk surface density $\Sigma(R)$, and is of the same order as $B_{\mathrm{NFW}}$
or $B_{\mathrm{H}}$.  Over the 
inner $95\%$ of the disk mass, the fractional deviation from
the $\Omega\propto\sqrt{\rho_{d}}$ correlation, normalized to the disk properties at one 
scale length, is only $B_{\mathrm{iso}}(0.355)/B_{\mathrm{iso}}(1)-1=1.04$ near the disk center and 
$B_{\mathrm{iso}}(4.75)/B_{\mathrm{iso}}(1)-1=0.37$ in the disk exterior.

\begin{figure}
\figurenum{15}
\epsscale{0.75}
\plotone{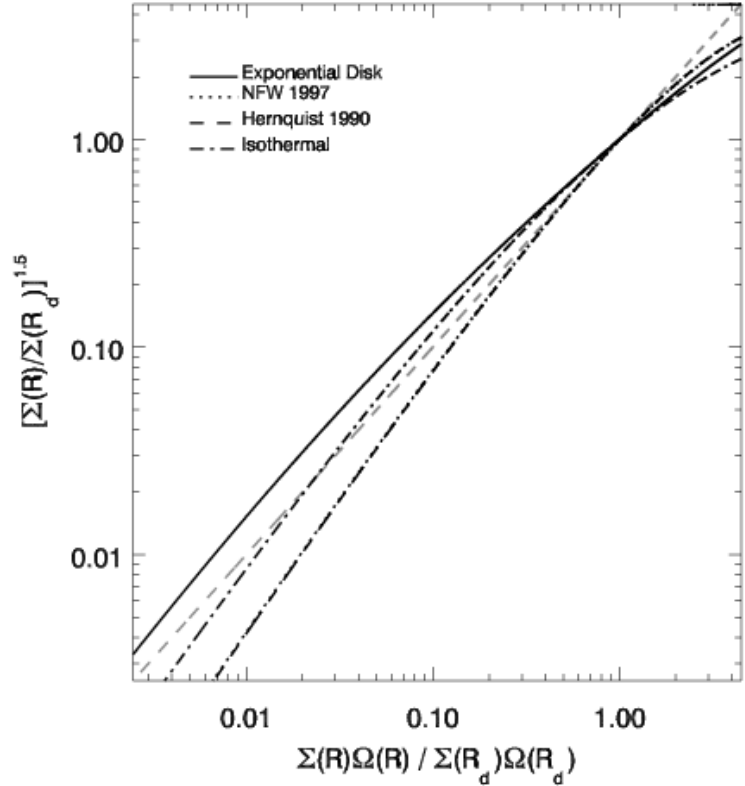}
\caption{\label{fig:rotation}
Relation between surface density $\Sigma$ and angular frequency $\Omega$
for a variety of disk galaxy model potentials. 
Shown is the quantity $\Sigma^{1.5}\propto\SigmaSFR$ plotted against
the product $\Sigma\Omega$ for exponential disks in a disk-dominated (solid
line), \cite{navarro1997a} halo-dominated (dotted line), 
\cite{hernquist1990a} halo-dominated 
(dashed line),
or isothermal halo-dominated (dash-dotted line) potential,
normalized to the disk properties at an exponential scale length.
Deviations from the $\Sigma^{1.5}\propto\Sigma\Omega$ correlation
(indicated with a dashed grey line) are typically $\sim\mathcal{O}(1)$
and are largest in the very center and exterior of the disk.  The
calculation suggests that the $\SigmaSFR\propto\Sigma\Omega$ may
reflect the scaling of the local density in exponential gas 
disks with the characteristic shape of the potentials that host them,
rather than an a fundamental connection between star formation 
and global disk processes.
\\
}
\end{figure}

\subsection{The $\SigmaSFR-\Sigma\Omega$ Correlation}
Figure \ref{fig:rotation} summarizes the behavior of the functions $\Omega_{\mathrm{disk}}$,
$\Omega_{\mathrm{NFW}}$, $\Omega_{\mathrm{H}}$, and  $\Omega_{\mathrm{iso}}$.
In the \cite{hernquist1990a} and \cite{navarro1996a} halo potential models, the disk scale
length-to-halo scale radius ratios were set to  $R_{d}/r_{s}=0.1$ appropriate
for observed spiral galaxies.
The quantity $\Sigma(R)^{1.5} \propto \SigmaSFR$ is plotted as a function of
the product $\Sigma(R)\Omega(R)$, normalized to the disk properties at the
scale length $R_{d}$ over the radial range $R=[0,5R_{d}]$, for the potentials
calculated in this Appendix.
The quantity $\Sigma\Omega$ closely tracks $\Sigma^{1.5}$ throughout the disk,
demonstrating that deviations from the
$\Omega\propto\sqrt{\rho_{d}}$ correlation are typically $\sim\mathcal{O}(1)$.
Hence, the $\SigmaSFR-\Sigma\Omega$ relation may be understood in terms of the
scaling of the local density in exponential disks with the angular frequency
of realistic galaxy potentials, rather than some more fundamental relation 
connecting star formation
processes in the ISM with global disk properties.  Regions of galaxies 
with large molecular gas fractions ($\SigmaHmol\sim\Sigmag$) are then expected
to follow the $\SigmaSFR-\Sigma\Omega$ relation, as demonstrated by our simulations
(see \S \ref{section:results:sfr_rotation}).
We note that while a different normalization would reduce the typical deviation
from the $\Sigma^{1.5}$ scaling, the high-density portion of the relation corresponding to the
region where the rotation curve is rapidly increasing should produce a shallower
relation than $\SigmaSFR\propto\Sigma\Omega$. Weak evidence for this may exist in
the available data \citep[e.g., the inner-disk region of Figure 9 of][]{kennicutt2007a}.

\end{document}